\documentclass[twocolumn,superscriptaddress,amsmath,amssymb,aps,prx,nofootinbib, 10pt,longbibliography,normalem]{revtex4-2}
  
\usepackage[normalem]{ulem}  
\usepackage{amsmath,amssymb,amsfonts,stmaryrd}
\usepackage{mathrsfs}
\usepackage[scr=boondox]{mathalpha}
\usepackage{color}
\usepackage{inputenc}
\usepackage[T1]{fontenc}
\usepackage{amsmath}
\usepackage{amssymb}
\usepackage{stmaryrd}
\usepackage{graphicx}
\graphicspath{ {./figures/} }
\usepackage{romannum}
\usepackage{physics}
\usepackage{comment}
\usepackage{mathtools}
\usepackage{import}
\usepackage{chris}
\usepackage{thmtools}
\usepackage{natbib}
\usepackage{algorithm}
\usepackage[noend]{algpseudocode}
\usepackage{ulem}
\usepackage{orcidlink}
\usepackage{suffix}

\definecolor{LightPink}{rgb}{0.858, 0.188, 0.478}
\definecolor{RED4}{rgb}{0.55,0,0}
\usepackage[unicode=true,bookmarks=true,bookmarksnumbered=false,bookmarksopen=false,breaklinks=false,backref=false,colorlinks=true]{hyperref}
\makeatletter

\@ifundefined{textcolor}{}
{
 \definecolor{BLACK}{gray}{0}
 \definecolor{WHITE}{gray}{1}
 \definecolor{RED}{rgb}{1,0,0}
 \definecolor{GREEN}{rgb}{0,1,0}
 \definecolor{BLUE}{rgb}{0,0,1}
 \definecolor{CYAN}{cmyk}{1,0,0,0}
 \definecolor{MAGENTA}{cmyk}{0,1,0,0}
 \definecolor{YELLOW}{cmyk}{0,0,1,0}
 \definecolor{navyblue}{rgb}{0.0, 0.0, 0.5}
}

\newcommand{\ra}{\rangle}
\newcommand{\la}{\langle}

\usepackage{amsfonts}
\usepackage{tabularx}
\usepackage{dcolumn}
\usepackage{bm}
\usepackage{graphicx}
\usepackage{epstopdf}

\usepackage[caption=false]{subfig}

\hypersetup{urlcolor=blue}

\usepackage{tensor}
\usepackage{braket}

\usepackage[capitalise,compress]{cleveref}
\usepackage{multirow}

\makeatother

\makeatletter
\newsavebox{\@brx}
\makeatother

\newcommand{\vi}{{\vec{\imath}}}%
\newcommand{\vj}{{\vec{\jmath}}}%

\begin{document}
\pagenumbering{arabic}

\hypersetup{linkcolor=black, urlcolor=blue, citecolor=blue, unicode=true}

\title{Disclinations, dislocations, and emanant flux at Dirac criticality}

\author{Maissam Barkeshli}
\affiliation{Department of Physics and Joint Quantum Institute, University of Maryland, College Park, MD, 20742, USA}

\author{Christopher Fechisin}
  \affiliation{Department of Physics and Joint Quantum Institute, University of Maryland, College Park, MD, 20742, USA}
  \affiliation{Joint Center for Quantum Information and Computer Science, NIST/University of Maryland, College Park, MD, 20742, USA}

  \author{Zohar Komargodski}
   \affiliation{Simons Center for Geometry and Physics, SUNY, Stony Brook, NY 11794, USA}
  \affiliation{C. N. Yang Institute for Theoretical Physics, Stony Brook University, Stony Brook, NY 11794, USA}

  \author{Siwei Zhong}
  \affiliation{Simons Center for Geometry and Physics, SUNY, Stony Brook, NY 11794, USA}
  \affiliation{C. N. Yang Institute for Theoretical Physics, Stony Brook University, Stony Brook, NY 11794, USA}

\date{\today}
\begin{abstract}
What happens when fermions hop on a lattice with crystalline defects? The answer depends on topological quantum numbers which specify the action of lattice rotations and translations in the low energy theory. One can understand the topological quantum numbers as a twist of continuum gauge fields in terms of crystalline gauge fields. We find that disclinations and dislocations -- defects of crystalline symmetries -- generally lead in the continuum to a certain ``emanant'' quantized magnetic flux. To demonstrate these facts, we study in detail tight-binding models whose low-energy descriptions are (2+1)D Dirac cones. Our map from lattice to continuum defects explains the crystalline topological response to disclinations and dislocations, and motivates the fermion crystalline equivalence principle used in the classification of crystalline topological phases. When the gap closes, the presence of emanant flux leads to pair creation from the vacuum with the particles and anti-particles swirling around the defect. We compute the associated currents and energy density using the tools of defect conformal field theory. There is a rich set of renormalization group fixed points, depending on how particles scatter from the defect. At half flux, there is a defect conformal manifold leading to a continuum of possible low-energy theories. We present extensive numerical evidence supporting the emanant magnetic flux at lattice defects and we test our map between lattice and continuum defects in detail. We also point out a no-go result, which implies that a single (2+1)D Dirac cone in symmetry class AII is incompatible with a commuting $C_M$ rotational symmetry with $(C_M)^M = +1$. 
\end{abstract}

{\begingroup
		\hypersetup{urlcolor=navyblue}
\maketitle
		\endgroup}

\tableofcontents

\section{Introduction}

A fundamental question in the study of quantum matter is to understand the effect of crystalline symmetries, such as lattice translations, rotations, and reflections, in quantum many-body systems. The past two decades have seen significant progress in our understanding of how crystalline symmetry can enrich the set of possible gapped quantum phases of matter (for a partial list of references see e.g.  \cite{wen2002quantum,hasan2010,barkeshli2012a,Essin2013SF,Benalcazar2014,ando2015,watanabe2015filling,Chiu2016review,Po2017symmind,song2017,Huang2017,Shiozaki2017point,Kruthoff2017TCI,Bradlyn2017tqc,schindler2018higher,Thorngren2018,Liu2019ShiftIns,Song2020,Li2020disc,manjunath2021cgt,manjunath2020FQH,Cano_2021,zhang2022fractional,zhang2023complete,zhang2023complete,zhang2022pol,manjunath2024Characterization,kobayashi2024crystalline}). These are typically referred to as crystalline topological phases and can be described by a topological quantum field theory (TQFT) in the infrared limit. The TQFT is characterized by crystalline topological invariants that determine the universal response to defects of the crystalline symmetry, such as lattice disclinations and dislocations. 

In this paper, we consider how crystalline symmetry enriches quantum critical points. We develop a general framework for studying crystalline defects at quantum criticality. As an example that is perhaps of most experimental interest, we focus on the case of two-dimensional lattice systems described at low energies by Dirac fermions. Along the way, we also show how our perspective can be used to explain many previously known facts about the gapped (insulating) phases, and the fermionic crystalline equivalence principle.

First, we explain how in the presence of an $M$-fold rotational symmetry, implemented by an operator $C_{M,o}$ about a fixed point $o$, each Dirac fermion has an angular momentum quantum number $s_o$, in a basis that diagonalizes $C_{M,o}$. If $(C_{M,o})^M = +1$, $s_o$ is quantized to a half-integer modulo $M$, whereas if $(C_{M,o})^M = (-1)^F$, where $F$ is the fermion number, then $s_o$ is quantized to an integer modulo $M$. The dependence of $s_o$ on $o$ is related to the lattice momentum of the Dirac point. 
More generally, we have a group homomorphism $\rho: G_\text{UV} \rightarrow G_\text{IR}$ from the lattice symmetry group $G_\text{UV}$ to the symmetry group of the IR quantum field theory, $G_\text{IR}$. The quantum numbers $\{s_o\}$ are part of the data parameterizing this group homomorphism. The constraints on $\{s_o\}$ provide an explanation of the fermionic crystalline equivalence principle for rotational symmetries, which has been used to understand crystalline topological phases of fermions~\cite{Thorngren2018,zhang2022real,manjunath2022mzm}. 

\begin{figure}[t]
    \centering
    \includegraphics[width=0.45\textwidth]{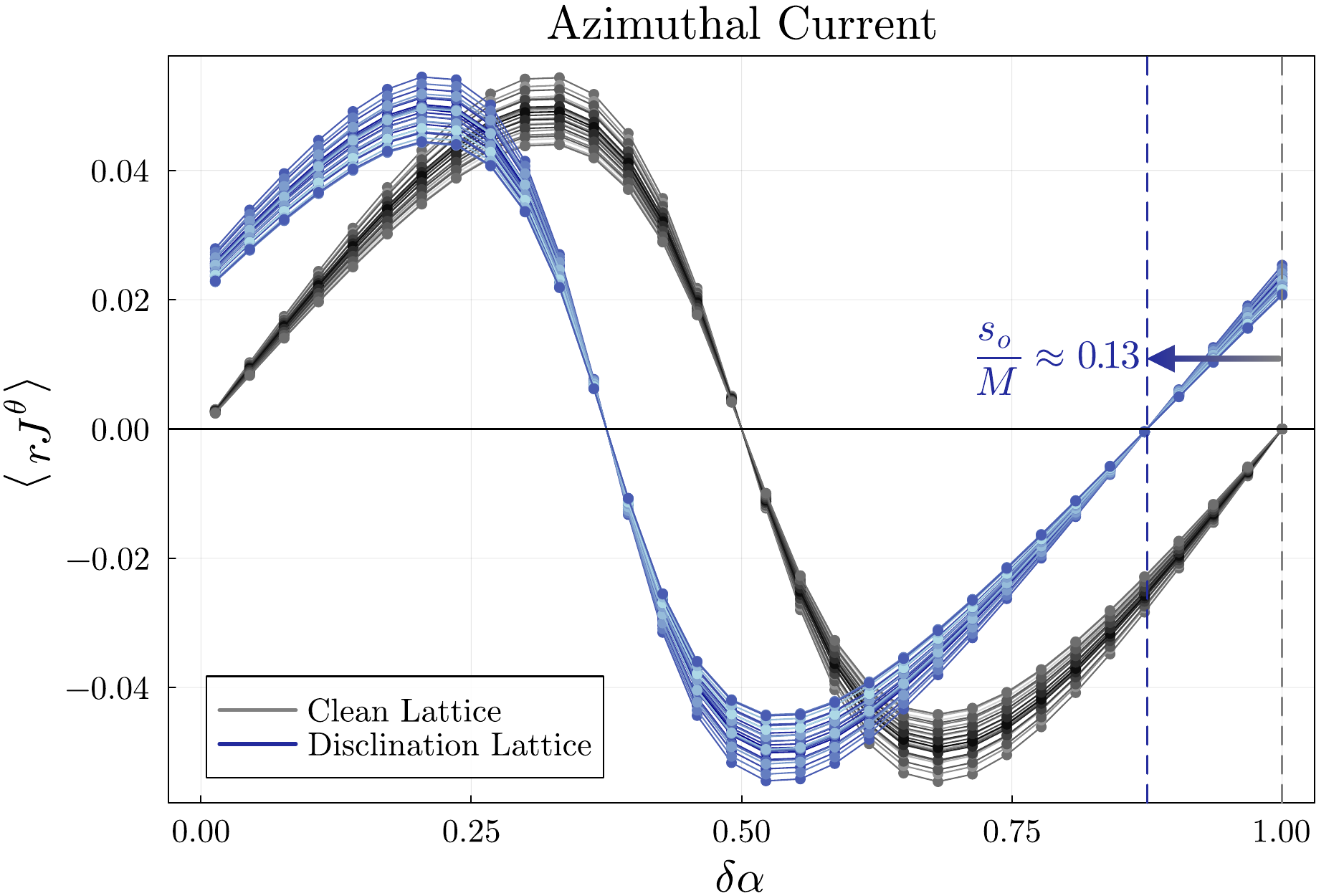}\\\vspace{0.7em}
\includegraphics[width=0.45\textwidth]{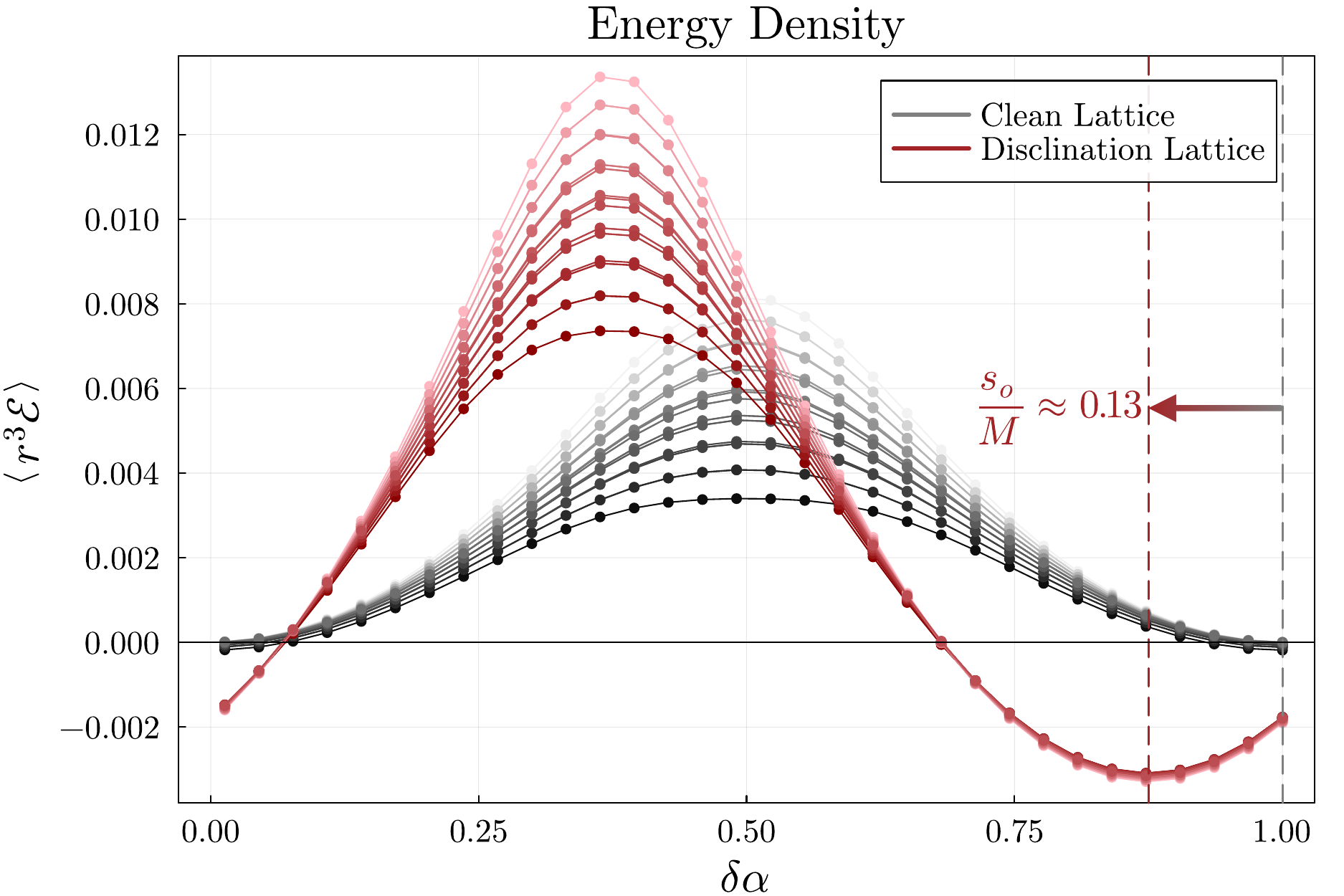}
    \caption{Observables measured in the QWZ model ground state at criticality on lattices with and without a crystalline defect, as a function of applied flux $\delta\alpha$ inserted at the rotation center. Notice that the defect lattice data is offset along the $\delta\alpha$ axis relative to the clean lattice data due to the \textit{emanant flux} generated by the crystalline defect. Each curve corresponds to a particular point in the bulk of the lattice, with lighter colors denoting curves closer to the rotation center. The magnitude of the clean lattice current (upper panel) is scaled for ease of comparison with the disclination lattice current. Remarkably, the magnitude of the minima and maxima of the energy density is larger with the disclination as compared to the clean case, which also matches the theoretical prediction in Sec. \ref{sec:universality_defects}.}
    \label{fig:intro_teaser}
\end{figure}

A central question that we address in this paper is how to detect $\{s_o\}$ and the group homomorphism $\rho$ more generally from universal physical observables at the critical point. The key point that enables this is that crystalline symmetry defects, labeled by the microscopic lattice symmetries $g \in G_\text{UV}$, should be modeled in the IR QFT by $\rho(g) \in G_\text{IR}$ defects, defined in terms of the symmetries of the infrared theory. In particular, we argue that this implies that lattice disclinations with disclination angle $2\pi/M$ should be modeled for each Dirac fermion in the IR by a conical defect with angle $2\pi/M$ together with a $U(1)$ holonomy $2\pi s_o/M$ (in the basis of Dirac fermions that diagonalizes the lattice rotation operator). We argue that lattice dislocations are modeled for each Dirac fermion in the IR by a $U(1)$ holonomy set by the momentum of the Dirac point and the Burgers vector (in the basis of Dirac fermions that diagonalizes the lattice translation operator).\footnote{Since one can always apply magnetic flux at the core of the lattice defect, it is important that the Hamiltonian in the presence of the lattice defect is determined, up to local operators at the core, by the lattice rotation and translation operators of interest.} 

Therefore, the continuum Dirac fermions see an \it effective \rm magnetic flux entirely due to the lattice defects, even if no magnetic flux is applied at the defect in the lattice model! We refer to this effective magnetic flux as an \it emanant \rm magnetic flux, following terminology introduced in \cite{cheng2023lieb}, for reasons we explain later. We demonstrate some consequences of this emanant flux, including particle creation from the vacuum, a universal shift in the dependence of the electric current and ground state energy density as a function of applied magnetic flux in the presence of the lattice defects (see Fig.\ref{fig:intro_teaser}), and various other interesting observables.

For gapped phases, we use our map from lattice to continuum defects to obtain topological invariants of crystalline systems. Indeed, in addition to the familiar Chern number of insulating bands, there are several topological invariants having to do with crystalline rotations and translations, leading to measurable physical effects, for instance, charge accumulation at a disclination \cite{manjunath2021cgt,zhang2022fractional,zhang2022pol}. We will show how several of the known topological crystalline invariants \cite{manjunath2021cgt,zhang2023complete} can be understood from well-known facts about continuum Dirac fermions together with the map $\rho$ from the lattice to the continuum. 

At the gapless point, the massless Dirac fermion obeys a conformal symmetry $so(3,2)$. Lattice defects flow to defect conformal theories, preserving the subalgebra $sl(2,\mathbb{R})\subset so(3,2)$. The continuum theory is therefore in the framework of Defect Conformal Field Theory (DCFT). (For recent reviews see~\cite{Billo:2016cpy, Andrei:2018die, soderberg2023defects, Chalabi:2023ohu}.) This applies to the long-distance limit of dislocations, disclinations, and Aharonov-Bohm fluxes. 

We characterize the $sl(2,\mathbb{R})$-invariant theory at long distances in all of these cases, focusing on its most important consequences. In particular, we study the particle-antiparticle creation due to a localized magnetic field,  which leads to a current and nonzero energy density in the vacuum, even away from the defect. This is to be contrasted with the Aharonov-Bohm effect, where non-relativistic particles and antiparticles circulate around the solenoid. Here particles are created from the vacuum and start circulating. The net particle number vanishes in our case in the Dirac field theory. The particle creation can be understood from the fact that to adiabatically turn on a localized magnetic field one has to create an electromotive force which will rip out electron-hole pairs from the vacuum and those will start circulating around the defect in opposite directions. 

An important quantity that needs to be specified is how the particles scatter from the defect. Generically, there are discrete choices that preserve the $sl(2,\mathbb{R})$ symmetry at long distances and correspond to fixed points of the RG. One special choice corresponds to a stable infrared fixed point and we expect it to agree with lattice models absent special fine tuning. 
An exception occurs at half flux, i.e., when the $U(1)$ flux is $\pi\ {\rm mod}  \ {2\pi}$. Then there is a continuum of choices of infrared stable fixed points with an exactly marginal operator (conformal manifold), see also~\cite{Bianchi:2021snj}. We remark that such a defect conformal manifold is particular for the Dirac criticality; it is absent, for example, for the free boson model studied in Appendix \ref{appendix_free boson model}. A conceptual way of understanding its appearance is by using the $g$-theorem~\cite{Cuomo:2021rkm}. It also has observable consequences for the fate of lattice model observables near $\pi$-flux, as we will discuss. 

By dialing the exactly marginal defect operator, the azimuthal current varies in a certain range while the energy density remains constant. 

Next, we develop the formalism of defect conformal field theory for conical singularities and discuss the absence of the displacement operator for such defects, even though they preserve $sl(2,\mathbb{R})$. Conical defects, unlike symmetry defects (monodromy defects) are therefore very slightly outside the usual framework of DCFT.

Matching results from the DCFT analysis to numerical calculations in critical tight-binding models, we are able to extract the quantum number $s_o$ of each Dirac fermion and verify our mapping between defects on the lattice and the continuum. We remark that the same methodology can be applied to extract the crystalline quantum numbers for systems whose UV completions are more complicated. 

\subsection*{Related work}

The emanant magnetic flux is reminiscent of the pseudomagnetic field that arises due to elastic strain in two-dimensional Dirac materials such as graphene \cite{castro2009}, except that for us it arises from topological defects of the lattice. The emanant magnetic flux induced by lattice disclinations has been noticed in earlier work on 
large fullerene molecules \cite{gonzalez1992continuum,gonzalez1993electronic}, graphitic cones \cite{krishnan1997graphitic,lammert2000topological,lammert2004graphene} and disordered graphene sheets \cite{gonzalez2001electron}. These works utilized this observation to compute electronic spectra and local density of states (LDOS) from continuum Dirac models. Ref. \cite{morpurgo2006intervalley} further used these observations, along with the emanant magnetic flux induced by lattice dislocations, to study weak localization in graphene. Our work builds on these results in the following ways:
\begin{itemize}
    \item We place the understanding of the emanant magnetic flux within a broader context that applies to strongly interacting systems. The emanant magnetic flux arising from disclinations and dislocations can be understood to be a consequence of the UV-IR homomorphism, which also exists for generic interacting systems beyond the single particle limit and other conformal fixed points. 
    \item As we explain, the fact that there must necessarily be a non-zero emanant magnetic flux at disclinations for certain theories is directly tied to the fact that the Dirac fermions have spin-1/2 while the lattice $2\pi/M$ rotational symmetry has order $M$, which is intimately related to the fermion crystalline equivalence principle.
    \item Our work focuses on identifying the universal observable consequences of the emanant magnetic flux and the UV-IR homomorphism, whereas the electronic spectra and LDOS are non-universal and depend on details of the defect core. In the gapped phases, this explains the crystalline topological invariants found in recent work. At the critical points, we identify the space of defect fixed points and RG flows between them for the simplest system of free Dirac cones. Our analysis extends existing results on the consequences of monodromy defects in conformal field theories. A collection of existing results about monodromy defects in conformal theories can be found in Refs.~\cite{Gaiotto:2013nva,Soderberg:2017oaa,Giombi:2021uae,Bianchi:2021snj,Gimenez-Grau:2021wiv,Dowker:2022mac, Dowker:2022mex, SoderbergRousu:2023pbe}. Our presentation of the subject will be self-contained and review some of the pertinent results.   
\end{itemize}

\section{UV and IR symmetries}\label{sec:UV_IR_Symm}

Let us consider a UV theory which could be a many-body lattice system or a QFT with relevant deformations. In the limit of long wavelengths, we assume an IR QFT describes the low-energy physics of this UV theory. A precise treatment of the RG flow that connects UV and IR is often too difficult in strongly coupled systems. Instead, in this section, we discuss how symmetries in the UV and IR are matched.

\subsection{Generalities}

Suppose $G_{\text{UV}}$ denotes the spatial and internal group symmetries of the UV theory, while $G_{\text{IR}}$ denotes that of the IR QFT. Then there is a group homomorphism
\begin{align}
\label{eq_def homomorphism}
    \rho: G_{\text{UV}} \rightarrow G_{\text{IR}} ,
\end{align}
which determines how symmetries in the UV map to symmetries in the IR. One has to define what the IR QFT is -- we will be working at a gapless point or its immediate neighborhood, so that only very light modes are kept and $G_\text{IR}$ is well-defined.

A corollary of this is that we have a map between symmetry defect configurations. Given a theory with a symmetry $G$, we can consider deforming the theory to create symmetry defects, labeled by $g \in G$. For example, these physically can correspond to inserting magnetic flux of a background gauge field, or creating lattice disclinations or dislocations. The group homomorphism $\rho$ can be thought of as determining a collection of RG flows, where the UV system with a symmetry defect $g \in G_\text{UV}$ is described in the IR QFT with a symmetry defect $\rho(g) \in G_\text{IR}$.
For completeness let us define what a symmetry defect is. Given a field theory with a symmetry $G$, whose elements $g\in G$ label topological operators, we can create a defect by taking the topological operator to be a semi-infinite plane and orienting the boundary in the time direction. Then in space, we have a topological line that ends at a point. At that point, there is curvature for the background gauge field. The idea of ``topological operators'' is more difficult to define for lattice quantum many-body systems. However, we can still define symmetry defects by introducing boundary conditions on the lattice where lattice fields are forced to undergo a symmetry transformation as they travel around the defect. 

Let us consider the case where the IR QFT is a fixed point of the RG flow, and $G_\text{IR}$ is the symmetry group of the fixed point QFT. In many cases, such as those of interest in this paper, there is a discrete set of possible choices of the group homomorphism $\rho$. In these cases, $\rho$ can be viewed as a \it topological invariant \rm of the UV theory, since it cannot be changed by local perturbations of the system without encountering a singularity in the ground state energy that fundamentally alters the nature of the RG flow. If we have two UV Hamiltonians $H_1$ and $H_2$ corresponding to two different homomorphisms $\rho_1$ and $\rho_2$, then $H_1$ cannot be continuously deformed into $H_2$ without encountering a singularity in the ground state energy. In the case where the IR QFT is critical, $\rho$ therefore parameterizes topologically distinct ways that quantum critical points can be enriched with the symmetry $G_\text{UV}$. We refer to these as distinct \it symmetry-enriched \rm quantum critical points.\footnote{More generally, deformation classes of $\rho$ should be viewed as invariants of the UV theory.} 

While it is not of direct concern for us in this paper, we note that in general $\rho$ is the bottom layer of a more complicated mathematical structure. It is expected that quantum field theories in $D$ space-time dimensions have a generalized symmetry that forms the structure of a higher category. Therefore, in general we expect that the UV and IR symmetries form the structure of higher categories, $\mathcal{G}_\text{UV}$ and $\mathcal{G}_\text{IR}$, and we have a higher functor between these higher categories, $\mathcal{G}_\text{UV} \rightarrow \mathcal{G}_\text{IR}$. These ideas are currently a topic of intense research and have only been fully developed in special cases. For example, (2+1)D topological phases of matter with symmetry group $G = G_\text{UV}$ can be described using $G$-crossed modular categories, which can be viewed as a 3-functor between 3-categories; for details see \cite{barkeshli2019,jones2020}. As another example, there is a fusion 2-category associated to (1+1)D quantum field theories, which is the fusion 2-category of topological defects of the bulk (2+1)D topological quantum field theory that hosts the (1+1)D theory at its boundary.

We now comment on a few basic properties of $\rho$. When $\rho$ has a non-trivial kernel, some UV internal symmetries become trivial in the low-energy 
effective theory. Physically, this can happen when massive degrees of freedom decouple from the spectrum, whose associated symmetries become invisible to the IR QFT. When $\rho$ has a non-trivial cokernel, it could be the case that the IR QFT is endowed with \it emergent \rm (accidental) symmetries, where symmetry-breaking operators are irrelevant and suppressed at low-energy. The \it emanant \rm symmetries studied in \cite{cheng2023lieb} provide an example where $\rho(g)$ and $g$ generate distinct groups; such symmetries are exact in the IR QFT even though the group generated by $\rho(g)$ may not be a subgroup of $G_{\text{UV}}$; the group generated by $\rho(g)$ is said to \it emanate \rm from the symmetry group generated by $g$. Such examples are common for lattice translation symmetries.

The primary focus of this paper is on cases where the UV theory is a fermion lattice model, and the IR theory consists of $N$ free massless or very light Dirac fermions. We thus have a symmetry group $G_{\text{UV}}^\text{f}$ that acts faithfully on fermionic operators. Let $\mathbb{Z}_2^f$ be the order-2 group generated by fermion parity $(-1)^F \in \mathbb{Z}_2^f$. The bosonic symmetry group is such that $G_{\text{UV}}^\text{b} \equiv G_{\text{UV}}^\text{f}/ \mathbb{Z}_2^\text{f}$. More specifically, we are concerned with lattice models with UV symmetry group
\begin{equation}
G_{\text{UV}}^\text{f} = U(1)^\text{f} \times (\mathbb{Z}^2 \rtimes \mathbb{Z}_M) ,  
\end{equation}
which is relevant for describing charged fermions on a lattice. $U(1)^\text{f}$ refers to a central extension of the bosonic global symmetry $U(1)$ by $\mathbb{Z}_2^\text{f}$. For the spatial symmetries, $\mathbb{Z}_M$ is the $M$-fold lattice rotational symmetry group, and $\mathbb{Z}^2$ is the lattice translation group. Physically, we have magnetic translations that commute up to a $U(1)$ transformation. 

We remark that lattice reflections are beyond the scope of this paper. Defects of reflection symmetry require non-local operations in the UV theory and require us to study the theory on non-orientable manifolds \cite{barkeshli2019tr}. The action of the crystalline reflection symmetries in the IR theory also contains several additional subtleties (see e.g. \cite{manjunath2022mzm,pace2024gauging}), which we leave for future work.

\subsection{Dirac fermion quantum numbers}
\label{sec:dirac_sym}

We consider a theory of $N$ massless (2+1)D Dirac fermions as the IR theory. For simplicity, we assume that they all have the same velocity (which we set to 1). The bosonic symmetry group is then 
\begin{align}
    G_\text{IR}^b = \frac{U(N)}{\mathbb{Z}_2} \times SO(3,2) ,
\end{align}
where $SO(3,2)$ is the conformal group associated to (2+1)D Minkowski space. The fermionic symmetry group is a $\mathbb{Z}_2^f$ central extension:
\begin{align}
    G_\text{IR}^f = [U(N)^f \times Spin(3,2)]/\mathbb{Z}_2,
\end{align}
where $U(N)^f$ refers to the group $U(N)$ that acts faithfully on the fermions and the $-1$ element corresponds to fermion parity. To obtain the bosonic symmetry group, we need to mod out by fermion parity, which leads to the bosonic internal symmetry group $U(N) / \mathbb{Z}_2$. Note that the IR theory also may contain charge conjugation, reflection, and/or time-reversal symmetries, which we ignore in this discussion.

For simplicity, we assume that the $U(1)$ symmetry in the UV acts as a $U(1)$ symmetry in the IR where each Dirac fermion carries charge $1$. If the UV model is a free fermion lattice model, this is the only possibility. If it includes interactions, then each IR Dirac fermion in principle could carry any odd integer charge.\footnote{Note that for concreteness we are discussing models with no dynamical gauge fields in the IR description.}

Let us consider a translation operator $T_{\vec{v}}$ acting on a Dirac fermion operator $\Psi=(\Psi_1,...,\Psi_N)^T$, where $\vec{x}$ is the spatial coordinate and $\Psi_i$, $i=1,\cdots, N$, is a single 2-component Dirac fermion. We have: 
\begin{align}
    T_{\vec{v}} : \Psi(t,\vec{x}) \rightarrow U_{T_{\vec{v}} }\Psi(t,\vec{x}+\vec{v}),
\end{align}
where $U_{T_{\vec{v}}} \in U(N)^f$. Lattice dislocations investigated in this paper are created by $|\vec{v}|$ on the order of the lattice spacing, which is taken to zero in the continuum limit. As we will discuss in subsequent sections, the long-distance physics of dislocations is therefore solely determined by $U_{T_{\vec{v}}}$. For a given lattice model, the $N$ Dirac cones will occur at momenta $\{\vec{k}^{(i)}_{\star}\}$ in the Brillouin zone. We can thus expect
\begin{align}
(U_{T_{\vec{v}}})_{ij} = \delta_{ij} e^{i \vec{k}_\star^{(i)} \cdot\vec{v}}~.
\end{align}

Now let us consider the action of $2\pi/M$ rotations around a high symmetry point $o$ in the lattice, implemented by an operator $C_{M, o}$ in the UV theory with $(C_{M,o})^M = 1$. We have:
\begin{align}
    C_{M,o} : \Psi(t,\vec{x}) \rightarrow U_{M,o} \Lambda(2\pi/M) \Psi(t,\vec{x'}),
\end{align}
where $\vec{x'}$ is obtained from $\vec{x}$ by a $2\pi/M$ rotation about the fixed point $o$. Here $\Lambda(2\pi/M)$ is a $2\pi/M$ spatial rotation in the IR field theory, which generates the spatial rotation group $SO(2)^f \subset Spin(3,2)$. $U_{M,o} \subset U(N)$ is an internal rotation. 

Therefore, we see that $C_{M,o}$ acts in the IR theory as a combination of the field theory spatial rotation together with an internal $U(N)$ rotation. This is sometimes referred to as a ``twisted rotation.'' Since the Dirac fermion is a spin-1/2 representation of the space-time symmetry group of the field theory, we have that $(\Lambda(2\pi/M))^M  = -1$ on the Dirac fermion. Since $(C_{M,o})^M = 1$, it follows that $(U_{M,o})^M = -1$. 
If we instead consider a $2\pi/M$ rotation that satisfies $(C_{M,o})^M = (-1)^F$, then we would have $(U_{M,o})^M = 1$. That is, the $(C_{M,o})^M$ would then have order $2M$, and act in the IR via an internal $U(N)$ operation that has order $M$. 

In general $C_{M,o}$ does not commute with translations. We can find a basis $\tilde{\Psi}$ where $C_{M,o}$ is diagonal:
\begin{align}
\label{eq_def topo number s}
    C_{M,o} : \tilde{\Psi}_i \rightarrow e^{2\pi i s_o^{(i)}/M} \Lambda(2\pi/M) \tilde{\Psi}_i.
\end{align}
We refer to the quantum numbers $s_o^{(i)}$ as the orbital angular momenta, which are defined modulo $M$. From the above discussion, we see that $s_o^{(i)}$ are necessarily half-odd integer in the case we consider, where $(C_{M,o})^M = 1$. If instead we consider rotation symmetry operators where $(C_{M,o})^M = (-1)^F$, then $s_o^{(i)}$ are integers. 

Rotations about different fixed points, such as vertex-centered vs plaquette-centered fixed points, are related by translations:
\begin{align}
    C_{M,o+\vec{v}} = T_{\vec{f}(\vec{v})} C_{M,o},
\end{align}
where $\vec{f}(\vec{v}) = (1 - R(2\pi/M)) \vec{v}$ and $R(2\pi/M)$ is a rotation by angle $2\pi/M$. If the Dirac fermion $\Psi_i$ occurs at a momentum in the Brillouin zone that is invariant under $2\pi/M$ rotations, then there is a simple relationship between the orbital angular momentum for rotations about different high symmetry points:
\begin{align}
\label{eq:so_dependence}
    s_{o+\vec{v}}^{(i)} = s_{o}^{(i)} + \frac{M}{2\pi}\vec{f}(\vec{v}) \cdot \vec{k}_\star^{(i)} \mod M.
\end{align}

\subsection{Relation to fermionic crystalline equivalence principle}

In the study of crystalline topological phases of matter, it has been pointed out that topological phases of fermions with a symmetry group $G$ that contains spatial symmetries can be effectively classified using the formalism for internal symmetries with a symmetry $G_{\text{eff}}$ \cite{Thorngren2018,debray2021invertible,zhang2022real,manjunath2022mzm}. The observation is that $M$-fold spatial rotation symmetries satisfying $(C_M)^M = +1$ should be described using internal symmetries that have order $2M$. Conversely, spatial rotation symmetries satisfying $(C_M)^M = (-1)^F$ should be described in terms of internal symmetries that have order $M$. The above discussion provides an explanation of this correspondence. Namely, if we want to classify gapped phases of fermions using topological quantum field theory, then the spin-statistics theorem tells us that the fermions in the TQFT should have half-odd integer spin under the continuous rotational symmetry of the IR TQFT. The discussion above then shows us how the $C_M$ operation must come with an additional internal symmetry that has order $2M$ in order to satisfy the relation $(C_M)^M = 1$. And, correspondingly, for the case where $(C_M)^M = (-1)^F$, the additional internal symmetry operation must have order $M$. Classifying topological phases with a $C_M$ rotational symmetry can then be done by classifying ways in which the corresponding internal symmetry acts in the IR theory. 

We can deduce that the crystalline equivalence principle could then break down in situations where the IR theory cannot be Lorentz invariant, for instance for gapped fracton models \cite{nandkishore2019fractons}, Fermi liquids, or anisotropic quantum critical points. 

Later we will see that using the twisted action of the lattice translations and lattice rotation on the continuum fermions, we will be able to derive several known topological invariants associated to crystalline symmetries. This is an important facet of the fermion crystalline equivalence principle that we can now illuminate.

\subsection{No-go results}

The existence of the group homomorphism $\rho$ can place strong constraints on UV completions for IR theories that are Lorentz invariant and contain fermions. Below we discuss two examples. 

\subsubsection{Constraints from time-reversal symmetry}

Time-reversal symmetry introduces additional constraints on the symmetry actions described above. For example, consider the symmetry group
\begin{align}
\label{Guvf1}
    G^f_\text{UV} = \frac{U(1)^f \rtimes \mathbb{Z}_4^{\mathcal{T},f}}{\mathbb{Z}_2} \times [\mathbb{Z}^2 \rtimes \mathbb{Z}_M]
\end{align}
Here $\mathbb{Z}_4^{\mathcal{T},f}$ is an order-4 group generated by time-reversal $\mathcal{T}$ such that $\mathcal{T}^2 = (-1)^F$, and the equivalence relation identifies the fermion parity elements from the $U(1)^f$ and $\mathbb{Z}_4^{\mathcal{T},f}$ factors. This can be thought of as symmetry class AII in the Altland-Zirnbauer classification, supplemented with the additional wallpaper group symmetry $\mathbb{Z}^2 \rtimes \mathbb{Z}_M$. For such a symmetry group, time-reversal and spatial rotations commute, $\mathcal{T} C_{M,o} = C_{M,o} \mathcal{T}$. 

Consider the case of a single Dirac cone. Then, from the above discussion, since $s_o \neq 0$, it follows that $\mathcal{T}$ and $C_{M,o}$ cannot commute. We see that a single Dirac cone is necessarily incompatible with the symmetry of Eq. \ref{Guvf1}. A similar argument shows that any odd number of Dirac fermions is inconsistent with commuting $\mathcal{T}$ and $C_{M,o}$ symmetries. 

If we instead have a spatial rotation symmetry $(C_{M,o})^M = (-1)^F$, then it is possible to have $s_o = 0$, in which case $C_{M,o}$ and $\mathcal{T}$ commute. 

Since a single Dirac fermion in class AII occurs at the surface of a (3+1)D topological insulator, we can further conclude that topological insulators in this symmetry class are incompatible with an order-$M$ $C_M$ rotational symmetry that commutes with $\mathcal{T}$. Instead, for $C_M$ to commute with $\mathcal{T}$, such systems must have a ``spinful'' spatial rotation symmetry, where $(C_M)^M = (-1)^F$. This result gives a more abstract perspective on the necessity of spin-orbit coupling for topological insulators \cite{hasan2010,qi2010RMP}, since spatial rotations must necessarily be supplemented with internal symmetry operations. 

\subsubsection{IR fixed points without internal symmetries}

Consider an IR theory containing half-odd-integer spin fermions. The above discussion shows that such a theory can only be compatible with an order-$M$ $C_M$ rotational symmetry in the UV ($(C_M)^M = 1$) if it contains a $\mathbb{Z}_{2M}^f$ internal symmetry in the IR. In other words, QFTs with half-odd-integer spin fermions without a $\mathbb{Z}_{2M}^f$ internal symmetry are fundamentally incompatible with a UV completion containing an $M$-fold rotational symmetry with $(C_M)^M = 1$. Any rotational symmetry in the UV would instead have to satisfy $(C_M)^M = (-1)^F$.  

\section{Symmetry eigenvalues in free fermion lattice models}\label{sec:symm_eigenvalues}

Here we review how to determine the UV-IR homomorphism described above for lattice models of free fermions. We also present an example, the QWZ model, which we use to numerically test our ideas throughout the rest of the paper. 

A free fermion lattice model is governed by a Hamiltonian
\begin{align}\label{eq:free_fermion_ham}
    H = \sum_{\vi,\vj,a,b} h_{\vi\vj}^{ab}c_{\vi,a}^\dagger c_{\vj, b} ,
\end{align}
where $\vi$, $\vj$ are 2-component vectors denoting lattice sites, $a,b$ denote flavor indices (which could stand for e.g. orbital, layer, spin), and $h_{\vi\vj}^{ab}$ is the first-quantized Hamiltonian. A $2\pi/M$ (magnetic) rotational symmetry about a high symmetry point $o$ acting on the fermion operators takes the general form 
\begin{equation}
    C_{M,o} c_{\vj,a} C_{M,o}^\dagger = e^{i\lambda_{\vj}^{M,o}}\sum_b V_{M,o}^{a,b} c_{R_{2\pi/M}^o(\vj),b},
\end{equation}
where $\lambda_{\vj}^{M,o}$ encodes a possible  $U(1)$ phase that is location dependent, $V_{M,o}$ encodes a rotation in the flavor space, and $R_{2\pi/M}^o(\vj)$ gives the lattice site obtained by rotating $\vj$ by $2\pi/M$ about the origin $o$. Note that we consider $( C_{M})^M=1$ by definition, so that the magnetic rotation generates a $\mathbb{Z}_{M}$ symmetry.

Similarly, a (magnetic) translation $ T_{\vec{v}}$ by the vector $\vec{v}$ is defined via
\begin{equation}
     T_{\vec{v}} c_{\vj,a} T_{\vec{v}}^\dagger = e^{i\lambda_{\vj}^{T_{\vec{v}}}}\sum_{b}V^{a,b}_{T_{\vec{v}}}c_{\vj+{\vec{v}},b},
\end{equation}
where $\lambda_{\vj}^{T_{\vec{v}}}$ encodes a possible position dependent $U(1)$ transformation and $V_{T_{\vec{v}}}$ encodes a rotation in the flavor space. These transformations, along with $U(1)$ charge conservation, generate the symmetry group of the lattice model on the infinite two-dimensional Euclidean plane,
\begin{equation}
    G_\text{UV}^f=U(1)^f\times(\Z^2\rtimes \Z_M),
\end{equation}
which we discussed extensively in \cref{sec:UV_IR_Symm}. In general this group must be modified to take into account the non-commutativity of translations when the unit cell contains nonzero total flux, but we will not encounter this complication in this work.

Given a free fermion model with symmetry $C_{M,o}$, which can be described at low energies by $N$ Dirac cones centered at momenta $\vec{k}_\star^{(i)}$, for $i = 1,\cdots, N$, we can explicitly determine the UV-IR homomorphism described above by determining the action of the symmetry operators on the single-particle eigenstates at the Dirac crossings. 
For example, for a critical point with a single Dirac cone, at $\vec{k}_\star$ we have a two-dimensional space of states, with corresponding annihilation operators $\Psi_{\vec{k}_\star} = (\psi_{\vec{k}_\star, \uparrow}, \psi_{\vec{k}_\star, \downarrow})$. We can pick a basis such that 
\begin{align}\label{eq:C_tranform_k_space}
    C_{M,o}: \Psi_{\vec{k}_\star} \rightarrow e^{i\frac{2\pi}{M}s_o} \Lambda(2\pi/M)
     \Psi_{\vec{k}_\star} ,
\end{align}
with $ \Lambda(2\pi/M) = e^{i \frac{1}{2} \frac{2\pi}{M} \sigma^z}$.
In other words, the eigenvalues of $C_{M,o}$ are given by $e^{i \frac{2\pi}{M}(s_o \pm 1/2)}$. Therefore the $s_o$ are directly related to the crystal angular momentum of the zero energy states at the Dirac crossing. If there are $N$ Dirac cones at a critical point, 
the procedure is analogous. We can pick a basis for the Dirac cones that diagonalizes $C_{M,o}$, such that $C_{M,o}$ acts by the $2N$-dimensional matrix $\oplus_{i=1}^N e^{i\frac{2\pi}{M_o}s_o^{(i)}}\Lambda(2\pi/M)$. 

Similarly, we can determine the invariants $\vec{k}_\star^{(i)}$ by measuring the eigenvalues of the translation operators along the primitive lattice vectors. For a single Dirac cone, we have 
\begin{equation}
    T_{\vec{v}}:\Psi_{k_\star}\rightarrow e^{i \vec{k}_*\cdot \vec{v}}\Psi_{\vec{k}_\star}
\end{equation}
The generalization to multiple fermions is straightforward.

While it is straightforward to determine $s_o^{(i)}$ and $\vec{k}_\star^{(i)}$ numerically given any free fermion lattice model, the primary purpose of this work is to understand its physical consequences and how it can be measured through physical observables. 

\subsection*{Example: QWZ model}\label{sec:QWZ_rot_eigs}

\begin{figure}[t]
\includegraphics[width=0.55\linewidth]{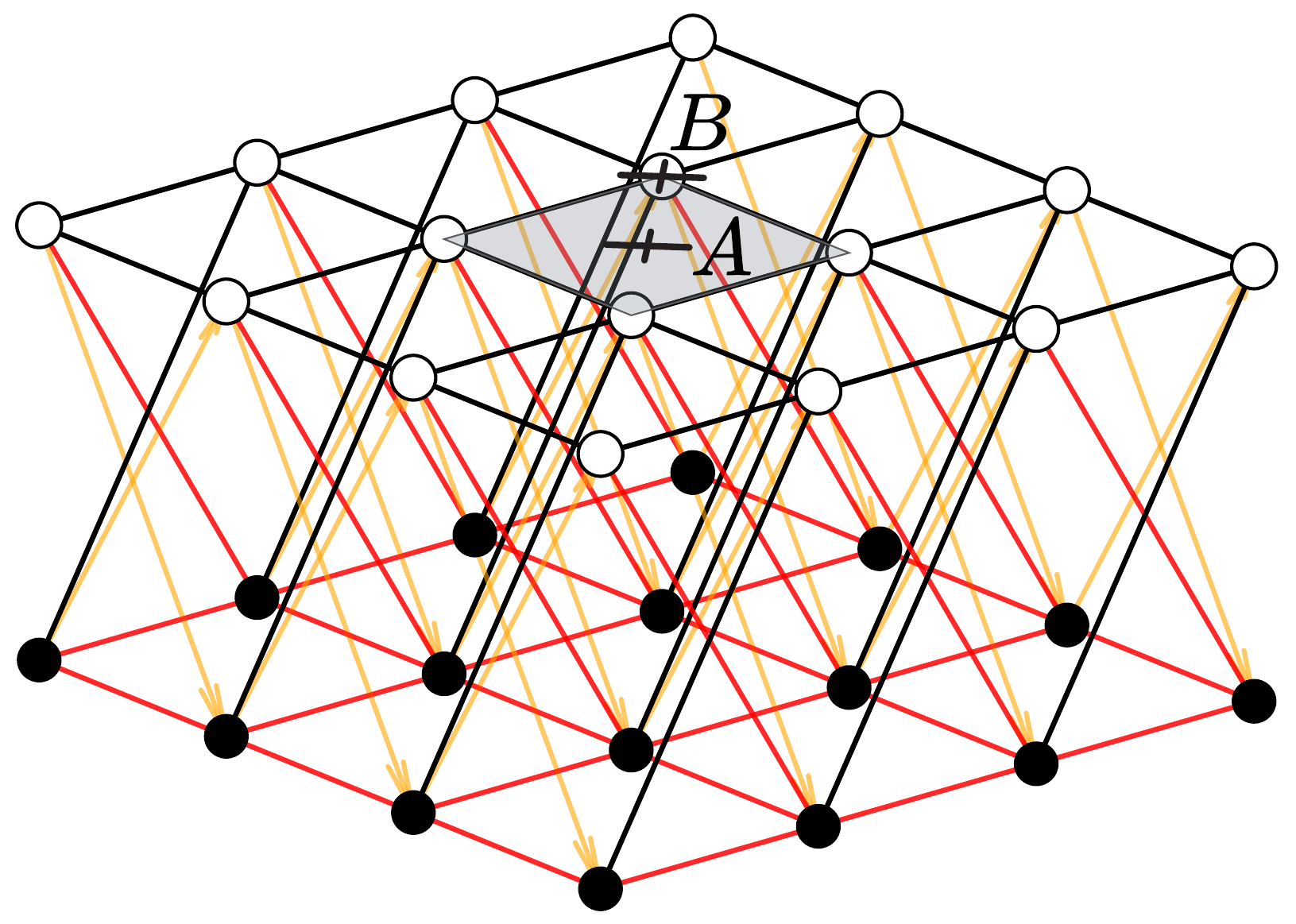}
    \caption{Real-space QWZ Hamiltonian, where black lines correspond to $h_{i,j}=\frac{1}{2}$, red lines to $h_{i,j}=-\frac{1}{2}$, and orange lines to $h_{i,j}=\frac{i}{2}$. White and black circles represent the on-site potentials $+2m$ and $-2m$, respectively. We have visualized the different flavors of fermion as different layers, with the top layer corresponding to the $\uparrow$ flavor. The rotationally invariant unit cell is shaded in gray, with the two inequivalent rotation centers marked with black crosses.}
    \label{fig:QWZ_ham}
\end{figure}

As a prototypical example, we will study the QWZ model \cite{qi_topological_2006-1}, which is a straightforward lattice regularization of a continuum Dirac theory. The model consists of two flavors of fermion hopping on a square lattice, which we denote as $\uparrow$ and $\downarrow$. The real-space Hamiltonian is given by
\begin{equation}\label{eq:QWZ_rs_ham}
\begin{gathered}
    H = \frac{1}{2} \sum_{\vj} 2m c_{j_x, j_y}^{\dagger} \sigma^z c_{j_x, j_y} 
+ c_{j_x+1, j_y}^{\dagger} \left[\sigma^z + i \sigma^x \right] c_{j_x, j_y} 
\\+ c_{j_x, j_y+1}^{\dagger} \left[ \sigma^z + i \sigma^y \right] c_{j_x, j_y} + \text{h.c.},
\end{gathered}
\end{equation}
where the $\sigma^i$ are Pauli matrices acting in the flavor space. This Hamiltonian is visualized in \cref{fig:QWZ_ham}. It can be expressed compactly in momentum space as
\begin{equation}
    H=\sum_{\vec{k}}(c_{\vec{k},\uparrow}^\dagger,c_{\vec{k},\downarrow}^\dagger)(\vec{d}_{\vec{k}}\cdot\vec{\sigma})\left(\begin{array}{c}
         c_{\vec{k},\uparrow} \\
         c_{\vec{k},\downarrow}
    \end{array}\right),
\end{equation}
where $\vec{\sigma}=(\sigma^x,\sigma^y)^T$ and
\begin{equation}
    \vec{d}_{\vec{k}}=\begin{pmatrix}
        \sin(k_x)\\
        \sin(k_y)\\
        2m+\cos(k_x)+\cos(k_y)
    \end{pmatrix}.
\end{equation}

Let us denote by $o = A$ and $o = B$ the plaquette and vertex centers, respectively. On a finite lattice with an odd number of sites per side, this model has a magnetic rotation symmetry about its central vertex given by
\begin{equation}\label{eq:QWZ_vert_sym}
     C_{4,B} c_{\vi,\uparrow} C_{4,B}^\dagger  = i c_{R^B_{\pi/2}(\vi),\uparrow},\quad C_{4,B} c_{\vi,\downarrow}  C_{4,B}^\dagger  = c_{R^B_{\pi/2}(\vi),\downarrow},
\end{equation}
where $R^B_{\pi/2}(\vi)$ is the lattice site obtained by rotating site $\vi$ about the central vertex by $\pi/2$. Similarly, on a finite lattice with an even number of sites per side, this model has a magnetic rotation symmetry about its central plaquette given by 
\begin{equation}\label{eq:QWZ_plaq_sym}
    C_{4,A} c_{\vi,\uparrow} C_{4,A}^\dagger  = i c_{R^A_{\pi/2}(\vi),\uparrow},\quad C_{4,A} c_{\vi,\downarrow}  C_{4,A}^\dagger  = c_{R^A_{\pi/2}(\vi),\downarrow},
\end{equation}
where $R^A_{\pi/2}(\vi)$ is the lattice site obtained by rotating site $\vi$ about the central plaquette by $\pi/2$. When placed on a square torus, the QWZ model is symmetric under both $C_{4,B}$ about any vertex and $C_{4,A}$ about any plaquette center. 
In general, the points about which the rotation symmetry is centered will be high-symmetry points of the rotationally invariant unit cell, as seen in \cref{fig:QWZ_ham}. In the QWZ case, the high symmetry points of the unit cell -- namely its center and corners -- correspond to the plaquette centers and vertices of the square lattice. For other models, this need not be the case. All symmetric rotation centers may be located at vertices of the lattice, for example, as we will see for another model in \cref{sec:critical_observables}.

The QWZ model also has a translation symmetry generated by a translation by one site in either the $x$- or $y$-direction when placed on a torus:
\begin{equation}
    T_{\hat x} c_{\vi,a} T_{\hat x}^\dagger=c_{\vi+\hat x,a},\quad T_{\hat y} c_{\vi,a} T_{\hat y}^\dagger=c_{\vi+\hat y,a}.
\end{equation}

Finally, the QWZ model is also endowed with an anti-unitary particle-hole symmetry $P$ which acts as
\begin{equation}
    P c_{\vi, a} P^\dagger = c_{\vi,\bar a}^\dagger,\quad P c_{\vi,a}^\dagger P^\dagger = c_{\vi,\bar a},
\end{equation}
where $\bar a$ denotes the flavor opposite $a$. This induces an effective transformation of the first-quantized Hamiltonian according to
\begin{equation}
    P:h_{\vi\vj}^{ab}\mapsto -\left(h_{\vi\vj}^{\bar a \bar b}\right)^*.
\end{equation}
This transformation remains a symmetry of the Hamiltonian on a clean lattice even when $\pi$ flux is inserted locally at the center of the lattice.

\begin{figure}
    \centering
\includegraphics[width=0.8\linewidth]{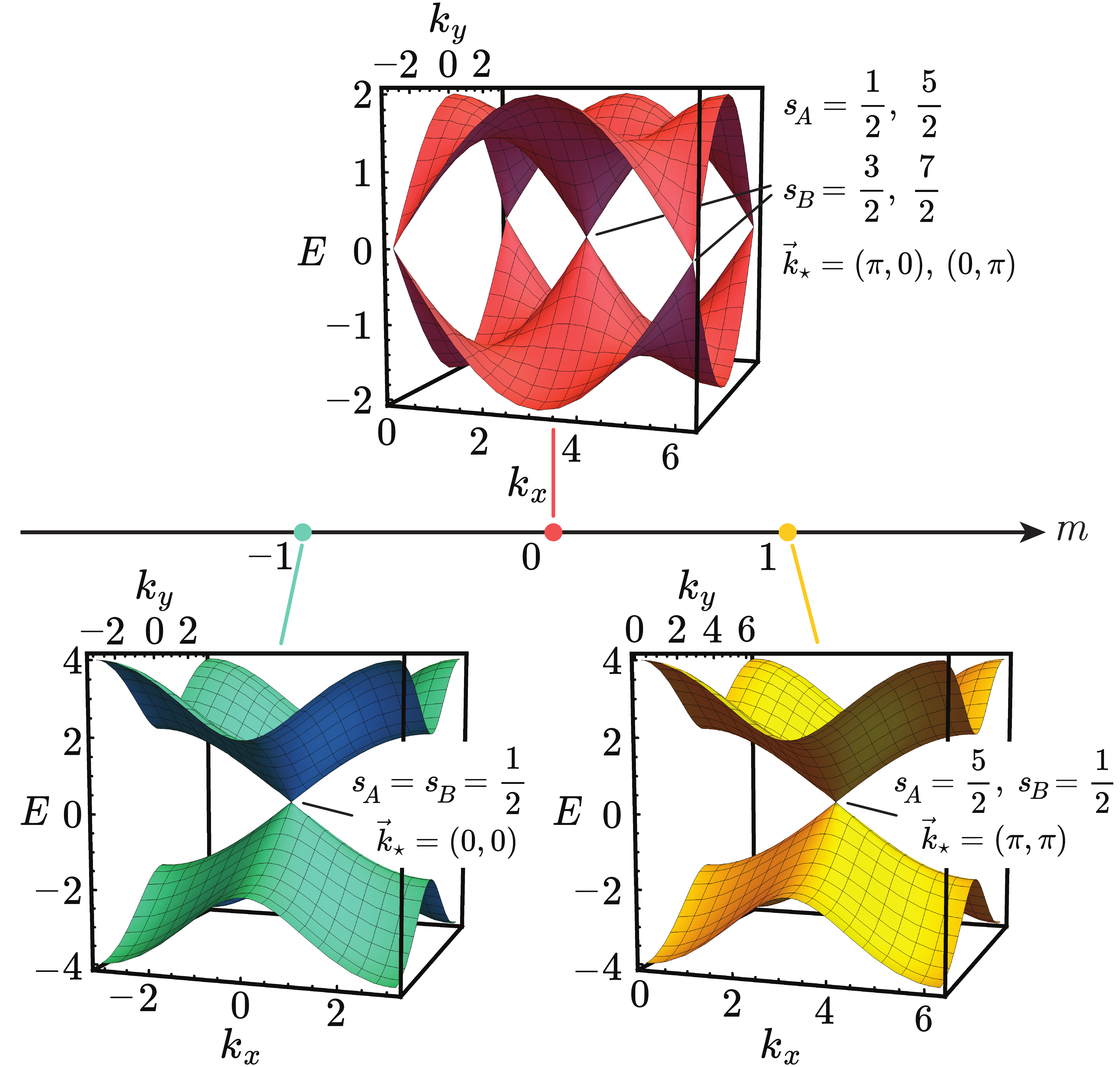}
    \caption{For each critical point of the QWZ model, we plot the band structure in the first Brillouin zone and label the angular momentum eigenvalues.}
    \label{fig:QWZ_invariants}
\end{figure}

The QWZ model exhibits four distinct gapped phases as the single parameter $m$ is tuned from $-\infty$ to $\infty$. The boundaries between these phases lie at $m=-1$, $0$, and $1$. The locations of the Dirac cones and values of $s_o^{(i)}$ at each critical point are shown in \cref{fig:QWZ_invariants}. At $m = +1$, there is a single Dirac cone at $\vec{k}_\star = (\pi,\pi)$, while at $m = -1$ there is a single Dirac cone at $\vec{k}_\star = (0,0)$. At $m = 0$ there are two Dirac cones at $\vec{k}_\star = (0,\pi), (\pi,0)$. The critical points all have different values of $\vec{k}_\star$ and $s_o$ and thus realize distinct crystalline symmetry-enriched Dirac critical points. 
We also show in \cref{ap:QWZ_sym_eigs} how the various $s_o$ appear analytically in the Dirac Hamiltonian treatment of the QWZ model.

\section{Continuum Defects from Lattice Defects}\label{sec:effective_lagr}

As we have seen, the QWZ lattice model (along with many other similar constructions) is described in the IR by massless Dirac fermions at various special values of the couplings. To fully determine the map from the lattice to the continuum we need more information. In particular, we will see how defects on the lattice map to defects in the continuum, which is governed by our homomorphism $\rho$. Equivalently, we will see that defects on the lattice lead to some nontrivial background gauge fields in the continuum. Therefore, our goal is to use the UV-IR homomorphism to determine the background gauge fields and spin connection of the IR theory in terms of the configuration of lattice defects in the UV. Before we proceed to do that, we have to review some basics concerning defects on the lattice.

\subsection{Lattice Defects and ``Crystalline Gauge Fields''} 

For the following discussion, it is not necessary, but very convenient, to describe lattice defects in terms of ``crystalline gauge fields.'' We use scare quotes because this formalism only makes sense for widely separated defects which can be clearly identified as disclinations and dislocations. (For more complicated lattice geometries, these gauge fields are subject to various constraints, but here we will only consider isolated elementary defects, for which the constraints on these gauge fields are unimportant.) Such ``crystalline gauge fields'' are nonetheless very useful for bookkeeping lattice defects. Having outlined this assumption, we drop the scare quotes below.  

Suppose we have a rotation operator $C_{M,o}$ around some high-symmetry point $o$ on the lattice which generates a $\mathbb{Z}_M$ symmetry. Disclinations are defined as $\mathbb{Z}_M$ defects, where we essentially cut out or add a wedge of angle $2\pi\over M$ and reconnect the lattice via the symmetry action. This is described in detail in \cite{zhang2022fractional,zhang2022pol}. We will employ a gauge field $\omega_o$ whose holonomies are $\int \omega_o \in {2\pi\over M} \mathbb{Z}$, such that $\int \omega_0 =0$ corresponds to no disclination, and $\int \omega_0 =\frac{2\pi k}{M}$ corresponds to removing $k$ wedges of angle $2\pi \over M$ for $k>0$ or adding $|k|$ wedges of angle $2\pi\over M$ for $k<0$. Since we cannot remove more than $M$ wedges, we have $k<M$. This should be interpreted as a constraint on the gauge field configurations realized by the cutting and gluing procedure we outlined.  

Further suppose we have a pair of translation operators $T_{\vec{v}_i}$ (for $\vec{v}_1$ and $\vec{v}_2$ the primitive vectors of the lattice) which together generate a $\mathbb{Z}^2$ symmetry. Dislocations are defects of this $\mathbb{Z}^2$ symmetry~\cite{manjunath2021cgt}. To record dislocations, we will employ a gauge field $\vec R$ whose holonomies are valued in $\mathbb{Z}^2$, ${1\over 2\pi}\oint \vec R\in \mathbb{Z}^2$. Physically, the holonomy $\frac{1}{2\pi} \oint_\gamma \vec{R}$ determines the Burgers vector for the region enclosed by the loop $\gamma$; elementary lattice dislocations correspond to the case where the Burgers vector is an elementary lattice vector. For widely separated dislocations it makes sense to discuss the gauge field $\vec R$. 

In general, since rotations and translations do not commute, each defect is labeled by a conjugacy class in $\mathbb{Z}^2 \rtimes \mathbb{Z}_M$. For instance, to describe the effect of disclinations centered at another point $o'$ in terms of $\omega_o$, we need to include $\vec{R}$ holonomies. As a corollary, a vertex-centered disclination and a plaquette-centered anti-disclination combine to a dislocation.

We emphasize that there are considerable subtleties in defining the gauge fields $\omega_o$ and $\vec{R}$ from the lattice geometry; we refer to \cite{zhang2022pol,zhang2024bdy} for detailed discussions. As we said, for the most part this will be unimportant for us as we will only discuss isolated elementary dislocations and disclinations. 

\subsection{UV - IR correspondence for symmetry defects}

To describe in the infrared a lattice system at or near its gapless points, we consider a (2+1)D theory of $N$ free Dirac fermions, coupled to a background $U(N)$ gauge field $A$, a spin connection $\Omega$ and mass matrix $\textbf{M}$:
\begin{align}\label{eq:Dirac_lagr}
    \mathcal{L} = \overline{\Psi} (i \slashed \partial + \slashed A + \frac{1}{2} \slashed \Omega_{\nu \rho} \sigma^{\nu \rho} + \textbf{M}) \Psi.
\end{align}
Here $\Psi = (\Psi_1, \cdots, \Psi_N)^T$ consists of $N$ 2-component Dirac fermions. The possible configurations of the parameters $A$, $\Omega_{\nu \rho}$, and $\textbf{M}$ depend on the lattice model. In particular, assuming no spontaneous symmetry breaking, they have to be compatible with the lattice symmetries (e.g. $\textbf{M}$ should commute with $U_{T_{\vec{v}}}$ and $U_{M,o}$ introduced in Section \ref{sec:dirac_sym}.)

We argued above that a $g \in G_\text{UV}$ symmetry defect in the UV is modeled in the IR by a $\rho(g) \in G_\text{IR}$ symmetry defect. It is therefore convenient to expand the $U(N)$ gauge field $A$ in terms of individual contributions from the UV symmetry defects. In other words, we need a map from the crystalline gauge fields $\omega_o$ and $\vec R$ to the infrared background fields $A$, $\Omega_{\nu \rho}$, and $\textbf{M}$.
We will restrict to writing the map at the gapless point $\textbf{M}=0$ and discuss the consequences of this for nonzero $M$ in the next section.

For simplicity, let us consider the case where $U_{T_{\vec{v}}}$ and $U_{M,o}$ are simultaneously diagonal, meaning the Dirac nodes occur at rotation-symmetric lattice momenta (we will relax this condition later). 

Our UV theory has some configuration of the $U(1)$ gauge field $A_\text{UV}$ along with lattice disclinations centered at $o$ and lattice dislocations with Burgers vector $\vec R$. 
We claim that each fermion $\Psi_i$ therefore sees a $U(1)$ gauge field

\begin{align}
\label{Auvir}
    A^{(i)} = A_\text{UV} + s_o^{(i)} \omega_o + \frac{1}{2\pi}\vec{k}_\star^{(i)} \cdot \vec{R} ,
\end{align}
where $s_o$ is the topological quantum number \eqref{eq_def topo number s} and $\vec{k}_\star$ is the lattice momentum of the Dirac mode. Recall that the holonomy of the $\omega_o$ gauge field around any loop $\gamma$ is quantized as $\oint_\gamma \omega_o \in \frac{2\pi}{M} \mathbb{Z}$, measuring the disclination angle of the underlying lattice in the interior of $\gamma$, and $\int \vec{R}\in 2\pi \mathbb{Z}^2$ keeps track of the Burgers vector of dislocations.

Furthermore, it is intuitively clear that a lattice disclination centered at $o$ maps in the continuum theory to a geometry with a conical deficit angle of $2\pi k/M$. Thus we set the spatial part of the spin connection to be
\begin{align}
    (\Omega_{xy})_\mu = \omega_{o,\mu}.
\end{align}
(We set the terms involving the time component to zero: $\Omega_{\mu t} = \Omega_{t \mu} = 0$, for $\mu = x,y$.) 

Equation~\eqref{Auvir} shows the map between the gauge field that couples to the Dirac fermions in the continuum description and the gauge field on the lattice, $A_\text{UV}$. The appearance of $s_o^{(i)} \omega_o$ means that disclinations on the lattice activate a magnetic field in the continuum description! Similarly, the term $\frac{1}{2\pi}\vec{k}_\star^{(i)} \cdot \vec{R}$ means that dislocations lead to a magnetic field!

We refer to these magnetic fields in the IR theory that are induced by the UV lattice defects as \it emanant \rm magnetic fields. This is because the corresponding group element of $U(1)$ arises from the image of translations or rotations under $\rho$. Even if the $U(1)$ charge conservation in the UV were broken by weak perturbations, the IR would have an exact symmetry generated by $U_{T_{\vec{v}}}, U_{M,o} \subset U(1)$ if the crystalline symmetry is preserved. Thus the $U(1)$ flux in the IR can be thought of as ``emanating" from the translation and rotation symmetry defects in the UV, analogous to the discussion of \cite{cheng2023lieb}. 

An important part of the mapping between the lattice and continuum that we discussed above, is that dislocations with Burgers vectors that are of the lattice scale activate a magnetic field in the continuum (with holonomy $\frac{1}{2\pi}\vec{k}_\star^{(i)} \cdot \vec{R}$) but carry no geometric consequences otherwise. The point, which we will explain in more detail later, is that the geometric response due to such dislocations is irrelevant in the continuum limit.  For Burgers vectors that remain finite in the continuum limit the story may well be very different and we do not discuss it here (indeed lattice scale translations map to internal symmetries in the continuum limit, while macroscopic translations map to non-trivial translations in the continuum and hence will have geometric imprints as well).

Finally, if the Dirac fermions do not occur at momenta that are symmetric under $2\pi/M$ rotations, then, as we discussed the transformation $U_{M,o}$ is not diagonal. In this case, the lattice disclinations induce a non-diagonal $U(N)$ monodromy. A lattice disclination corresponding to the operator $C_{M,o}$ induces a monodromy 
\begin{align}
    \mathcal{P} e^{i \oint_\gamma A} = e^{i \oint_\gamma A_\text{UV}} U_{M,o},
\end{align}
where we have also included the $U(1)$ gauge field $A_\text{UV}$ and $\mathcal{P}$ denotes path ordering. 
 
 In summary, we have provided a prescription of how to compute the infrared background fields corresponding to certain lattice defects. A lattice disclination should be described in the IR in terms of a conical defect with $U(N)$ monodromy, while a lattice dislocation should be described in the IR purely in terms of a $U(N)$ monodromy. It is important that lattice dislocations and disclinations generically lead to emanant magnetic fields at long distances. 

\section{Crystalline Invariants of Gapped Phases}

Our perspective of realizing the long-distance limit of lattice defects as standard gauge field configurations in the continuum~\eqref{Auvir}  
allows us to study the gapped (insulating) phases. In the continuum, this is straightforward and boils down to integrating out the massive fermions at one loop. Once the result is expressed in terms of the original lattice gauge fields this leads to many predictions, as we will momentarily see. 
In particular we will show how the crystalline topological invariants predicted in \cite{manjunath2021cgt,manjunath2020FQH,zhang2022pol,zhang2022fractional,zhang2023complete} can be understood in terms of the Dirac fermion quantum numbers. 

These results provide another way of deducing the Dirac fermion quantum numbers $s_o^{(i)}$ in terms of the crystalline topological invariants of the nearby gapped phases. They also demonstrate how crystalline symmetry-enriched quantum critical points between different crystalline topological phases can be distinguished and how the crystalline topological invariants can jump due to the appearance of gapless modes. 
Remarkably, we will see that the results match the results of the symmetry eigenvalue analysis of the preceding sections, and also the results for the current at the gapless points presented in the subsequent sections. 

\subsection{From Dirac Fermions to crystalline topological invariants}\label{sec:effective_lagr}

In this section, we will always assume that $U_{M,o}$ and $U_{T_{\vec{x}}}$ are simultaneously diagonalized, so that the Dirac cones occur at the $2\pi/M$ symmetric momenta.  

Suppose that we consider a diagonal mass matrix, $\textbf{M} = \text{diag}(m_1, \cdots, m_N)$. Integrating out the fermions then gives an effective Lagrangian\footnote{As we discuss later, this Lagrangian should be understood as determining changes in the couplings relative to a reference. Strictly speaking. Eq. \ref{CStheory} should have suitable counterterms, which specify the reference, in order to make this term well-defined in general.} 
\begin{align}
\label{CStheory}
    \mathcal{L} = \sum_{i=1}^N \frac{\text{sgn}(m_i)}{2} \left( \frac{1}{4\pi} A^{(i)} \wedge d A^{(i)} - \frac{1}{48 \pi} \omega_o \wedge d \omega_o \right),
\end{align}
where the second term is the contribution from the gravitational Chern-Simons term. Expanding $A^{(i)}$, which is the $U(1)$-connection of the $i$-th Dirac fermion, using Eq.~\eqref{Auvir} we get: 

\begin{align}
\label{eq:top_response_theory}
    \mathcal{L} = &\frac{C}{4\pi}A_\text{UV}\wedge dA_\text{UV} +\frac{\mathscr{S}_o}{2\pi}A_\text{UV} \wedge d\omega_o 
    +\frac{\tilde{\ell}_o}{4\pi} \omega_o \wedge d\omega_o 
    \nonumber \\
    &+ \frac{\vec{\mathscr{P}}_o}{2\pi}\cdot A_\text{UV} \wedge d \vec{R} + \frac{\vec{\mathscr{Q}}_{o}}{2\pi}\cdot \omega_o \wedge d \vec{R} + \frac{1}{4\pi} \vec{R} \cdot \Pi \wedge d \vec{R}
\end{align}

The coefficients determine the quantized topological invariants that characterize the gapped state with crystalline symmetry, expressed using the gauge fields that can be defined on the lattice, $A_\text{UV}$, as well $\omega_o,\vec R$ which capture disclinations and dislocations. The Chern number of the topological insulator is
\begin{equation}\label{eq:C_coeff}
    C=\sum_{i=1}^N \frac{\operatorname{sgn}\left(m_i\right)}{2},
    \end{equation}
    which sets the quantized Hall conductivity. The discrete shift is
    \begin{equation}\label{eq:S_coeff}
    \mathscr{S}_o=\sum_{i=1}^N \frac{\operatorname{sgn}\left(m_i\right) s_o^{(i)}}{2},
    \end{equation}
    which gives a fractional quantized contribution to $U(1)$ charge in the vicinity of a lattice disclination and determines the angular momentum of magnetic flux \cite{zhang2022fractional}. Note \cite{zhang2022fractional} demonstrated the constraint that $\mathscr{S}_o = \frac{C}{2} \mod 1$. In our current description, we see that this constraint is a consequence of the fact that the orbital angular momenta $s_o^{(i)}$ are half-odd integers. Next, we have
\begin{align}\label{eq:ls_coeff}
\begin{split}
    \tilde{\ell}_o &= \ell_o - \frac{c_-}{12}\\
    \ell_o &= \sum_{i=1}^N   \frac{\operatorname{sgn}\left(m_i\right) (s_o^{(i)})^2}{2},
\end{split}
\end{align}    
where $c_- = \sum_{i=1}^N \frac{\text{sgn}(m_i)}{2} = C$ is the chiral central charge. $\ell_o$ can be determined in a quantum many-body system from ground state expectation values of partial rotation operations \cite{zhang2023complete}. \cite{manjunath2024Characterization} demonstrated the constraint $\ell_o = \frac{C}{4} \mod \text{gcd}(M,2)$, which can also be understood from the above equation and that $s_o^{(i)}$ are half-odd integers. 

The electric polarization is
\begin{align}
\label{eq:Po}
    \vec{\mathscr{P}}_o = \frac{1}{2\pi}\sum_{i=1}^N \frac{\text{sgn}(m_i)}{2}\vec{k}_\star^{(i)},
\end{align}
which can be detected in the gapped phase through a variety of methods, including quantized contributions to the electric charge in the vicinity of lattice dislocations and boundaries and the magnetic flux-dependence of the momentum of the ground state \cite{zhang2022pol,zhang2024bdy}. The $o$-dependence of the RHS is implicit in $k_\star^{(i)}$. For example, for free fermion systems, the Bloch functions have an implicit choice of origin in real space, implying that $k_\star^{(i)}$ has an implicit choice of real space origin, which is related to the choice $o$ above; the detailed relationship is left to future work \cite{zhang2025}. 

Finally, we have
\begin{align}
\label{eq:Pso}
    \vec{\mathscr{Q}}_{o} &= \frac{1}{2\pi} \sum_{i=1}^N \frac{\text{sgn}(m_i)}{2} s_o^{(i)} \vec{k}_\star^{(i)},
    \nonumber \\
    \Pi_{\alpha \beta} &= \frac{1}{4\pi^2} \sum_{i=1}^N \frac{\text{sgn}(m_i)}{2} k_{\star,\alpha}^{(i)} k_{\star,\beta}^{(i)} . 
\end{align}
The terms in Eq. \eqref{eq:top_response_theory} above explain the origin of the crystalline topological gauge theory presented in \cite{manjunath2021cgt} in terms of the quantum numbers of Dirac fermions.  

\begin{table*}
\def\arraystretch{1.4}
    \subfloat{\begin{tabular}{rccccccccccccc}
         \hline\hline  & $\text{ }$ & $s_A^{(i)}$ & $s_B^{(i)}$ & $\operatorname{sgn}(m_i)$ & $\vec{k}^{(i)}_\star$ & $\text{ }$ & $\Delta C$ & $\Delta \mathscr{S}_A$ & $\Delta \mathscr{S}_B$ & $\Delta\ell_A$ & $\Delta \ell_B$ & $\Delta \vec{\mathscr{P}}$ \\\hline
         $m=-1$ && $\frac{1}{2}$ & $\frac{1}{2}$ & $\textcolor{Green}{+}\to\textcolor{purple}{-}$ & $(0,0)$ && $-1$ & $\frac{7}{2}$ &  $\frac{7}{2}$ & $\frac{15}{4}$ & $\frac{15}{4}$ & $(0,0)$ \\\hline
         \multirow{2}{*}{$m=0$} && $\frac{1}{2}$ & $\frac{3}{2}$ & $\textcolor{purple}{-}\to\textcolor{Green}{+}$ & \multirow{2}{*}{$(\pi,0)$,$(0,\pi)$} && \multirow{2}{*}{$2$} & \multirow{2}{*}{$3$} &  \multirow{2}{*}{$1$} & \multirow{2}{*}{$\frac{5}{2}$} & \multirow{2}{*}{$\frac{5}{2}$} & \multirow{2}{*}{$(\frac{1}{2},\frac{1}{2})$}\\
          && $\frac{5}{2}$ & $\frac{7}{2}$ & $\textcolor{purple}{-}\to\textcolor{Green}{+}$ &  && &\\\hline
         $m=1$ && $\frac{5}{2}$ & $\frac{1}{2}$ & $\textcolor{Green}{+}\to\textcolor{purple}{-}$ & $(\pi,\pi)$ && $-1$ & $\frac{3}{2}$ & $\frac{7}{2}$ & $\frac{7}{4}$ & $\frac{15}{4}$ &$ (\frac{1}{2},\frac{1}{2})$\\
         \hline\hline
    \end{tabular}}
    \caption{Changes in topological invariants and parameters in the Dirac fermion effective Lagrangian for the QWZ model. The values of $s_o^{(i)}$, $k_\star^{(i)}$, and $\text{sgn}(m_i)$ are deduced from matching the changes of the topological invariants to \eqref{eq:C_coeff}, \eqref{eq:S_coeff}, \eqref{eq:ls_coeff}, and \eqref{eq:Po}. Results agree with those reported in Fig. \ref{fig:QWZ_invariants}, showing that these independent methods give the same result.}
    \label{tab:QWZ_invar}
\end{table*} 

The crystalline topological gauge theory of \cite{manjunath2021cgt} includes two additional terms involving the charge and angular momentum per unit cell. These are outside of the Dirac theory, and must be included separately from the outset, since the Dirac theory does not include information about the charge and angular momentum per unit cell of the underlying lattice system. 

The results above should be understood as providing us with the quantum critical theory that describes critical points between gapped phases where the crystalline topological invariants change their values. In particular, as we go through a transition where the mass of $N$ Dirac fermions goes through zero, the changes in the invariants are given by the above formulas, with $\text{sgn}(m_i)$ replaced by the change in the sign of the masses across the transition, $\Delta \text{sgn}(m_i)$.  
If we know how the crystalline invariants change across a phase transition, we can invert the above equations to obtain a minimal massless Dirac theory with quantum numbers $\vec{k}_\star^{(i)}$, $s_o^{(i)}$. 

The above result yields a simple understanding of some non-trivial properties of the crystalline invariants. The non-trivial relationship between $\mathscr{S}_o$ , $\ell_o$, and $C$ was already mentioned above. 
Moreover, the independent invariants of the gapped phase correspond to certain combinations of the above coefficients. For example, consider $M = 4$. Across a phase transition where the Dirac masses change sign, we have the independent invariants \cite{manjunath2024Characterization}:
\begin{align}
    \Delta I_1 &:= \Delta \mathscr{S}_o - \Delta \ell_o - \Delta c_-/4 \in \mathbb{Z}_8,
    \nonumber \\
    \Delta I_2 &:= \frac{1}{2} (\Delta \ell_o - \Delta c_-/4) \in \mathbb{Z}_2,
\end{align}
which generate a $\mathbb{Z}_8 \times \mathbb{Z}_2$ classification. The fact that these are the appropriate independent invariants and their quantization rules and equivalence relations can be understood from the Dirac theory using the equivalences $s_o^{(i)} \sim s_o^{(i)} + M$ for any given $i$. Similarly, the equivalence relation $\vec{k}_\star^{(i)} \sim \vec{k}_\star^{(i)} + \vec{K}$, where $\vec{K}$ is a reciprocal lattice vector, gives the equivalence relation for the electric polarization. 

Under the assumption that all of the Dirac fermions occur at rotation-symmetric momenta, we can use the origin dependence of $s_o^{(i)}$ in \eqref{eq:so_dependence} to determine the $o$-dependence of $\Delta \mathscr{S}_o$. We find 
\begin{align}
\label{Eq:Soshift}
    \Delta \mathscr{S}_{o + \vec{v}} &= \Delta \mathscr{S}_o + M \vec{f}(\vec{v}) \cdot \Delta \vec{\mathscr{P}}_o \mod M .
\end{align}
Remarkably, \eqref{Eq:Soshift} is consistent with the results of \cite{zhang2022pol}, which were derived using completely different methods. From \cite{zhang2022pol}, we know that \eqref{Eq:Soshift} is valid in general, and does not require the Dirac cones to be centered at rotation-symmetric momenta in the Brillouin zone. Therefore a more general derivation than that presented above should be possible using the Dirac theory.  

A thorough understanding of $\vec{\mathscr{Q}}_o$, $\Pi$, and the $o$-dependence of $\ell_o$ is more complicated and beyond the scope of this paper. Here we simply note that $\vec{\mathscr{Q}}_o$, $\Pi$ combine to give the angular momentum polarization that was defined in \cite{manjunath2021cgt,manjunath2024Characterization}.

Note that, using the results of \cite{zhang2022pol}, $\Delta \vec{\mathscr{P}}_o$ is independent of $o$ when the charge per unit cell remains unchanged through the transition, a fact which we will use below, which is indeed the case in these transitions where the Dirac mass changes sign. 

In cases where the Dirac fermions do not occur at $2\pi/M$ rotation symmetric momenta, we can use the above equations for $\mathscr{S}_o$, $\vec{\mathscr{P}}_o$, and $\ell_o$ by considering the response in terms of $\omega_o$ and $\vec{R}$ separately. However, the equation for $\vec{\mathscr{Q}}_{o}$ cannot be used directly.

\subsection{Example: QWZ model}\label{sec:QWZ_invariants}

Using the method of partial rotations discussed in \cite{zhang2023complete,manjunath2024Characterization} and reviewed in \cref{ap:partial_rotations}, we measured the topological invariants $C$, $\mathscr{S}_o$, and $\ell_o$, in the QWZ model. As we discussed in the previous section, the Dirac theory unambiguously captures changes in the topological invariants rather than their absolute values. We therefore report the changes of these topological invariants across each of the three critical points in \cref{tab:QWZ_invar}, along with $\Delta \vec{\mathscr{P}}$ which is obtained using \cref{Eq:Soshift}. We then deduce the Dirac theory with the minimal number of massless fermions and their corresponding values of $s_o^{(i)}$, $\operatorname{sgn}(m_i)$, and $\vec{k}^{(i)}_\star$ which reproduce the changes in these invariants. We present the results in \cref{tab:QWZ_invar}. 

Remarkably, the results agree precisely with those found in \cref{sec:QWZ_rot_eigs} using completely different methods, through eigenvalues of rotation and translation operators! In particular, the number of massless fermions at each critical point and topological quantum numbers  $s_o^{(i)}$ and $k_\star^{(i)}$ match precisely. This correspondence is a non-trivial prediction of the relationship between the UV-IR homomorphism and the field theory analysis discussed above. 

\section{Defect conformal field theory at Dirac criticality}\label{sec:universality_defects}

In the preceding sections, we saw that critical lattice models with crystalline defects map onto continuum theories with conical singularities and with excess magnetic field flux. Our discussion was for gapped theories and we concentrated on the topological response, extending the Chern number of the insulator by the invariants in~\eqref{eq:top_response_theory}. 

It is of both theoretical and experimental interest to understand how this excess flux and conical singularities influence observables at the critical point. In order to guide the search for suitable observables, we turn to the framework of defect conformal field theory. 
The massless bulk fermions preserve the conformal algebra in 2+1 dimensions $so(3,2)$, which consists of the Lorentz symmetries along with scaling (the fermion field scales like mass) and special conformal transformations. Conical singularities and magnetic flux defects are point-like in space and by the general philosophy of the renormalization group must flow to infrared fixed points, preserving the $so(1,2)\times so(2)$ subalgebra of the conformal algebra, where the $so(2)$ factor corresponds to rotations transverse to the defect while the $so(1,2) \simeq sl(2,\mathbb{R})$ factor corresponds to time translations, scaling symmetry about the location of the defect, and a special conformal transformation. (Naively, one only expects scale invariance at long distances, but the appearance of the full conformal group is common and in many cases can be proven to occur. For defects, this was discussed in~\cite{Nakayama:2012ed}.) 

For a thorough introduction to the kinematics of these symmetries see~\cite{Billo:2016cpy, Andrei:2018die, soderberg2023defects, Chalabi:2023ohu}. For our present purposes, it would be most important to identify the fixed points of the defect RG flows and then infer their salient predictions for bulk physics. 

We will thus identify some relevant observables at the gapless points, understand the scaling of correlation functions, and predict their amplitudes as a function of flux. The results of \cref{sec:universality_defects} determine what observables we measure in lattice models and help us to interpret the numerical results we find.

Crystalline defects modify the effective metric that emerges in the infrared description of the system (see e.g.~\cite{kupferman2015metric}.) We have already discussed the fact that crystalline disclinations lead to conical singularities in space. More generally, one can parameterize a local geometric defect via the curvature 
\begin{equation}
\label{eq_ricci scalar}
R=4\pi(1-\beta)\delta^2(x)+\ell(\vec{b}\cdot \partial) \delta^2(x)+O\left(\ell^2\right)~.
\end{equation}
where $\ell$ is the ultraviolet scale associated with lattice spacing. The first term of $\delta^2(x)$ corresponds to a conical defect and emerges from lattice disclinations with a deficit angle $2\pi (1-\beta)$. 

The second term, proportional to $\ell \vec{b}$, could arise from lattice dislocations~\cite{kupferman2015metric} (having two opposite disclinations next to each other does not necessarily imply there is a Burgers vector, but it can. That is why $\vec{b}$ could but does not have to correspond to the Burgers vector on the lattice). 
Due to the explicit factor of $\ell$ at the long-distance limit, the term proportional to $\ell \vec{b}$ is irrelevant and is therefore ``screened'' in the infrared. This is why lattice-scale dislocations do not have a geometric imprint at long distances in the continuum description. Therefore, the spatial metric is determined solely by the conical singularity and our Dirac cones are seeing the continuum effective metric 
\begin{equation}
\label{eq_cone metric}
(ds)^2=-(dt)^2+(dr)^2+\beta^2 r^2 (d\theta)^2~,
\end{equation}
where $r$ is the distance to the defect location and $\theta\in \mathbb{S}^1$ is the angular polar coordinate. Remarkably, the metric \eqref{eq_cone metric} admits conformal Killing vectors leading to the conformal symmetry algebra  $so(1,2)\times so(2)\subseteq so(3,2)$. This is precisely the symmetry algebra of infrared fixed points of defects in gapless systems in 2+1 dimensions. Therefore, the conical defect can be viewed as a conformal defect. (There is an important technical difference between the defect conformal field theory of a conical singularity and other, more familiar conformal defects. We will discuss it later.) 
A simple way to see that the metric~\eqref{eq_cone metric} admits $so(1,2)\times so(2)$ conformal Killing vectors is to rewrite \begin{align}\label{AdStwo}(ds)^2&=r^2\left({-(dt)^2+(dr)^2\over r^2}+\beta^2  (d\theta)^2\right)\cr & \sim {-(dt)^2+(dr)^2\over r^2}+\beta^2  (d\theta)^2~,\end{align}
where $\sim$ means Wely equivalence that is up to a conformal factor. We recognize the factor ${-(dt)^2+(dr)^2\over r^2}$ as $\text{AdS}_2$ (A Minkowski version of the Poincar\'e disk) and $\beta^2  (d\theta)^2$ is just a circle of radius $\beta$. Therefore the conical defect is conformally equivalent to $\text{AdS}_2\times \mathbb{S}^1_{\beta}$ and the $so(1,2)\times so(2)$ act on the two factors, respectively. For the calculations below we will analytically continue to Euclidean signature. The conical singularity is then at the boundary of the Poincar\'e disk. Of course, these various conformal transformations are allowed only in the massless Dirac theory, which is invariant under the conformal group.  

The defect conformal field theory depends on the conical singularity as well as the magnetic fluxes at $r=0$. In the $\text{AdS}_2\times \mathbb{S}^1_{\beta}$ coordinates the flux at the defect point translates to a holonomy of the gauge field (chemical potential) along the $\mathbb{S}^1_{\beta}$. We will go back and forth between the $\text{AdS}_2\times \mathbb{S}^1_{\beta}$ conventions and the original description~\eqref{eq_cone metric} when convenient. 

\subsection{DCFT spectrum}

In this section, we focus on a single Dirac cone. We also provide the analysis for the free boson in appendix \ref{appendix_free boson model}, which has several qualitatively distinct properties that we highlight throughout the discussion.

The low-energy description of the single Dirac cone with the effective conical metric \eqref{eq_cone metric} reads 
\begin{equation}
\label{eq_fermion action}
    \begin{aligned}
S_\text{Dirac}=&\int_{\text{Cone}} d^3 X  \Bar{\Psi}\slashed{\nabla}\Psi~.
    \end{aligned}
\end{equation}
In \eqref{eq_fermion action}, $d^3 X$ and $\slashed{\nabla}=i\slashed \partial+\slashed A+ \frac{1}{2} \slashed \Omega_{\nu \rho} \sigma^{\nu \rho}$ are the volume element and Dirac operator on the space \eqref{eq_cone metric}, respectively. We also activate a static connection $A_\mu=\alpha \delta_{\mu \theta}$, with $\alpha\in [0,1)$ that corresponds to the Aharonov-Bohm flux at $r=0$. The explicit form of the Dirac operator is~\footnote{Our convention for gamma matrices is $\gamma_t=i \sigma_z$, $\gamma_r=\cos{(\theta)} \sigma_x+\sin{(\theta)}\sigma_y$, and $\gamma_\theta=\beta r(-\sin{(\theta)}\sigma_x+\cos{(\theta)}\sigma_y
)$.}
\begin{equation}
    \begin{aligned}
\slashed \nabla= \begin{pmatrix}
-i\partial_t & e^{-i\theta}(\partial_r+\frac{-i\partial_\theta+\alpha-\frac{\beta}{2}}{r\beta})\\
e^{i\theta}(\partial_r+\frac{i\partial_\theta-\alpha-\frac{\beta}{2}}{r\beta }) &  i\partial_t\\
\end{pmatrix}~.
    \end{aligned}
\end{equation}

Familiar point-like defects such as the Kondo defect or charge defects can often be dynamically screened, namely, their effect on the infrared is only through irrelevant operators and the $so(1,2)\times so(2)$ invariant defect theory is trivial.
The conical defect and the monodromy defect (the monodromy defect is the Aharonov-Bohm flux, which leads to an additional phase $e^{i\oint A}=e^{2 \pi i \alpha}$ when the Dirac particle $\Psi$ is transported along a contour that encircles the origin) cannot be screened. This is because they are attached to ``branch cuts'' that go all the way to infinity. Indeed, the monodromy defect can be viewed as the boundary of the $U(1)$ open symmetry surface while the conical defect is the boundary of the open rotations surface. One can say that the conical and monodromy defects are attached to topological surfaces that go all the way to infinity and that is why they cannot be screened.

Next, we need to find the renormalization group fixed points of the defect RG flows. This problem can be approached by first analyzing the near-defect behavior of the field $\Psi$ at $r=0$. 
An essential point is that there are different possible boundary conditions at the origin $r=0$. Without choosing a concrete boundary condition there is no consistent quantum system to discuss. One has to identify the boundary conditions that respect the defect conformal symmetry and then analyze the flows between different defect fixed points. The analysis of the modes in $\text{AdS}_2\times \mathbb{S}^1_{\beta}$ coordinates vs. the original coordinates are essentially identical, so we proceed with the $\text{AdS}_2\times \mathbb{S}^1_{\beta}$ coordinates.

We can perform a mode expansion over the $\mathbb{S}^1_\beta$ and think about the theory as consisting of infinitely many modes in $\text{AdS}_2$. We denote these modes by  $\tilde{\Psi}_s(t,r)$ and they are related to the original fermions in flat space via 
\begin{equation}
\Psi(t,r,\theta)=\sum_{s\in\mathbb{Z}+\frac{1}{2}}\frac{e^{i \left(s-\frac{\sigma_z}{2} \right)\theta}}{\sqrt{2\pi}r}e^{\frac{i \pi}{4}\sigma_y}\tilde{\Psi}_s(t,r)~.
\end{equation}
We note that the $e^{-i\frac{\sigma_z}{2}\theta}$ dependence is necessary to reflect the different spins of the two fermion components, and the $e^{\frac{i \pi}{4}\sigma_y}$ factor will be important to diagonalize the action of $\tilde{\Psi}_s(t,r)$ in $\text{AdS}_2$. The monodromy can also be understood in terms of the phonon $\sigma\in \mathbb{S}^1$ dual to the connection $A_\mu$, such that the local operator $(e^{i\sigma} \Psi)\to e^{2\pi i\alpha }(e^{i\sigma} \Psi)$ upon $\theta\to \theta+2\pi$.

Letting $d^2\Tilde{X}$ and $\tilde{\slashed{\nabla}}$ be the volume element and Dirac operator of $\text{AdS}_2$, we find the action for our modes $\tilde \Psi_s(t,r)$  is
\begin{equation}
\label{eq_fermion map to AdS}
    \begin{aligned}
S_\text{Dirac}
=\beta\sum_{s}\int_{\text{AdS}_2}d^2 \Tilde{X}\Bar{\tilde{\Psi}}_s\left(\tilde{\slashed{\nabla}}-\frac{s+\alpha}{\beta}\right)\tilde{\Psi}_s~.
    \end{aligned}
\end{equation}
We see that our modes are standard massive fermions on the Poincar\'e disk, with mass $(s+\alpha)/\beta$. 
If we now analyze the $r\to 0$ behavior of the wave functions (which is the boundary of the Poincar\'e disk), we find two distinct consistent possibilities (See Eq. \eqref{eq_appendix_ads fermion green function asy 1}): 
\begin{equation}
\label{eq_std&alt quantization}
    \begin{aligned}
&\Tilde{\Psi}_s= r^{\Delta_s^\pm} \begin{pmatrix}
 \psi_s^\uparrow(t)\\
 \psi_s^\downarrow (t)
\end{pmatrix}+O\left(r^{\Delta_s^\pm+1}\right)~,\\
&\text{where $\Delta_s^\pm=\frac{1}{2}\pm \frac{|s+\alpha|}{\beta}$} ~.
    \end{aligned}
\end{equation}
The interpretation of having two distinct possible modes near the defect is crucial to understand. The same issue with boundary conditions was discussed at great length in the context of charged impurities in~\cite{Aharony:2022ntz, Aharony:2023amq}. 

As a helpful digression, we note that something similar happens in the non-relativistic quantum mechanics of a particle with the following Hamiltonian (see for instance~\cite{landau2013quantum, Kaplan:2009kr}) $H = -{d^2\over dr^2 }+{h \over r^2}$, which allows wave functions that behave for $r\to 0$ as either $r^{\half \pm\sqrt{h+{1\over 4}}}$. 
For instance, if $-\frac{1}{4}< h<0$ then both choices are normalizable and appear physically admissible. The meaning of this ambiguity is well known: the Hamiltonian is not Hermitian and requires an extension, which in physical terms means a choice of the $r\to 0$ boundary condition. Two choices which preserve the scale invariance of the differential operator 
$-{d^2\over dr^2 }+{h\over r^2}$ are to disallow the $r^{\half +\sqrt{h+{1\over 4}}}$ or to disallow the $r^{\half -\sqrt{h+{1\over 4}}}$ mode. The former is a fine-tuned fixed point (unstable fixed point) while the latter is a stable fixed point. 

The discussion of the meaning of the two solutions of \eqref{eq_std&alt quantization} is entirely analogous. The modes in~\eqref{eq_std&alt quantization} are admissible as long as  
\begin{equation}\label{ubound}\Delta_s^\pm \geq 0~.\end{equation} 
This positivity constraint on scaling dimensions is a consequence of unitarity. 
The solution ($+$) is known as the standard choice (or standard quantization) and the solution ($-$) is known as the alternative choice (or alternative quantization) \cite{burges1986supersymmetry,d2004supersymmetric,Giombi:2021uae}. Of course, as in the quantum mechanical digression, one can also choose mixed boundary conditions, but those that correspond to fixed points of the defect renormalization group flows are the ones where one of the modes is set to zero. The flow is implemented by integrating bilinears involving $\psi_s^\uparrow$ and $\psi_s^\downarrow$ on the defect; we will discuss these bilinears later. The scaling dimension of $\psi_s$ is $\Delta_s^{\pm}$. The $so(2)$ quantum numbers of $\psi_s^\uparrow$ and $\psi_s^\downarrow$ are  $s+\alpha+\frac{1}{2}$ and $s+\alpha-\frac{1}{2}$, respectively. A special case worth mentioning is the flat space with no monodromy defect, where $\alpha=0$, $\beta=1$ and $\Delta^-_{\pm\frac12}=0$ saturates the bound~\eqref{ubound}. However, an operator of dimension 0 has to be decoupled from the bulk and therefore the corresponding modes furnish the Hilbert space of a decoupled qubit on the defect. At $\alpha=0$ the RG flow to the standard (trivial) fixed point corresponds to gapping the qubit. More generally we can describe the alternative fixed point for generic values of $\alpha$ by coupling a qubit to the standard fixed point. (This is a fermionic analog of the Hubbard-Stratonovich trick.)
This discussion is perfectly parallel to that of~\cite{Nagar:2024mjz}, though the setup is a little different. 

As we demonstrate in the appendix \eqref{eq_appendix_ads fermion green function asy 1} and \eqref{eq_appendix_ads fermion green function asy 2}, whether we choose standard or alternative quantization is crucial and leads to entirely different physical predictions. Of course, it is more natural to choose the boundary conditions that correspond to stable infrared fixed points of the defect renormalization group flows. 
 
Finally, we comment on the time-reversal symmetry $\mathcal{T}$ that changes the flux $\mathcal{T}:\alpha\to 1-\alpha$. Naively, one expects the DCFT to be $\mathcal{T}$-symmetric when $\alpha=0$ or $\alpha=\frac{1}{2}$. As we will see momentarily, that is true for $\alpha=0$ but is more involved for $\alpha=\frac{1}{2}$. The subtlety can be traced to the bilinear form of the defect spinor $\psi_s = (\psi_s^\uparrow, \psi_s^\downarrow)$, which flips its sign under time-reversal $\mathcal{T}: \bar{\psi}_s\psi_s\to -\bar{\psi}_s\psi_s$.

\subsection{One-point functions and defect parameters}

In this subsection, we continue the study of the conical and magnetic flux defect in the free fermion theory~\eqref{eq_fermion action}. At the standard fixed point $\Delta_{s}=\frac{1}{2}+\frac{|s+\alpha|}{\beta}$ for every $s\in \mathbb{Z}+\frac{1}{2}$, we note that all defect deformations are irrelevant for $\alpha\neq \frac{1}{2}$. (Recall that $\alpha\in [0,1)$ is restricted to the fundamental domain.) Therefore, the standard fixed point is stable in the infrared and we expect it to match with the lattice theory in the long-distance limit. Corrections are controlled by the cut-off scale of the field theory, which is typically identified with the lattice spacing $\ell$. While bulk irrelevant operators lead to corrections suppressed by integer powers of $\ell$, those from defect operators such as the defect Dirac masses $\int dt \bar{\psi}_{s}\psi_{s}$ scale as $O(\ell^{2|s+\alpha|/\beta})$. Thus, when $|\alpha-\frac{1}{2}|<\frac{\beta}{2}$, finite-size defect corrections dominate over those of the bulk. The defect deformation $\int dt \bar{\psi}_{-\frac{1}{2}}\psi_{-\frac{1}{2}}$ approaches marginality as $\alpha\to \frac{1}{2}$, and it leads to large finite-size corrections. 

Note, that there are no defect operators that can change the magnetic flux $\alpha$ or the conical angle $\beta$, since both of these are measurable at infinity. Defect operators, roughly speaking, can only change the boundary conditions at the defect itself.

The fact that at $\alpha=\frac{1}{2}$ there is a marginal operator in the continuum $\int dt \bar{\psi}_{-\frac{1}{2}}\psi_{-\frac{1}{2}}$ means that the large volume limit and the half-flux limit $\alpha \rightarrow \frac{1}{2}$ do not commute. If we start at $\alpha\neq \frac{1}{2}$ and slowly change $\alpha$ we will land on a point on the conformal manifold that does not necessarily coincide with the point on which the lattice theory lands. 

Quadratic deformations $\int dt \bar{\psi}_{s}\psi_{s}$ are analogous to double trace deformations as in~\cite{gubser2003universal}, and they lead to one-loop exact defect RG flows in certain scheme. For $|s+\alpha|<\frac{\beta}{2}$, such RG flows connect the alternative fixed point with the IR-stable standard fixed point. As we said, for $\alpha=\frac{1}{2}$ there is a defect conformal manifold. 

The observables that we are going to focus on in the infinite-volume continuum theory are one-point functions, especially the expectation value of the current in the vacuum and the expectation value of energy density. These are interesting theoretically and also natural to compare with lattice calculations. The one-point function of the $U(1)$ current $\langle J_{\mu} \rangle$ at defect fixed points is fixed by  $sl(2,\mathbb{R}) \times so(2)$ symmetry and conservation up to a constant $C_J$ \cite{Billo:2016cpy}, such that 
\begin{equation}
\label{eq_def azimuthal current}
    \begin{aligned}
\langle J_\mu \rangle\equiv& \langle i \Bar{\Psi} \gamma_\mu \Psi \rangle=\frac{\delta_{\mu \theta}}{2\pi r}C_{J}~.
    \end{aligned}
\end{equation}
There is no charge density away from the defect but there can be a current. That is because particles spin around the defect and anti-particles (holes) are spinning in the opposite direction. 

The constant $C_J$ depends on the flux $\alpha$ and the conical singularity $\beta$. Similarly, the stress-energy tensor one-point function is fixed by conservation, tracelessness, and the conformal symmetry up to a single coefficient (our metric is as in~\eqref{eq_cone metric}, $g_{\mu\nu}=\text{diag}\left(-1,1,\beta^2r^2\right)$)
\begin{equation}
\label{eq_def energy density}
\langle T_{\mu \nu } \rangle\equiv \langle \Bar{\Psi}  \gamma_\mu \partial_\nu \Psi \rangle=\text{diag}\left(1,-1,2\beta^2r^2\right)\frac{C_T}{4\pi r^3}~.
\end{equation}
$C_T$ in \eqref{eq_def energy density}, as a function of $\alpha$ and $\beta$, is commonly referred to as the defect conformal weight \cite{kapustin2006wilson,hung2014twist}. When $\beta=1$, we also interpret the component $\langle T_{tt} \rangle$ as the local energy density. Note that there are no off-diagonal components in~\eqref{eq_def energy density}, and there is therefore no angular momentum density, which aligns with the physical picture that holes and anti-holes are rotating around the defect in opposite directions. Finally, we also comment on the charge conjugation $\mathcal{C}:A_\mu\to -A_{\mu}$ and time-reversal $\mathcal{T}$. The $\mathcal{C}\mathcal{T}$-symmetry is preserved by the flux presence, and charged observables vanish identically if it is not spontaneously broken. Indeed, we find that the bulk fermionic condensate $\la\bar{\Psi} \Psi\ra=0$. (Note that in the free boson model, there is a condensate as is shown in \eqref{appendix_eq_boson one point functions}.) 
 
$C_J$ and $C_T$ can be obtained by regularizing certain infinite sums,  \begin{equation}
\label{eq_coincident point decomposition}
    \begin{aligned}
C_J=&-\beta\sum_{s\in \mathbb{Z}+\frac{1}{2}}\langle\Bar{\tilde{\Psi}}_s\tilde{\Psi}_s\rangle~,\\
C_T=&-\frac{1}{\beta}\sum_{s\in \mathbb{Z}+\frac{1}{2}}(s+\alpha)\langle\Bar{\tilde{\Psi}}_s\tilde{\Psi}_s\rangle~.
    \end{aligned}
\end{equation}
The detailed computation can be found in the appendix \eqref{eq_appendix_ads fermion coincident point function}. At the standard fixed point, we obtain for $\alpha<\frac{1}{2}$ and $\beta=1$ that: 
\begin{equation}
\label{eq_CJ&CT at beta=1}
    \begin{aligned}
C_{J}(\alpha)=&\frac{\tan (\pi  \alpha )}{8} \left(1-4 \alpha ^2\right)~,\\
C_{T}(\alpha)=&\frac{\tan (\pi  \alpha )}{12} \alpha\left(1-4 \alpha ^2\right)~.
    \end{aligned}
\end{equation}
For $\alpha>\frac{1}{2}$, it follows from the time-reversal symmetry $\mathcal{T}$ that $C_J(\alpha)=-C_J(1-\alpha)$ and $C_T(\alpha)=C_T(1-\alpha)$. The $s=-\frac{1}{2}$ alternative fixed point can be reached by fine-tuning the defect Dirac mass $\int dt \bar{\psi}_{-\frac{1}{2}}\psi_{-\frac{1}{2}}$. From the decomposition \eqref{eq_coincident point decomposition}, we find for the multi-critical fixed point
\begin{equation}
\label{eq_alter CJ&CT at beta=1}
    \begin{aligned}
C'_{J}=&C_J+(\alpha-\frac{1}{2})\tan{(\pi \alpha)}\\
C'_{T}=&C_T+(\alpha-\frac{1}{2})^2\tan{(\pi \alpha)}\\
    \end{aligned}
\end{equation}
where $\alpha<\frac{1}{2}$ and $\beta=1$.
\begin{figure}[tb]
\centering
\includegraphics[width=.3\textwidth ]{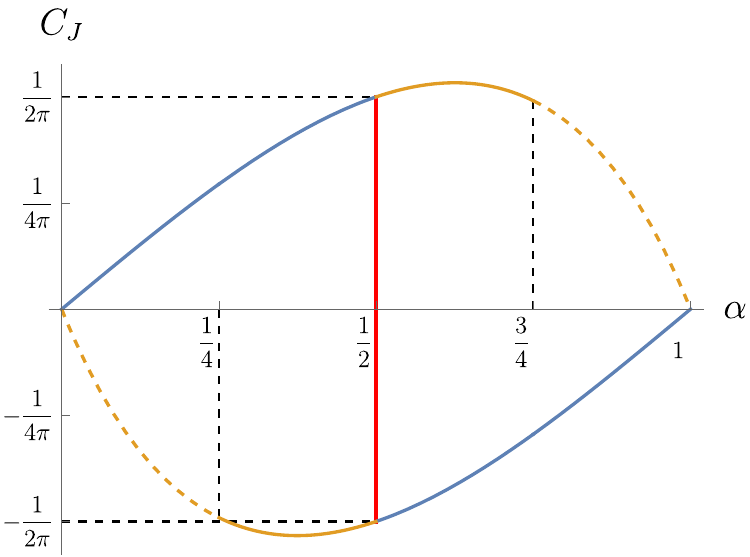}
\hspace{.01 \textwidth }
\includegraphics[width=.3\textwidth ]{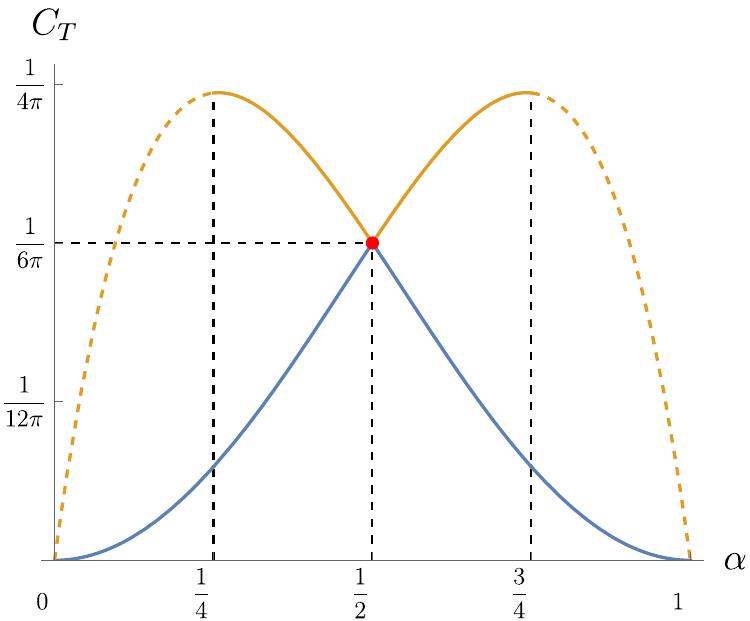}
  \caption{\label{pic_beta=1 one point functions} Azimuthal current $C_J$ and energy density $C_T$ of the defect without conical singularity ($\beta=1$). Throughout this paper, we use blue curves to denote values at the standard fixed point, while orange curves denote those at the alternative fixed points. The dashed part of the orange curve represents where the defect quartic interaction becomes relevant. The red line and point mark the values at the defect conformal manifold.}
\end{figure}

The conformal manifold at $\alpha=\frac{1}{2}$ is due to the $s=-\frac{1}{2}$ $\text{AdS}_2$ mode in \eqref{eq_fermion map to AdS} becoming massless. As a consequence, the alternative and the standard quantization are both part of a continuum of possible boundary conditions, all preserving the conformal symmetry. We find that as we scan over the boundary conditions, the azimuthal current can take values $-\frac{1}{2\pi}\leq C_J \leq \frac{1}{2\pi}$ \footnote{The Zamolodchikov distance on the conformal manifold is $\arcsin{(2\pi C_J)}$. We thank A. Sharon for a helpful discussion.} while the conformal weight $C_T=\frac{1}{6\pi}$ is single valued. Different points of the conformal manifold are reached by the operator $\int dt \bar{\psi}_{-\frac{1}{2}}\psi_{-\frac{1}{2}}$. We remark that in general, the defect Dirac mass deformation $\int dt \bar{\psi}_{-\frac{1}{2}}\psi_{-\frac{1}{2}}$ breaks the time-reversal symmetry $\mathcal{T}$. There is one point on the defect conformal manifold where $C_J=0$, where $\mathcal{T}$ is preserved. Results regarding the one-point functions $C_J$ and $C_T$ are summarized in \cref{pic_beta=1 one point functions}. We remark that in the free boson model where the defect conformal manifold is absent, $C_J$ and $C_T$ in \eqref{appendix_eq_boson one point functions} as functions of flux are analytic for $\alpha\in [0,1)$.

An important characteristic of line defects is the $g$-function, which is a monotone of defect RG flows~\cite{Cuomo:2021rkm, Casini:2022bsu, Casini:2023kyj} (see also~\cite{Kobayashi:2018lil} for earlier, important work).
For monodromy defects in flat space ($\beta=1$), it is easy to prove that at defect fixed points, 
\begin{equation}
\label{eq_g function formula}
    \partial_\alpha \log g=2\pi C_J(\alpha)~.
\end{equation}
Therefore we can compute the $g$ function at fixed points if only we know the current, which for the standard and alternative fixed points is given by \cref{eq_CJ&CT at beta=1}, and \cref{eq_alter CJ&CT at beta=1}, respectively. From \cref{pic_beta=1 one point functions} we see that $g\geq 1$ for $\alpha\in [0,1)$ of the standard quantization.  Finally, for alternative quantization $g(\alpha=0)=2 $ which allows for a very simple interpretation of the alternative fixed point in the absence of the defect, namely, it is just a decoupled qubit. Finally, $g$ has to be constant on conformal manifolds, as follows from it being a monotone of the renormalization group. Indeed it can be explicitly verified that it is a constant on the conformal manifold at $\alpha=\frac{1}{2}$. 

\subsection{Flux displacement}

\begin{figure*}
\centering
\includegraphics[width=.3\textwidth ]{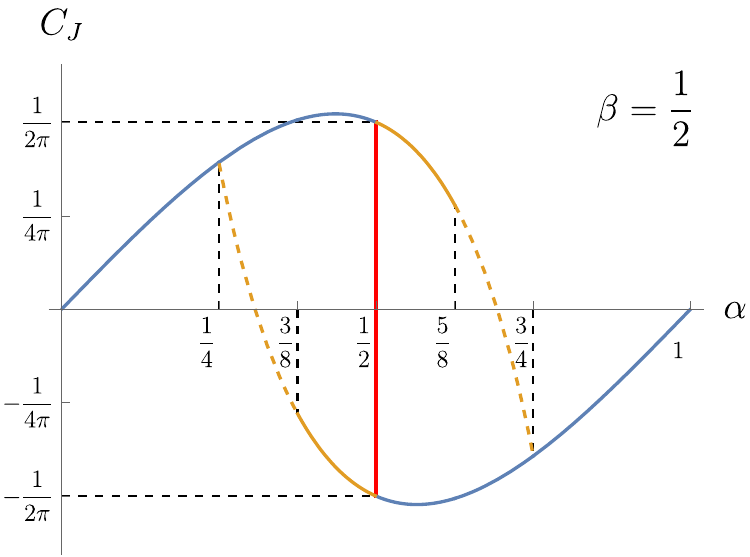}
\hspace{.01 \textwidth }
\includegraphics[width=.3\textwidth ]{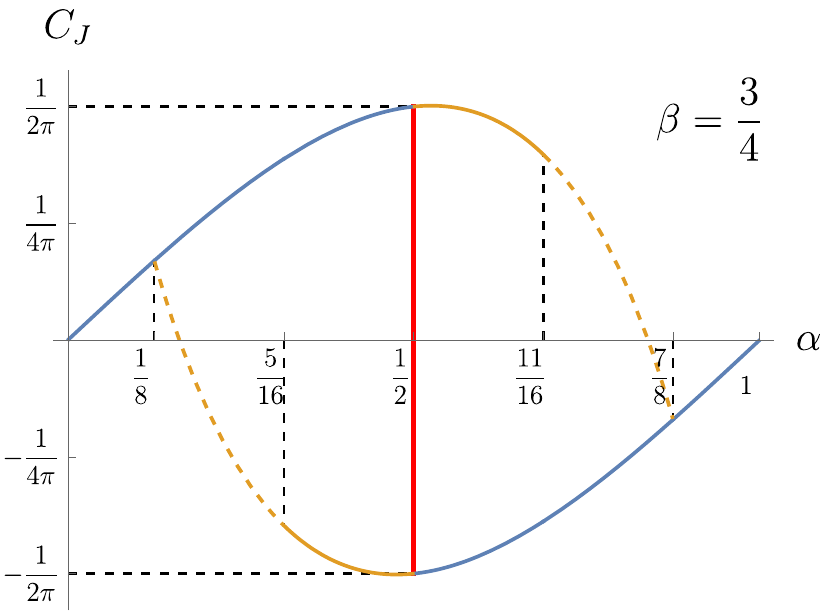}
\hspace{.01 \textwidth }
\includegraphics[width=.3\textwidth ]{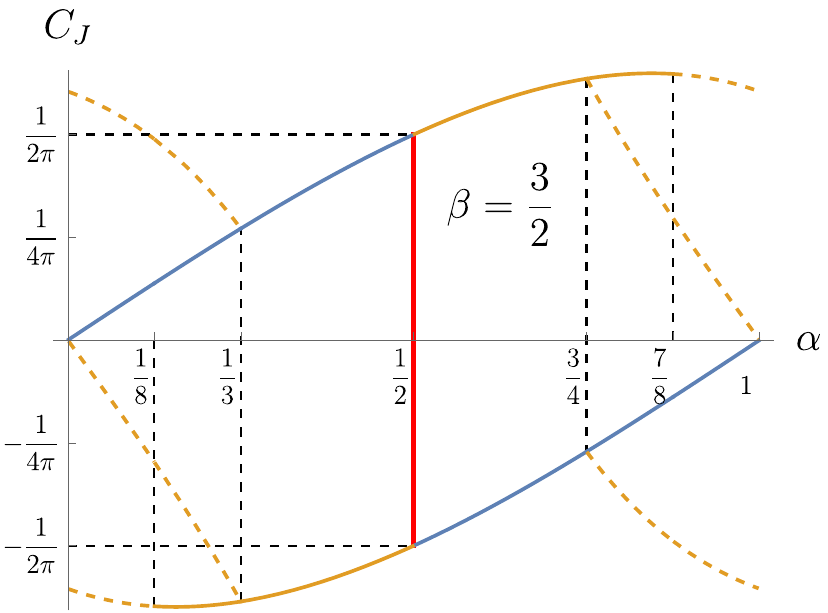}
\includegraphics[width=.3\textwidth ]{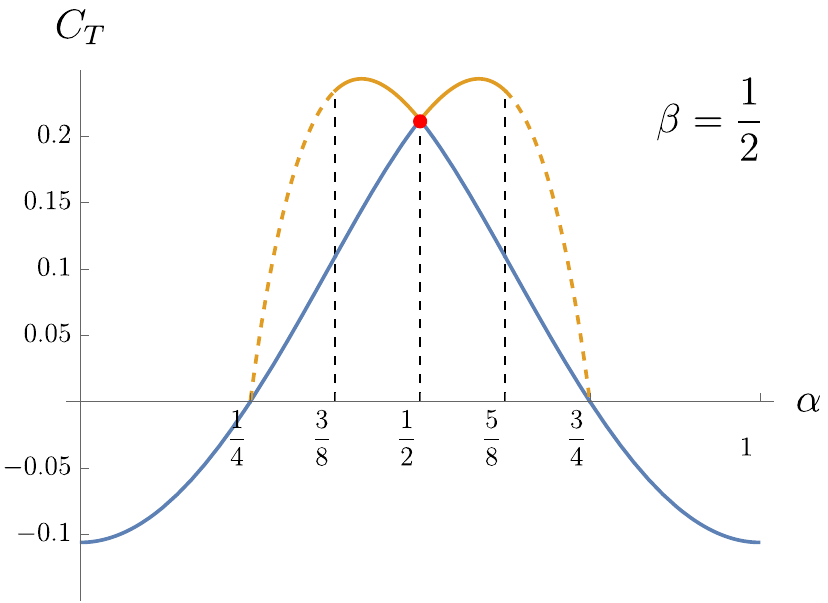}
\hspace{.01 \textwidth }\
\includegraphics[width=.3\textwidth ]{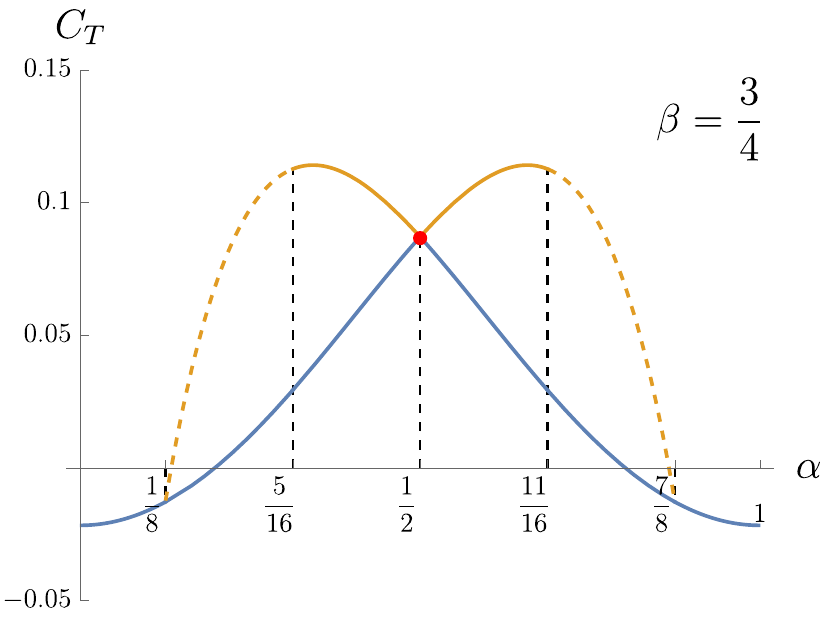}
\hspace{.01 \textwidth }
\includegraphics[width=.3\textwidth ]{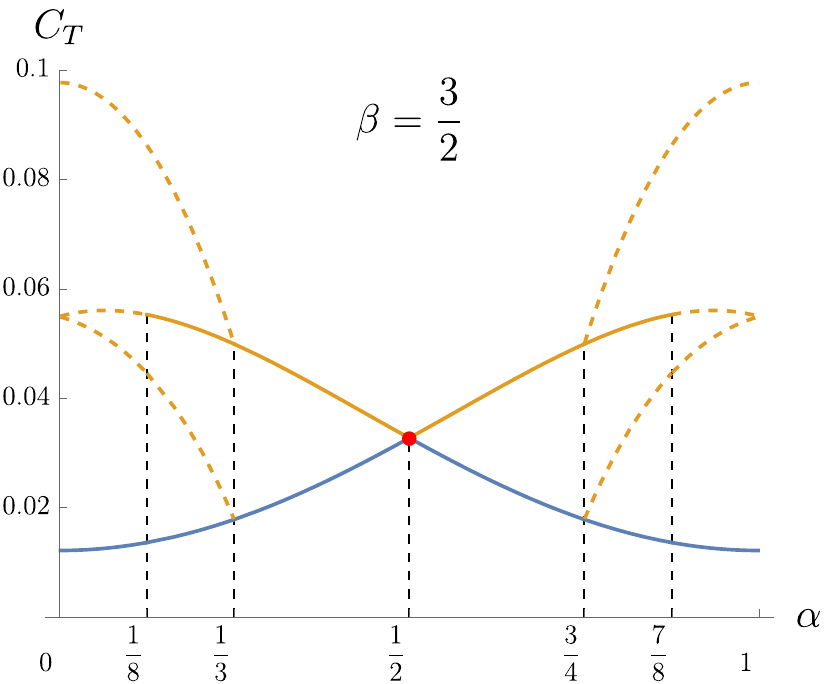}
  \caption{\label{pic_beta=1/2 and beta=3/2 one point functions} Azimuthal current $C_J$ and conformal weight $C_T$ of the defects with conical singularity. From the left to the right: defects with a conical deficit ($\beta=\frac{1}{2}$ and $\beta=\frac{3}{4}$) and defects with a conical excess ($\beta=\frac{3}{2}$).}
\end{figure*}

An important hallmark of flat space defect conformal field theory is the existence of the displacement operator~\cite{McAvity:1993ue} (see also~\cite{McAvity:1995zd, Liendo:2012hy, Jensen:2015swa}). The physical reasoning is clear -- there must be an operator that moves the location of the defect. Let us consider a local perturbation $\delta X_{\perp}^i(t)$ to the line defect location, where $i$'s are the transversal directions. By definition, the displacement operator $D^i$ reads
\begin{equation}\delta S=\int dt D^i(t)\delta X_{\perp}^i(t)~, \end{equation}
such that it characterizes the system's response to $\delta X_{\perp}^i$. From dimensional analysis, $D^i$ is a defect operator of scaling dimension 2 and it clearly has transverse spin 1. 

A monodromy defect can be moved around, so we certainly expect that the defect fixed points admit a displacement operator. Indeed, at the standard fixed point with $\beta=1$, $D^i$ is identified with the lowest-lying spin-$1$ defect operators $\bar{\psi}_{-\frac{1}{2}}\psi_{\frac{1}{2}}$ and $\bar{\psi}_{\frac{1}{2}}\psi_{-\frac{1}{2}}$, whose scaling dimension is exactly as required $\Delta(\bar{\psi}_{-\frac{1}{2}}\psi_{\frac{1}{2}})=\Delta(\bar{\psi}_{\frac{1}{2}}\psi_{-\frac{1}{2}})=2$. 

The conical singularity is different. It cannot be moved around by manipulations in the region of the defect\footnote{We thank J. Maldacena for a discussion of this topic.}. Technically the difference stems from the the fact that the metric and not the spin connection is the fundamental variable, while in gauge theories the gauge connection is fundamental. The metric deformation required to move around a conical defect is non-normalizable, unlike the gauge field deformation. Remarkably, this is the continuum manifestation of the well-known fact that disclinations are immobile in crystalline solids (see e.g. \cite{gromov2024colloquium} and references therein). 

Indeed, at the standard fixed point with $\beta\neq 1$, we find from \eqref{eq_std&alt quantization} that $\Delta(\bar{\psi}_{-\frac{1}{2}}\psi_{\frac{1}{2}})=\Delta(\bar{\psi}_{\frac{1}{2}}\psi_{-\frac{1}{2}})=1+\frac{1}{\beta}$. Such operators are the lowest-lying ones in the bulk-to-defect OPE of the $U(1)$ current operator\footnote{In special cases where $\beta\in \mathbb{Z}$ is an integer, there exist defect operators with spin $1$ and scaling dimension $2$. However, it is only when $\beta=1$ that they are the lowest-lying spin $1$ operators.}. Physically, they describe perturbation to the location of the magnetic flux while keeping the conical singularity unmoved.

\subsection{Current and energy density for conical defects}

We now move on to computing the current and energy density in the presence of a conical defect. The details are in appendix \ref{appendix_AdS Propagators}. For $C_J$ in \eqref{eq_def azimuthal current} at the standard fixed point, we find an integral representation  
\begin{equation}
\label{eq_conical CJ integral}
C_J=-\frac{1}{\pi}\left[\int_{0}^1\left(\frac{u^{\frac{\alpha}{\beta}}-u^{-\frac{\alpha}{\beta}}}{u^{\frac{1}{2\beta}}-u^{-\frac{1}{2\beta}}}\right)\frac{u^{\frac{v}{2}-1}du}{(1-u)^v}\right]_{v\to 2}~,
\end{equation}
where the integral $\int du$ is evaluated as the analytical continuation of the index $v\to 2$. Similarly, for $C_T$ in \eqref{eq_def energy density} we find for the standard fixed point that
\begin{equation}
\label{eq_conical CT integral}
\mathtoolsset{multlined-width=0.88\displaywidth}
\begin{multlined}
C_T=\frac{1}{\pi \beta^2}\left[\int_0^1 \left(\frac{(\frac{1}{2}-\alpha)(u^{\frac{2\alpha+2}{2\beta}}+u^{-\frac{2\alpha+1}{2\beta}})}{\left(u^{\frac{1}{2\beta}}-u^{-\frac{1}{2\beta}}\right)^2} \right. \right.\hfill\\
\hfill\left.\left. +\frac{(\frac{1}{2}+\alpha)(u^{\frac{2\alpha-1}{2\beta}}+u^{\frac{1-2\alpha}{2\beta}})}{\left(u^{\frac{1}{2\beta}}-u^{-\frac{1}{2\beta}}\right)^2}\right)\frac{u^{\frac{v}{2}-1}du}{(1-u)^v}\right]_{v\to 2}~.
\end{multlined}
\end{equation}
As a consistency check, \eqref{eq_CJ&CT at beta=1} is reproduced by taking $\beta=1$. Denoting the spins for which we impose alternative quantization by ${s}'\in \mathbb{Z}+\frac{1}{2}$, the generalization of \eqref{eq_alter CJ&CT at beta=1} reads
\begin{equation}
\label{eq_alternaive fixed point current and energy}
\begin{aligned}
{C}'_J=&C_J+\sum_{{s}'}\frac{{s}'+\alpha}{\beta}\cot(\pi\frac{|{s}'+\alpha|}{\beta})~,\\
{C}'_T=&C_T+\sum_{{s}'}\frac{({s}'+\alpha)^2}{\beta^3}\cot(\pi\frac{|{s}'+\alpha|}{\beta})~.
    \end{aligned}
\end{equation}
Remember that the $sl(2,\mathbb{R})$ unitarity bound requires that $|{s}'+\alpha|<\frac{\beta}{2}$. Some examples of results for $C_J$ and $C_T$ can be found in \cref{pic_beta=1/2 and beta=3/2 one point functions}.

From \eqref{eq_conical CJ integral} we find $C_J(\alpha=0)=0$ (as expected, there is no current in the absence of an Aharonov-Bohm flux) and, remarkably, $C_J(\alpha=\frac{1}{2})$ ranges from $-\frac{1}{2\pi}$ to $\frac{1}{2\pi}$, regardless of the cone angle $2\pi \beta$. 

Time-reversal symmetry imposes $C_J=0$ at $\alpha =0$ and $\alpha=\frac12$. In this case, we cannot distinguish the half flux and the zero flux using the current only but we can still measure a difference in the energy density:
\begin{equation}
\label{eq_fermion energy difference}
\mathtoolsset{multlined-width=0.88\displaywidth}
\begin{multlined}
C_T(\alpha=\frac{1}{2})-C_T(\alpha=0)=\\
\frac{1}{4\pi \beta^2}\left[2+\int_{0}^1\left(\frac{u^{\frac{1}{2\beta}}-1}{u^{\frac{1}{2\beta}}+1}\right)^2\frac{du}{(u-1)^2}\right]>0~,
\end{multlined}
\end{equation}
 which, among many other observables, distinguishes the time-reversal symmetric defects. (Notably, a similar inequality \eqref{eq_appendix_boson energy difference} also holds for the free boson model.)

\section{Observables at criticality in lattice models}\label{sec:critical_observables}

In this section, we will discuss how to numerically measure currents and energy densities in lattice models with a disclination or dislocation. We will then present numerical results for several free fermion models at criticality, including the QWZ model we have already begun to discuss. Our numerical results confirm the following non-trivial theoretical predictions of the defect CFT:
\begin{itemize}
    \item We find a universal shift in the azimuthal current and energy density as a function of applied flux due to the emanant flux of the disclination and dislocations. This provides a direct way to measure the invariants $\{s_o^{(i)}, k_\star^{(i)}\}$ at criticality and confirms the existence of the emanant flux. 
    \item 
    We find that the absolute values of the maxima and minima of $C_T(\alpha)$ are larger in the presence of the disclination as compared to without the disclination. 
    \item We find qualitative agreement for the shape of $C_T(\alpha)$ and $C_J(\alpha)$ as a function of the applied flux. 
    \item The azimuthal current with and without particle-hole symmetry provides evidence for the existence of the conformal manifold at $\alpha = 1/2$. 
\end{itemize}
We have not extracted the scaling exponents $\Delta_s^\pm$, which we leave to future work. 

\subsection{Numerical procedure generalities}

As we have discussed, we expect a disclination or dislocation to contribute an additional $U(1)$ flux to the continuum fermions proportional to $s_o^{(i)}$ or $\vec{k}_\star^{(i)}\cdot\vec{b}$, respectively. This can be understood as a consequence of the $s_o^{(i)} \omega_o$ and $\frac{1}{2\pi}\vec{k}_\star^{(i)}\cdot\vec{R}$ terms in the effective Lagrangian as well as the UV-IR homomorphism. The discussion of \cref{sec:universality_defects} points to the local energy density $\mathcal{E}_\vi$ and the azimuthal current $J^\theta_\vi$ as universal observables sensitive to this emanant flux. These quantities can be directly measured in lattice models with or without a crystalline defect.

At critical points where there is only a single Dirac cone, these microscopic observables capture the contribution due to the corresponding massless Dirac fermion. The microscopic current is defined via the continuity equation 
\begin{equation}
    \frac{\partial n_\vi}{\partial t}+\vec{\nabla}\cdot \vec{J}_\vi=0,
\end{equation}
where $n_\vi=\sum_\sigma c_{\vi,\sigma}^\dagger c_{\vi,\sigma}$ is the number operator on site $\vi$, $\vec{J}_\vi$ is the current at site $\vi$, and the divergence is defined on the lattice. The time derivative can be evaluated via Heisenberg evolution of the number operator, and the resulting equation can be solved for the component of the current in the $\vec{r}$ direction:
\begin{equation}
    J_\vi^{\vec{r}}=
    2i\sum_{a,b}h_{\vi,\vi+\vec{r}}^{a,b}c^\dagger_{\vi,a}c_{\vi+\vec{r},b}-\left(h_{\vi,\vi+\vec{r}}^{a,b}\right)^*c^\dagger_{\vi+\vec{r},a}c_{\vi,b}.
\end{equation}
To obtain the azimuthal current in which we are interested, we project $J_\vi^{\vec{r}}$ along the $x$- and $y$-axes to obtain $J_\vi^x$ and $J_\vi^y$. We can finally obtain $J_\vi^\theta$ via the linear combination $J_\vi^\theta = -i_y J^x_\vi + i_x J^y_\vi$. The local energy density is defined as
\begin{equation}
    \mathcal{E}_{\vi}=\sum_{\vj,a,b} H_{\vi\vj}^{ab}=\sum_{\vj,a,b} h_{\vi\vj}^{ab} c^\dagger_{\vi,a}c_{\vj,b}.
\end{equation}
We are interested in the expectation values of these operators in the many-body ground state.

The current as obtained above requires us to define spatial $x$- and $y$-directions. This presents no difficulty on the clean lattice or lattice with a dislocation. However, for a lattice with a disclination, these directions cannot be defined globally. We can still define them locally and use this local definition to measure the current numerically in a particular patch of the lattice.

In the case where there are multiple massless Dirac fermions at a single critical point (i.e. multiple valleys), the microscopic observables are no longer a good proxy for the one-point functions of the individual Dirac fermions. In some cases, it is possible to explicitly define a valley current and energy density in terms of microscopic creation and annihilation operators, but this can be a subtle process due in part to the absence of a valley symmetry in the UV. We will discuss an example where we were able to define valley observables in \cref{sec:pi_flux_no_potential}.

In order to extract the emanant flux due to a lattice defect, we insert flux $\delta\alpha\in[0,1]$ (in units of $2\pi$) at the defect core and measure the azimuthal current $J^\theta_\vi$ and energy density $\mathcal{E}_\vi$ at various spatial points in the bulk as a function of $\delta\alpha$. We plot the quantities $rJ^\theta_\vi$ and $r^3\mathcal{E}_\vi$, where $r$ is the distance from the defect. Because $J^\theta_\vi$ and $\mathcal{E}_\vi$ are predicted to scale as $\frac{1}{r}$ and $\frac{1}{r^3}$, respectively, this quantity extracts the universal amplitude of the observables and demonstrates the finite-size deviation from the fixed point prediction.  

\begin{figure}[t]
    \centering
    \includegraphics[width=0.9\linewidth]{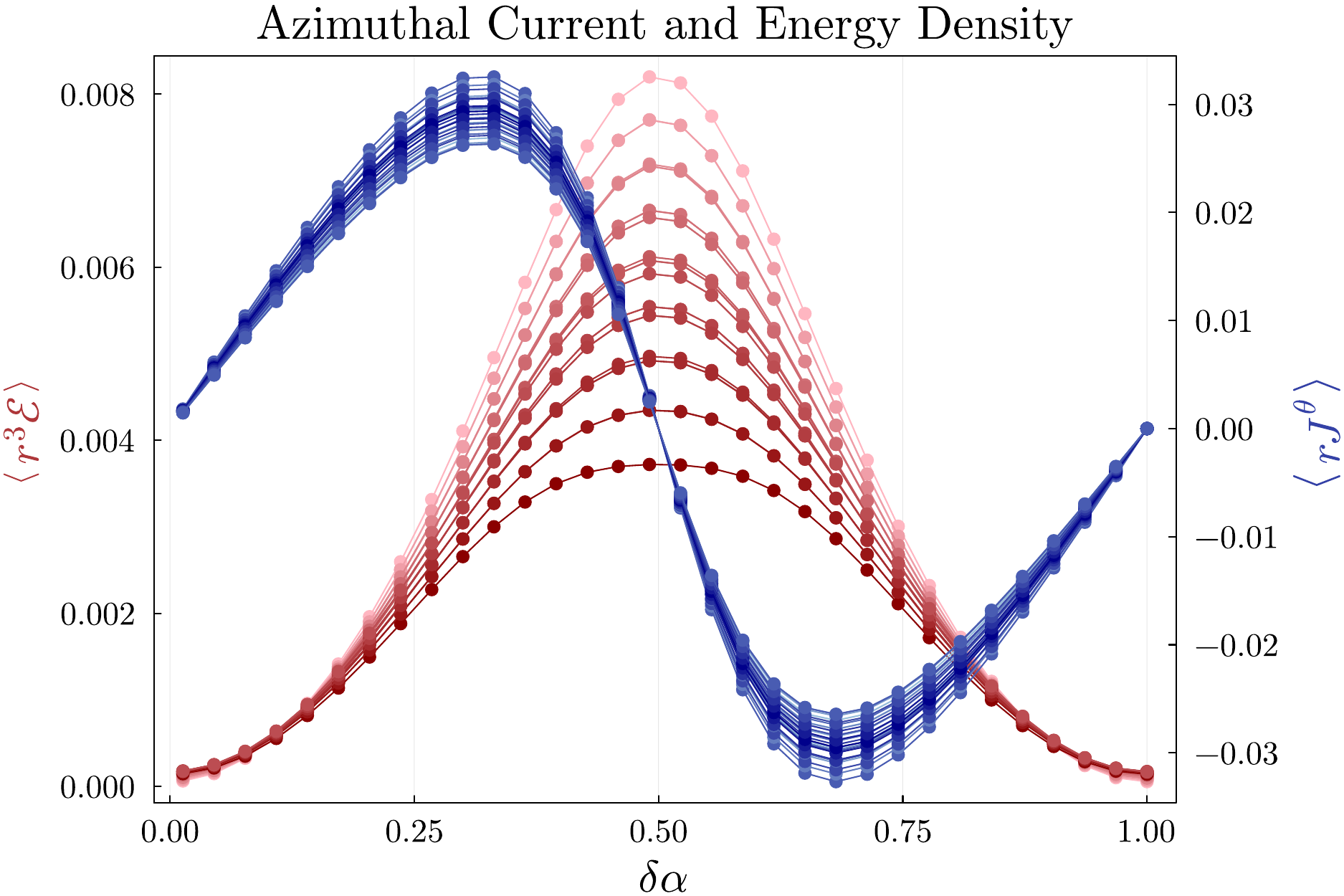}
    \caption{QWZ model, clean lattice. We plot observables at the $m=-1$ critical point on a lattice without crystalline defects, scaling by $r$ and $r^3$ respectively in order to extract the universal amplitude data. Each curve gives the values of the observables at a particular spatial position in the bulk of the lattice. The shading of the curve corresponds to the distance of the spatial point from the flux insertion center, with lighter colors closer.}
    \label{fig:QWZ_clean_current}
\end{figure}

In a model without defects, we expect that $J^\theta=0$ and $\mathcal{E}$ is minimized when $\delta\alpha=0$. As seen in \cref{fig:QWZ_clean_current} for the QWZ model, this occurs when $J^\theta$ crosses from negative to positive values for the flux insertion convention we have chosen. To measure the excess flux due to a crystalline defect, we measure the shift in this point away from $\delta\alpha=0$. The zero crossing in $J^\theta$ and minimum in $\mathcal{E}$ will always be at total flux $\alpha=0$, where 
\begin{align}
    \alpha=\delta\alpha+\alpha_\text{em}
\end{align}
and $\alpha_\text{em}$ is the emanant flux due to the defect. This means that if this point is shifted to some nonzero $\delta\alpha$, then the flux due to the defect is $\alpha_\text{em}=-\delta\alpha$. In each case, we have calculated the average $\delta\alpha$ at which (1) the two zero crossings in $J^\theta$ and (2) the minimum and maximum of $\mathcal{E}$ occur for the plotted data. These measured values are plotted along with the predicted values based on the expected excess flux for each defect. 

\begin{figure*}[t]
        \subfloat[Lattice with $N=5766$ sites and an $A$-centered $\beta=3/4$ disclination]{
            \includegraphics[width=.35\linewidth]{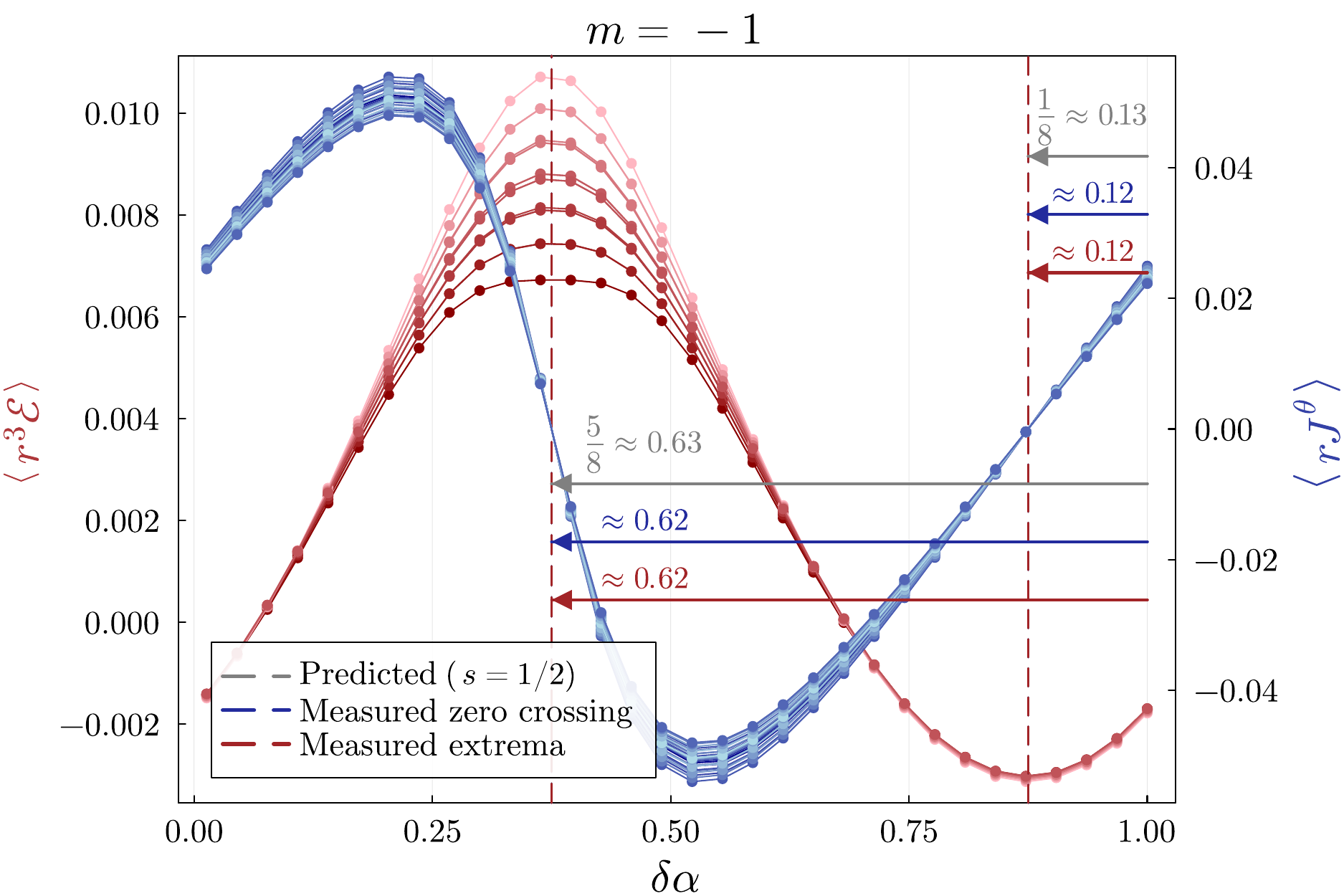}
            
    \qquad\includegraphics[width=.35\linewidth]{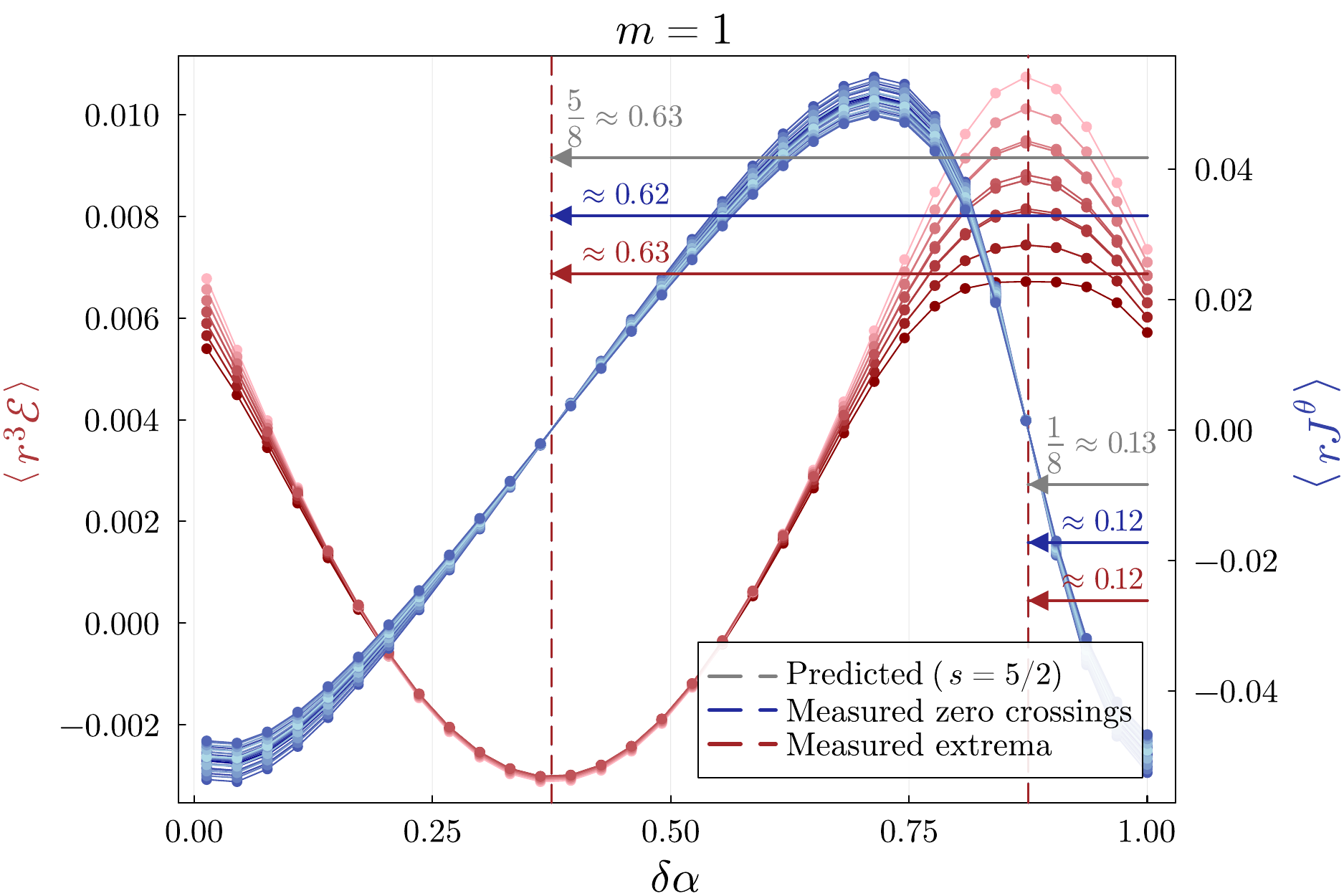}
        }\\
        \subfloat[Lattice with $N=5582$ sites and a $B$-centered $\beta=3/4$ disclination]{
            \includegraphics[width=.35\linewidth]{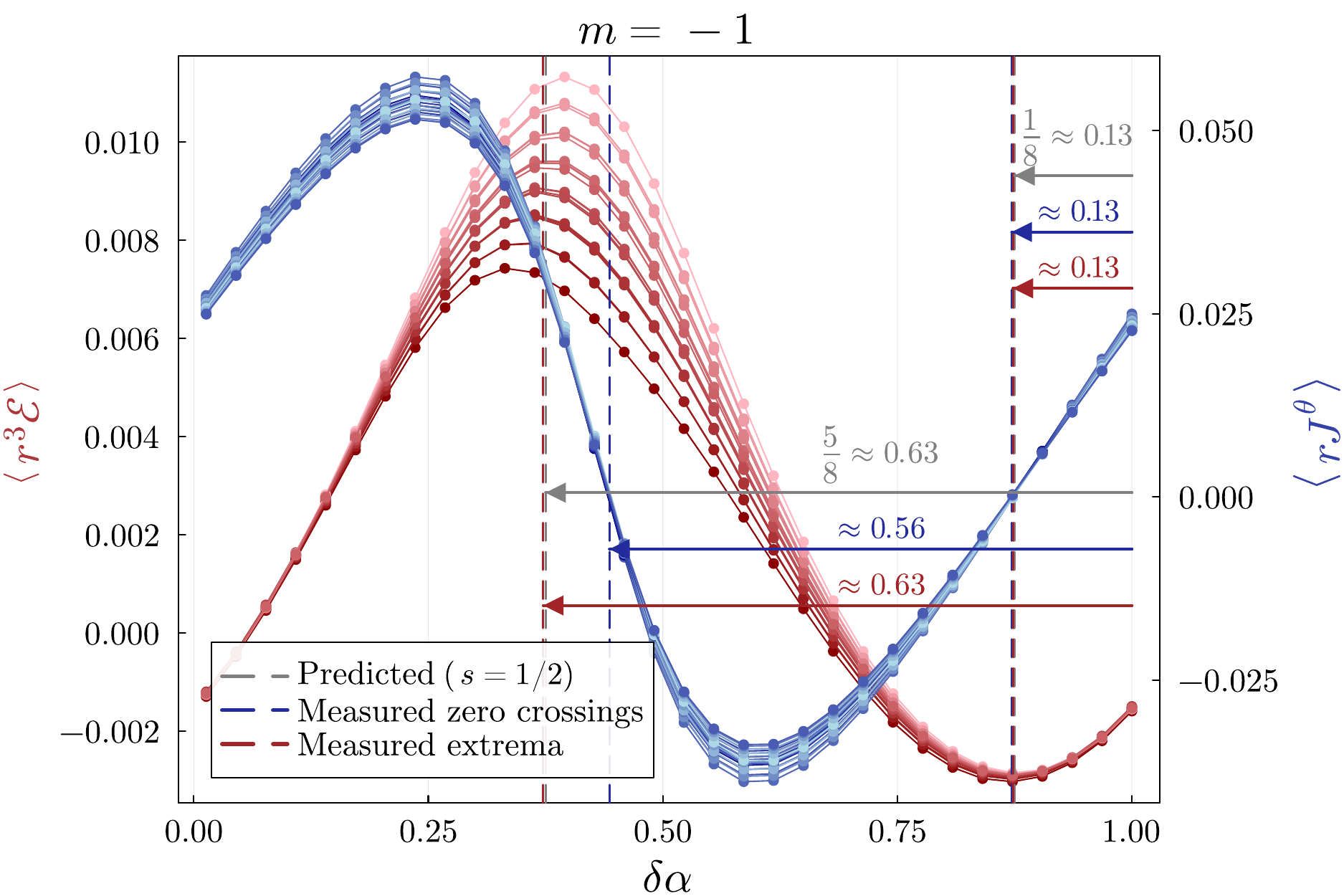}
            \qquad\includegraphics[width=.35\linewidth]{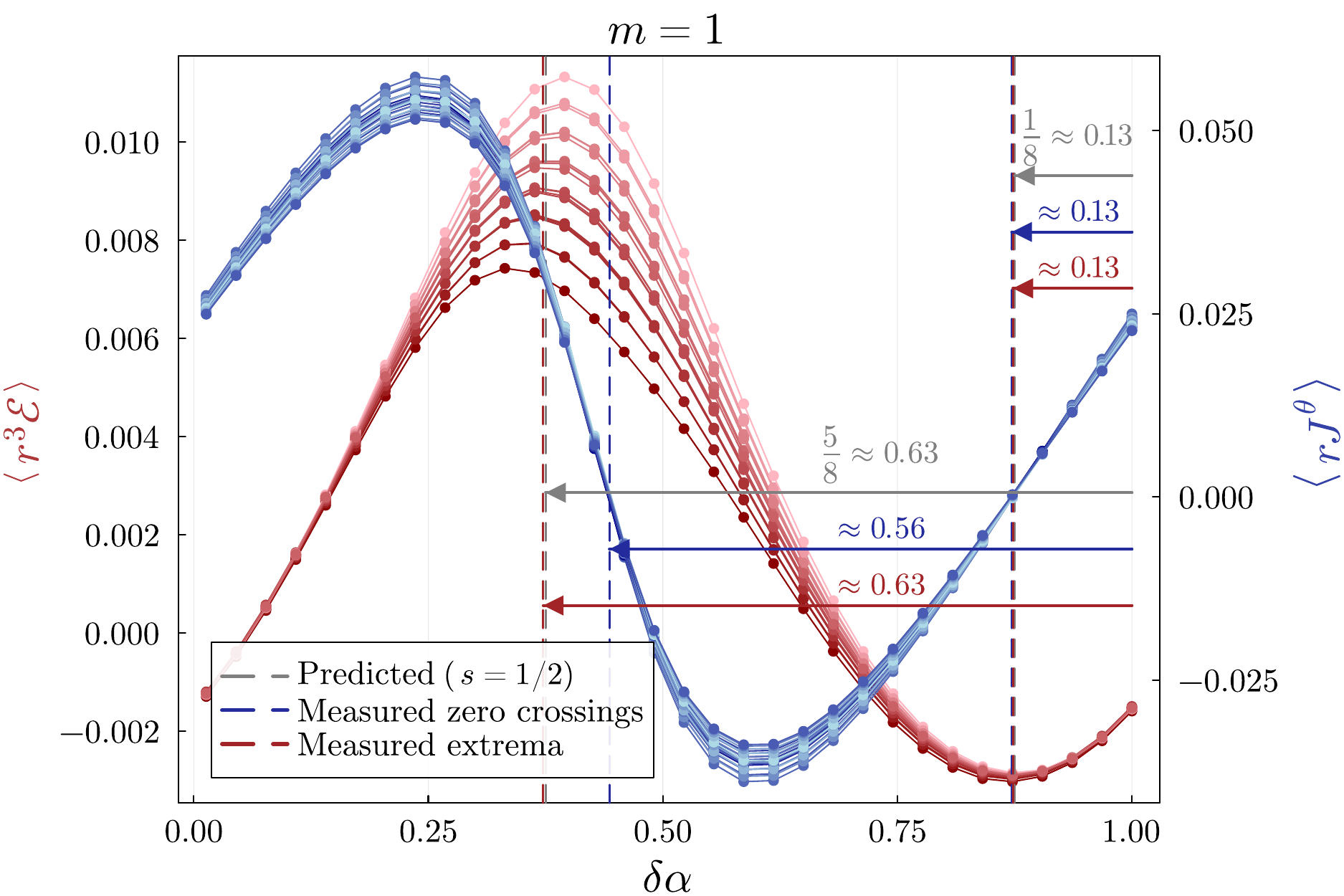}
        }\\
        \subfloat[Lattice with $N=5460$ sites and a $\vec{b}=(-1,0)$ dislocation]{
            \includegraphics[width=.35\linewidth]{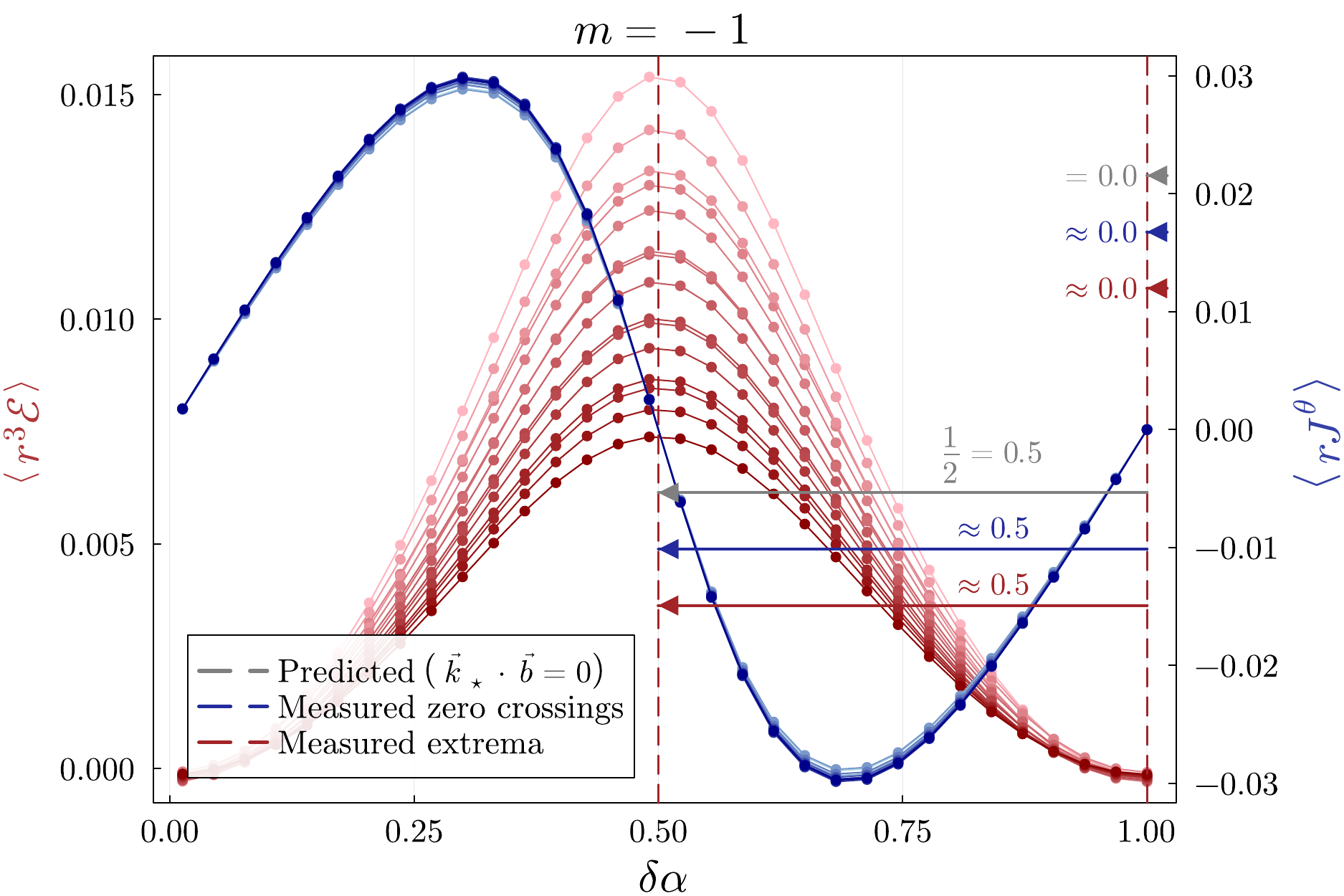}
            \qquad\includegraphics[width=.35\linewidth]{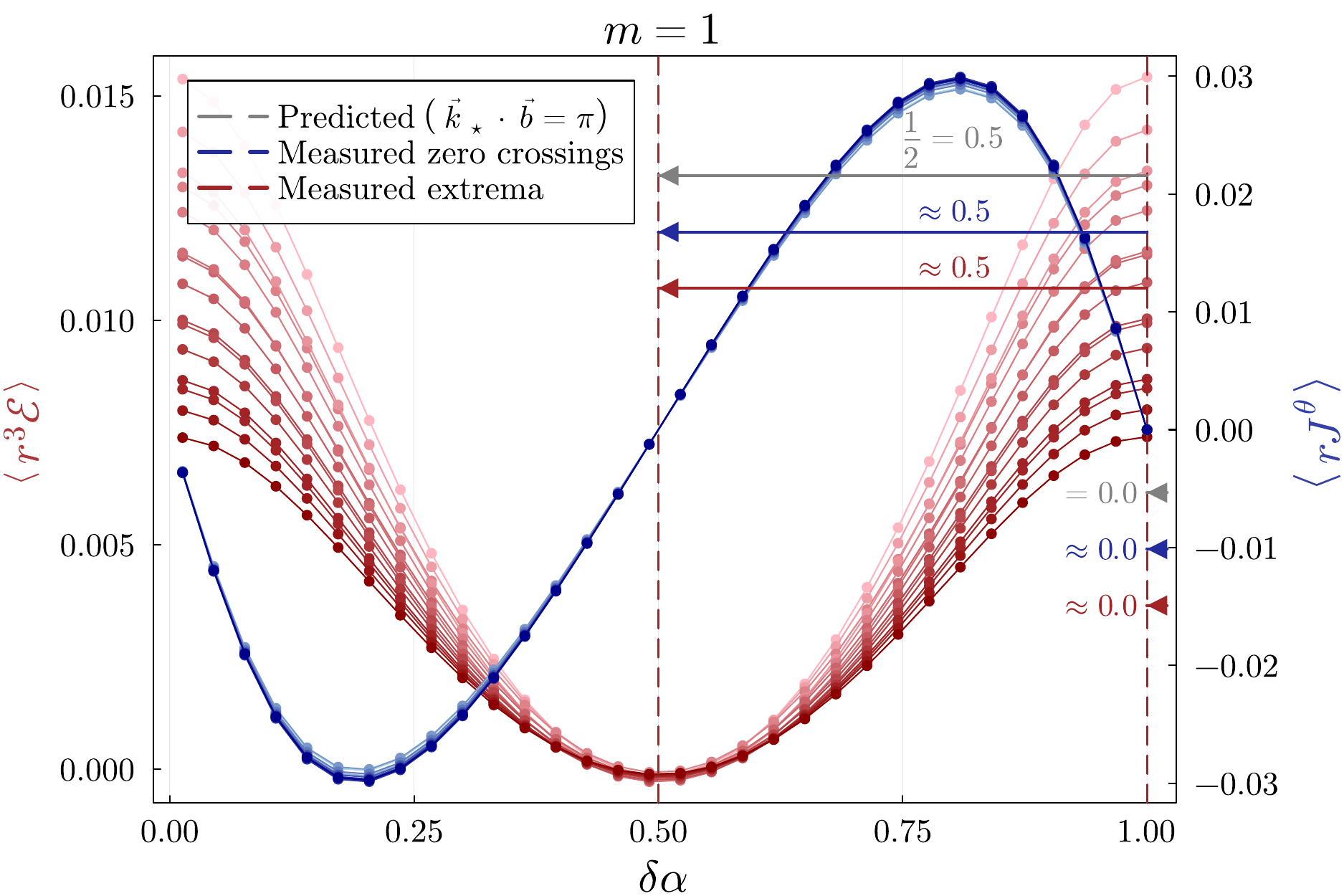}
        }
        \caption{QWZ model with crystalline defects. Here we plot the scaled current and energy density in the QWZ model for the $m=\pm 1$ critical points and for several defects. The shading of each curve in the plot indicates the distances from the defect, with lighter curves corresponding to lattice sites closer to the defect.}
        \label{fig:QWZ_currents}
    \end{figure*}

\subsection{QWZ model}\label{sec:QWZ_observables}

In this section, we will continue with the QWZ model example which was discussed in \cref{sec:QWZ_rot_eigs,sec:QWZ_invariants}. Because there are two inequivalent choices of fourfold rotation center in the QWZ model, there are two distinct disclinations which can be inserted into the lattice. These disclination Hamiltonians are constructed using the lattice rotation operators $C_{4,A}$ and $C_{4,B}$ using the procedure discussed in \cite{zhang2022fractional}. 

The $m=\pm1$ critical points each have a single Dirac cone, so we can use the microscopic quantities as a proxy for the field theory quantities associated with the Dirac fermion. The $m=0$ critical point has two Dirac cones and therefore requires a more complicated treatment, so we do not study it here. The measured observables are reported in \cref{fig:QWZ_currents} for several crystalline defects.

Notice that there is strong agreement between the measured and predicted zero crossing from negative to positive values of $J_\theta$, but not always for the zero crossing from positive to negative $J_\theta$. The former corresponds to $\alpha=0$, while the latter is around $\alpha=\frac{1}{2}$. This phenomenon can be understood in terms of the analysis of \cref{sec:universality_defects}. In that section, we saw that there is a conformal manifold of fixed points for $\alpha=\frac{1}{2}$. This is precisely the set of field theories corresponding to the situation where $\alpha=\frac{1}{2}$ on the lattice. This has the implication that -- unlike at $\alpha=0$ -- the current is \textit{not} forced to be zero at $\alpha=\frac{1}{2}$. In other words, the UV theory at $\alpha=\frac{1}{2}$ does not flow to the IR theory with $J_\theta=0$ in these cases, but to another theory within the conformal manifold at $\alpha=\frac{1}{2}$. This causes the second zero crossing to be near, but not necessarily at, $\alpha=\frac{1}{2}$. This understanding is quite general and can be applied to any of the models that we study.

In contrast, the measured minima \textit{and} maxima of $\mathcal{E}$ are generally quite close to $\alpha=0$ and $\frac{1}{2}$. This can be understood from the fact that the energy density is invariant throughout the conformal manifold at $\alpha=\frac{1}{2}$, so the conformal manifold is not a source of ambiguity. However, in a finite-size system, numerical results receive corrections from irrelevant operators. When $\alpha\to \frac{1}{2}$, the dominant correction are those from the defect fermion bilinear $\int dt \bar{\psi}_{-\frac{1}{2}}\psi_{-\frac{1}{2}}$, which is also the deformation that connects the IR-stable standard fixed point with the multi-critical alternative one.

Despite the flexibility which is in principle afforded by the conformal manifold, the current may be pinned to zero at $\alpha=\frac{1}{2}$ due to additional symmetry. For example, this occurs in the QWZ model due to the presence of the particle-hole symmetry discussed in \cref{sec:QWZ_rot_eigs}. This particle-hole symmetry is preserved by the $A$-centered disclination and $\vec{b}=(-1,0)$ dislocation for $\alpha=0$ or $\frac{1}{2}$, but not by the $B$-centered disclination. As a result, $J^\theta$ is pinned to zero at $\alpha=0$ and $\frac{1}{2}$. Essentially, the symmetry picks out a particular fixed point in the conformal manifold, namely that which satisfies $J^\theta=0$ at $\alpha=\frac{1}{2}$.

We also emphasize that the vertical profiles of the energy density curves plotted in \cref{fig:QWZ_currents} are consistent with several predictions from the previous section. In particular, we see that the minimum values are below zero and the range of energies is larger than on the clean lattice, confirming the theoretical prediction from the defect CFT. While the offset of the energy density along the vertical axis may seem arbitrary because we are free to redefine the zero point energy, there is actually a preferred zero point because we are scaling the energy density by $r^3$. We find that energy density curves corresponding to different lattice sites all nearly coalesce at a particular raw value of the energy density. We expect $\langle r^3 \mathcal{E}\rangle$ to be invariant across lattice sites up to finite-size corrections, so we want to preserve this coalescence when we scale by $r^3$. This is done by defining $\mathcal{E}=0$ to be at the point of coalescence. This allows us to talk about positive and negative values of $r^3\mathcal{E}$ in an absolute sense.

\subsection{$\pi$-flux models}\label{sec:pi_flux_models}
We will discuss several $\pi$-flux models defined on the square lattice which include nearest neighbor and next-nearest neighbor hoppings as well as possibly non-zero on-site potentials. They are called $\pi$-flux models because the hoppings are chosen such that $\pi$ flux pierces each plaquette of the square lattice. The $\pi$-flux models are of the form
\begin{equation}\label{eq:pi_flux_ham}
\begin{gathered}
    H=\sum_{\langle \vi,\vj\rangle}e^{iA_{\vi,\vj}}c^\dagger_\vi c_\vj+\frac{t'}{4}\sum_{\llangle \vi,\vj\rrangle}e^{iA_{\vi,\vj}}c^\dagger_\vi c_\vj+\Delta\sum_{\vi}\mu_\vi c^\dagger_\vi c_\vi,
\end{gathered}
\end{equation}
where single and double brackets indicate sums over nearest and next-nearest neighbors, respectively. $A_{i,j}$ is a static gauge configuration which inserts $\pi$ flux through each plaquette and $\mu_\vi$ specifies the configuration of the on-site potential. The three Hamiltonians we study share the same hopping terms and differ only in their on-site potentials. The exact Hamiltonians are presented in
\cref{fig:pi_flux_clean_hams}.

\subsubsection{No on-site potential}\label{sec:pi_flux_no_potential}

\begin{figure}[t]
    \centering
    \includegraphics[height=0.35\linewidth]{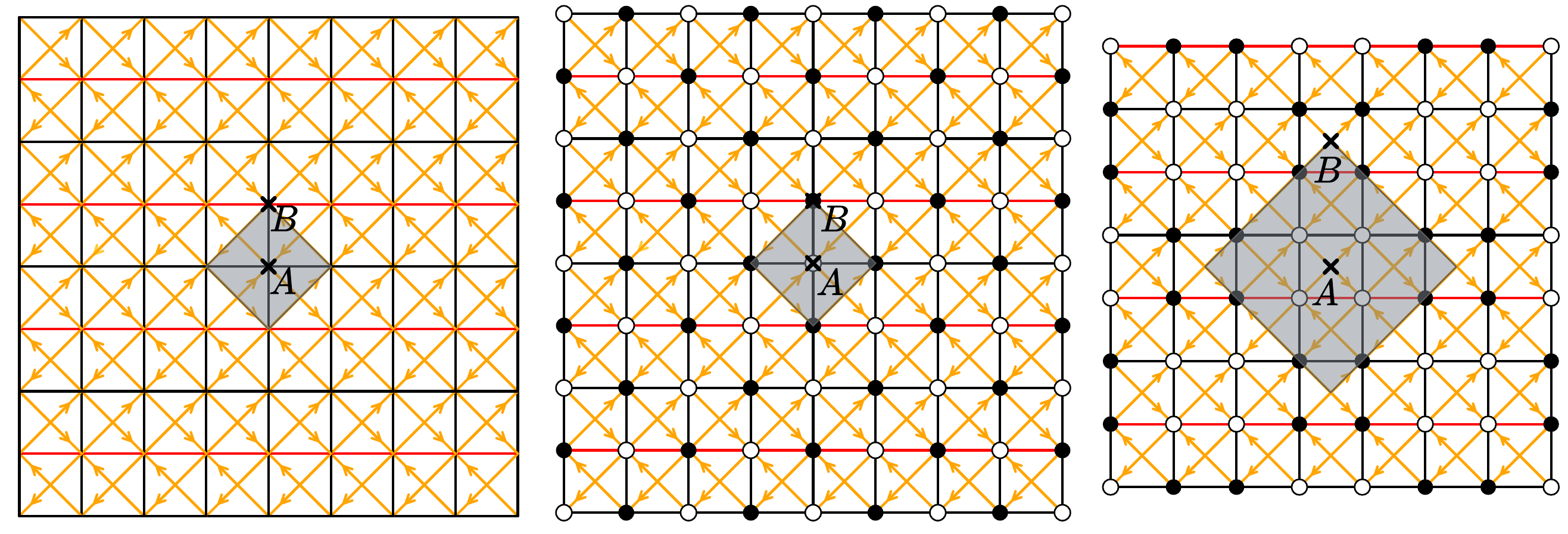}
    \caption{Real-space $\pi$-flux Hamiltonians with no on-site potential and one- and two-site staggered potentials. Black lines correspond to $h_{i,j}=1$, red lines to $h_{i,j}=-1$, and orange lines to $h_{i,j}=it'/4$. White and black circles represent the on-site potentials $+\Delta$ and $-\Delta$, respectively. The rotation centers are marked with black crosses and labeled.}
    \label{fig:pi_flux_clean_hams}
\end{figure}

The simplest $\pi$-flux model is given by setting $\mu_\vi=0$ everywhere in \cref{eq:pi_flux_ham}. This model has a critical point at $t'=0$ with a pair of Dirac cones. We consider two inequivalent rotation centers, located at the points labeled $A$ and $B$ in
\cref{fig:pi_flux_clean_hams}. The gauge transformation accompanying each rotation is given by
\begin{equation}
\begin{split}
    \lambda_{\vi}^{4,A}&=\begin{cases}
        -\pi/2 & i_x,i_y\text{ even}\\
        \pi/2 & \text{otherwise}
        \end{cases},\\
    \lambda_{\vi}^{4,B}&=\begin{cases}
        0 & i_x,i_y\text{ odd}\\
        \pi & \text{otherwise}
    \end{cases},
\end{split}
\end{equation}
where in each case we take the origin to be at a site with both coordinates odd. This model also exhibits a translation symmetry along its two primitive lattice vectors: $\vec{v}_1=(1,1)$ and $\vec{v}_2=(-1,1)$. The accompanying gauge transformations are given by
\begin{equation}
    \lambda_{\vi}^{T_{\vec{v}_1}}=-\lambda_{\vi}^{T_{\vec{v}_2}}=\begin{cases}
        -\pi/2 & i_x\text{ even}\\
        \pi/2 & \text{otherwise}
        \end{cases}.
\end{equation}
These transformations must be chosen in such a way that the resulting operators are symmetries and that they form a faithful representation of the spatial symmetry group.

As we discussed in \cref{sec:symm_eigenvalues}, one method to extract topological data at the critical point is to measure the eigenvalues of the symmetry operators. Repeating this analysis in the current model at $t'=0$, we find 
\begin{equation}\label{eq:symm_eigs_pi_flux_none}
\begin{split}
    s_A^{(i)}=\dfrac{1}{2},\text{ }\dfrac{3}{2},\text{ } s_B^{(i)}=\dfrac{1}{2},\text{ }\dfrac{7}{2},\text{ }
     \vec{k}_\star^{(i)}=(0,0),\text{ }(\pi,\pi).
\end{split}
\end{equation}
Alternatively, we can deduce $s_o^{(i)}$ and $\vec{k}_\star^{(i)}$ by first measuring topological invariants in the gapped phases using partial rotations, then constructing the effective Lagrangian \cref{eq:Dirac_lagr} governing the critical point, as discussed in \cref{sec:effective_lagr}. We report in \cref{tab:pi_flux_none_invariants} the changes in each topological invariant across the transition as well as the parameters appearing in the effective Lagrangian which reproduces them. (The absolute values of the measured invariants are given in \cref{sec:abs_invariants}.) Notice that the values of the critical point invariants $s_o^{(i)}$ and $\vec{k}_\star^{(i)}$ obtained via the effective Lagrangian exactly match those obtained in \cref{eq:symm_eigs_pi_flux_none} by symmetry eigenvalues. This again shows the remarkable agreement between two independent methods of extracting these invariants. It is also interesting to note that the symmetry eigenvalue method involved only measurements at the critical point, while the effective Lagrangian method involved only measurements in the gapped phases, so that it is possible to obtain $s_o^{(i)}$ and $\vec{k}_\star^{(i)}$ in an either restricted setting.

Because this model has two Dirac cones at its critical point, the microscopic current and energy density are no longer a good proxy for the field theory quantities discussed in the previous section. However, we can define a valley current and valley energy density by writing the Hamiltonian in the vicinity of each Dirac point in the Dirac basis, $H=\Psi^\dagger\vec{p}\cdot \vec{\sigma}\Psi$. Using this $\Psi$, we can express the field theory current and energy density operators in terms of annihilation operators on the lattice and measure it directly. The results, which are presented in \cref{fig:valley_results}, are consistent with the predicted emanant flux.

\subsubsection{One-site staggered on-site potential}
The next $\pi$-flux model we will discuss is given by setting $\mu_\vi=(-1)^{i_x+i_y}$ in \cref{eq:pi_flux_ham}. This model has critical points at $t'=\pm\Delta$, each of which has a single Dirac cone. We consider two inequivalent rotation centers, located at the points labeled $A$ and $B$ in \cref{fig:pi_flux_clean_hams}. Notice that both $A$ and $B$ are lattice sites in this case. The gauge transformation accompanying each rotation is given by
\begin{equation}
\begin{split}
    \lambda_{\vi}^{4,A}=\begin{cases}
        \frac{\pi}{2} & i_x,i_y\text{ even}\\
        -\frac{\pi}{2}& \text{otherwise}
    \end{cases},\\
    \lambda_{\vi}^{4,B}=\begin{cases}
        0 & i_x,i_y\text{ odd}\\
        \pi & \text{otherwise}
    \end{cases},
\end{split}
\end{equation}
where once again we take the origin to be located at odd coordinates in each case. This model also exhibits a translation symmetry along the primitive lattice vectors $\vec{v_1}=(1,1)$ and $\vec{v}_2=(-1,1)$. The accompanying gauge transformation is given by
\begin{equation}
    \lambda_{\vi}^{T_{\vec{v}_1}}=-\lambda_{\vi}^{T_{\vec{v}_2}}=\begin{cases}
       -\frac{\pi}{2} & i_x\text{ odd} \\
        \frac{\pi}{2} & \text{otherwise}
        \end{cases}.
\end{equation}

\begin{table*}
\def\arraystretch{1.3}
    \subfloat[No on-site potential\label{tab:pi_flux_none_invariants}]{
    \begin{tabular}{ccccccccccccccccccccccccccccc}
         \hline\hline & $\text{ }$ & $s_A^{(i)}$ & $s_B^{(i)}$ & $\operatorname{sgn}(m_i)$ & $\vec{k}_\star^{(i)}$ & $\text{ }$ & $\Delta C$ & $\Delta \mathscr{S}_A$ & $\Delta\mathscr{S}_B$ & $\Delta\ell_A$ & $\Delta\ell_B$ & $\Delta\vec{\mathscr{P}}$  \\\hline
         \multirow{2}{*}{$t'=0$} && $\frac{1}{2}$ & $\frac{1}{2}$ & $\textcolor{Green}{+}\to\textcolor{purple}{-}$ & $(0,0)$ && \multirow{2}{*}{$-2$} & \multirow{2}{*}{$2$} & \multirow{2}{*}{$0$} & \multirow{2}{*}{$\frac{3}{2}$} & \multirow{2}{*}{$\frac{7}{2}$} & \multirow{2}{*}{$\left(\frac{1}{2},\frac{1}{2}\right)$}
        \\
        && $\frac{3}{2}$ & $\frac{7}{2}$ & $\textcolor{Green}{+}\to\textcolor{purple}{-}$ & $(\pi,\pi)$ &&
        \\\hline\hline
    \end{tabular}}

\subfloat[One-site staggered potential\label{tab:pi_flux_vert_invariants}]{\begin{tabular}{cccccccccccccccccccc}
         \hline\hline &$\text{ }$& $s_A$ & $s_B$ & $\operatorname{sgn}(m)$ & $\vec{k}_\star$ & $\text{ }$ & $\Delta C$ & $\Delta\mathscr{S}_A$ & $\Delta\mathscr{S}_B$ & $\Delta\ell_A$ & $\Delta\ell_B$ & $\Delta\vec{\mathscr{P}}$ \\
         \hline
          $t'=-\Delta$ && $\frac{7}{2}$ & $\frac{7}{2}$ & $\textcolor{Green}{+}\to\textcolor{purple}{-}$ & $(0,0)$ && $-1$ & $\frac{1}{2}$ & $\frac{1}{2}$ & $\frac{15}{4}$ & $\frac{15}{4}$ & $\left(0,0\right)$  \\
          $t'=\Delta$ && $\frac{5}{2}$ & $\frac{1}{2}$ & $\textcolor{Green}{+}\to\textcolor{purple}{-}$ & $(\pi,\pi)$ && $-1$ & $\frac{3}{2}$ & $\frac{7}{2}$ & $\frac{7}{4}$ & $\frac{15}{4}$ & $\left(\frac{1}{2},\frac{1}{2}\right)$ \\
          \hline\hline 
    \end{tabular}}

\subfloat[Two-site staggered potential\label{tab:pi_flux_plaq_invariants}]{\begin{tabular}{cccccccccccccccccccc}
         \hline\hline &$\text{ }$& $s_A$ & $s_B$ & $\operatorname{sgn}(m)$ & $\vec{k}_\star$ & $\text{ }$ & $\Delta C$ & $\Delta\mathscr{S}_A$ & $\Delta\mathscr{S}_B$ & $\Delta\ell_A$ & $\Delta\ell_B$ & $\Delta\vec{\mathscr{P}}$ \\
         \hline
          $t'=-\Delta$ && $\frac{7}{2}$ & $\frac{7}{2}$ & $\textcolor{Green}{+}\to\textcolor{purple}{-}$ & $(0,0)$ && $-1$ & $\frac{1}{2}$ & $\frac{1}{2}$ & $\frac{15}{4}$ & $\frac{15}{4}$ & $(0,0)$  \\
          $t'=\Delta$ && $\frac{3}{2}$ & $\frac{3}{2}$ & $\textcolor{Green}{+}\to\textcolor{purple}{-}$ & $(0,0)$ && $-1$ & $\frac{5}{2}$ & $\frac{5}{2}$ & $\frac{7}{4}$ & $\frac{7}{4}$ & $(0,0)$\\
          \hline\hline 
    \end{tabular}}
    \caption{Topological invariants in each phase of the three $\pi$-flux models discussed in \cref{sec:pi_flux_models}.}
\end{table*}

The values of the invariants extracted via eigenvalues of the crystalline symmetry operators at the critical points are given by
\begin{equation}\label{eq:symm_eigs_pi_flux_vert}
\begin{split}
    t'=-\Delta:&\quad s_A=\frac{7}{2},\text{ } s_B=\frac{7}{2},\text{ }\vec{k}_\star=\left(0,0\right),\\
    t'=\Delta:&\quad s_A=\frac{5}{2},\text{ } s_B=\frac{1}{2},\text{ }\vec{k}_\star=(\pi,\pi).\\
\end{split}
\end{equation}
The changes in topological invariants across the transition as measured through partial rotations and deduced effective Lagrangian parameters are given in \cref{tab:pi_flux_vert_invariants}. Notice once again that the values of $s_o$ and $\vec{k}_\star$ obtained via our two independent methods are identical. We present results for the azimuthal current and local energy density in the presence of lattice defects in \cref{fig:pi_flux_4_currents}. We find that the curves are shifted along the $\delta\alpha$ axis by the predicted emanant flux, namely $s_o/4$ and $\frac{1}{2\pi}\vec{k}_\star\cdot\vec{b}$.

\subsubsection{Two-site staggered on-site potential}

The final $\pi$-flux model we will study is given by setting $\mu_\vi=1.68\times(-1)^{\left\lfloor\frac{i_x+i_y}{2}\right\rfloor}$ in \cref{eq:pi_flux_ham}, where the scale is chosen so that the critical points remain at $t'=\pm\Delta$. This model exhibits a fourfold magnetic rotation symmetry about two inequivalent rotation centers, located at the points labeled $A$ and $B$ in \cref{fig:pi_flux_clean_hams}. Notice that both $A$ and $B$ are plaquette centers in this case. The rotation operators about each center are defined with the following gauge transformation:
\begin{equation}
    \lambda_{\vi}^{4,A}=\lambda_{\vi}^{4,B}=\begin{cases}
        5\pi/4 & i_x,i_y\text{ even}\\
        \pi/4 & \text{otherwise}
    \end{cases},
\end{equation}
where we take the site directly to the lower left of each rotation center to have both coordinates even. This model has a translation symmetry along its two primitive lattice vectors: $\vec{v}_1=(2,2)$ and $\vec{v}_2=(-2,2)$, with trivial accompanying gauge transformation.

Extracting the topological data through symmetry eigenvalues, we find
\begin{equation}\label{eq:symm_eigs_pi_flux_plaq}
\begin{split}
    t'=-\Delta:&\quad s_A=s_B=\frac{7}{2},\quad \vec{k}_\star=(0,0),\\
    t'=\Delta:&\quad s_A=s_B=\frac{3}{2},\quad \vec{k}_\star=(0,0).\\
\end{split}
\end{equation}
The values of the topological invariants which are instead measured by performing partial rotations and forming the effective Lagrangian are given in \cref{tab:pi_flux_plaq_invariants}. Once again, the values obtained via the two methods coincide. We present results for the azimuthal current and local energy density in the presence of lattice defects in \cref{fig:pi_flux_8_currents}. We find that the curves are shifted along the $\delta\alpha$ axis by the predicted emanant flux, namely $s_o/4$.

\subsection{Haldane model}
\label{sec:honeycomb_models}

\begin{table*}
\def\arraystretch{1.3}
    {\begin{tabular}{cccccccccccccccccccccccc}
         \hline\hline & $\text{ }$ & $s_A$ & $s_B$ & $s_C$ & $\operatorname{sgn}(m)$ & $\vec{k}_\star$ & $\text{ }$ & $\Delta C$ & $\Delta\mathscr{S}_A$ & $\Delta\mathscr{S}_B$ & $\Delta\mathscr{S}_C$ & $\Delta\ell_A$ & $\Delta\ell_B$ & $\Delta\ell_C$ & $\Delta \vec{\mathscr{P}}$ \\\hline
          $t'=-\Delta$ && $\frac{1}{2}$ & $\frac{5}{2}$ & $\frac{3}{2}$ & $\textcolor{Green}{+}\to\textcolor{purple}{-}$ & $(\frac{2\pi}{3},\frac{4\pi}{3})$ && $-1$ & $\frac{5}{2}$ & $\frac{1}{2}$ & $\frac{3}{2}$ & $\frac{11}{4}$ & $\frac{11}{4}$ & $\frac{3}{4}$ & $\left(\frac{2}{3},\frac{1}{3}\right)$ \\
          $t'=\Delta$ && $\frac{5}{2}$ & $\frac{1}{2}$ & $\frac{3}{2}$ & $\textcolor{Green}{+}\to\textcolor{purple}{-}$ & $\left(\frac{4\pi}{3},\frac{2\pi}{3}\right)$ && $-1$ & $\frac{1}{2}$ & $\frac{5}{2}$ & $\frac{3}{2}$ & $\frac{11}{4}$ & $\frac{11}{4}$ & $\frac{3}{4}$ & $\left(\frac{1}{3},\frac{2}{3}\right)$ \\
          \hline\hline
    \end{tabular}}
    \caption{Topological invariants at each critical point of the Haldane model discussed in \cref{sec:honeycomb_models}.}
\label{tab:honeycomb_staggered_invariants}
\end{table*}

\begin{figure}[t]
    \centering
    \includegraphics[height=0.5\linewidth]{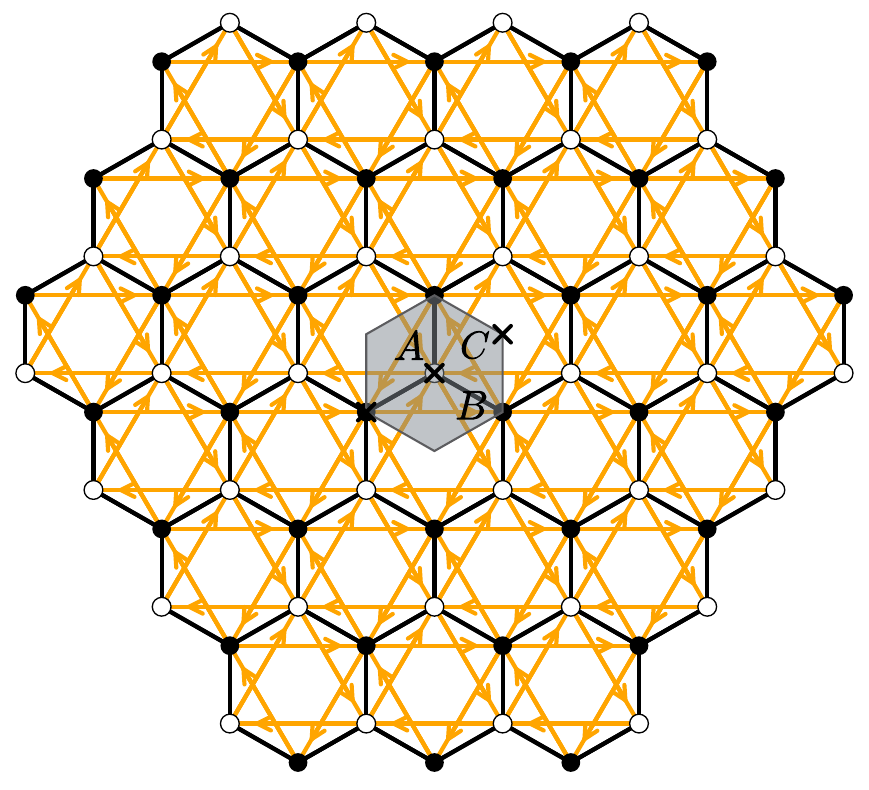}
    \caption{Real-space Haldane model Hamiltonian where black lines correspond to $h_{i,j}=1$ and orange lines to $h_{i,j}=it'/4$. White and black circles represent the on-site potentials $+\Delta$ and $-\Delta$, respectively. The rotationally invariant unit cell is shaded in gray and inequivalent rotation centers are marked with black crosses and labeled.}
    \label{fig:honeycomb_clean_hams}
\end{figure}

\begin{figure*}[t]
        \subfloat[Lattice with $N=4761$ sites and $A$-centered $\beta=2/3$ disclination]{%
            \includegraphics[width=.38\linewidth]{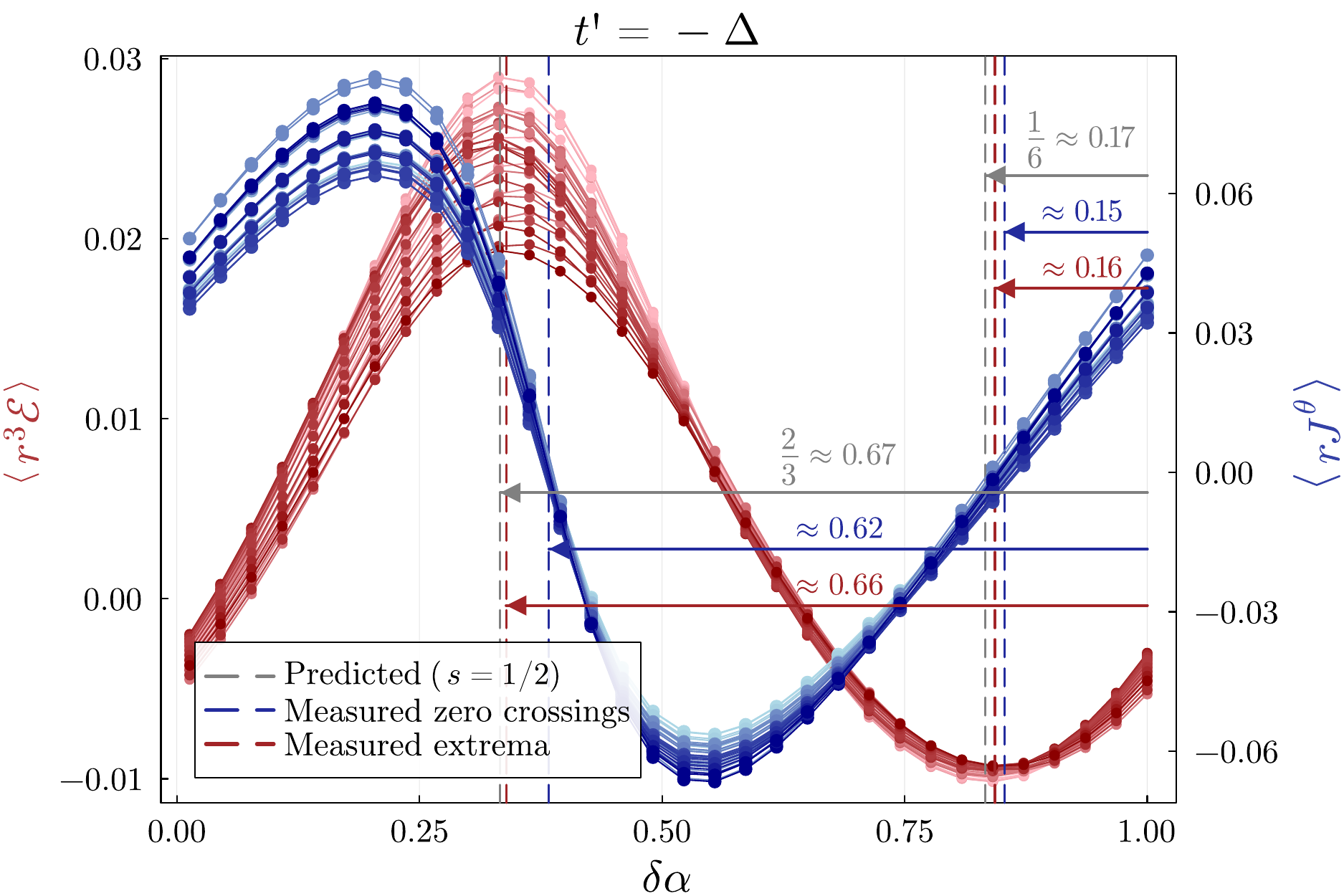}%
        \qquad\includegraphics[width=.38\linewidth]{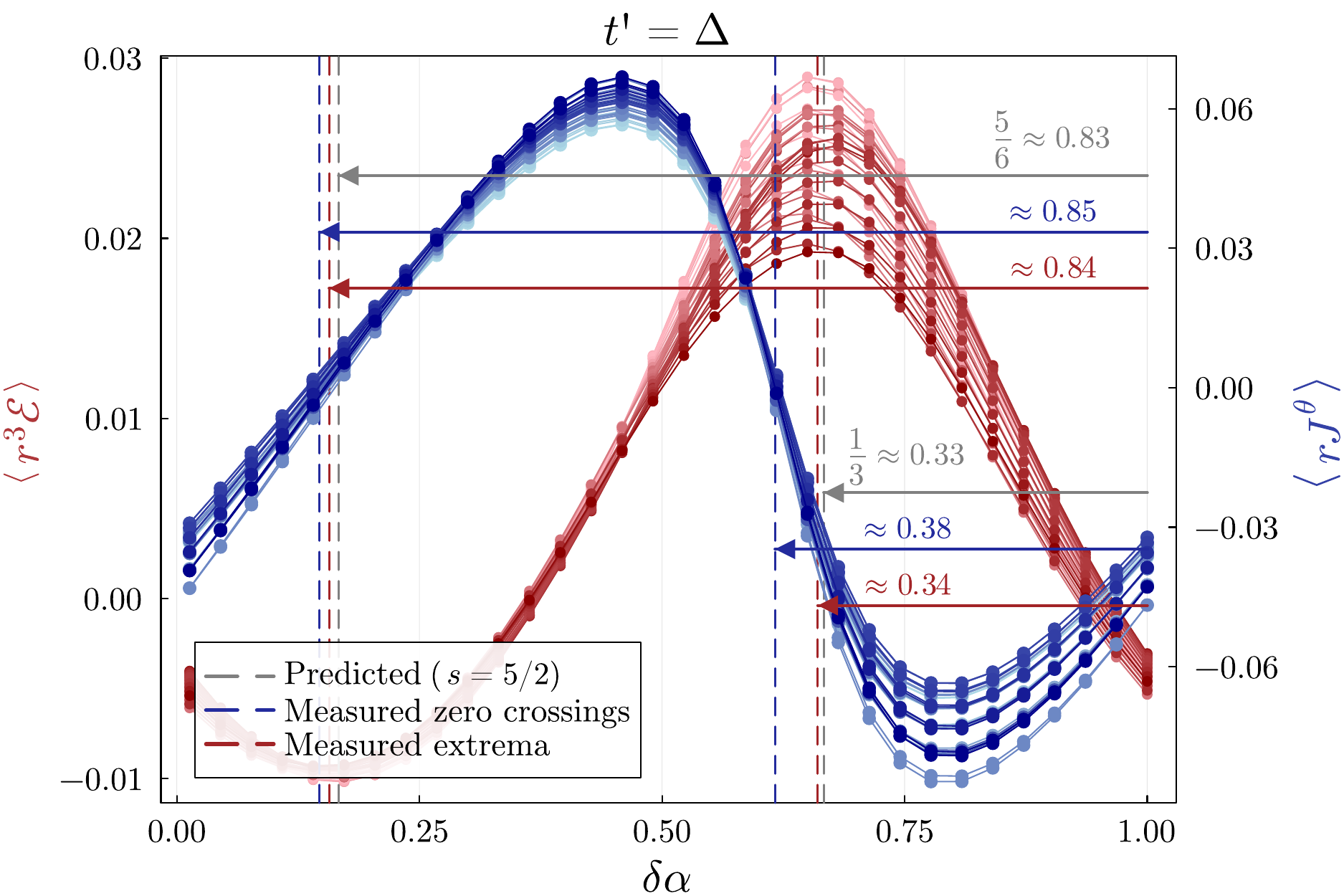}%
        }\\
        \subfloat[Lattice with $N=9521$ sites and $A$-centered $\beta=4/3$ disclination]{%
            \includegraphics[width=.38\linewidth]{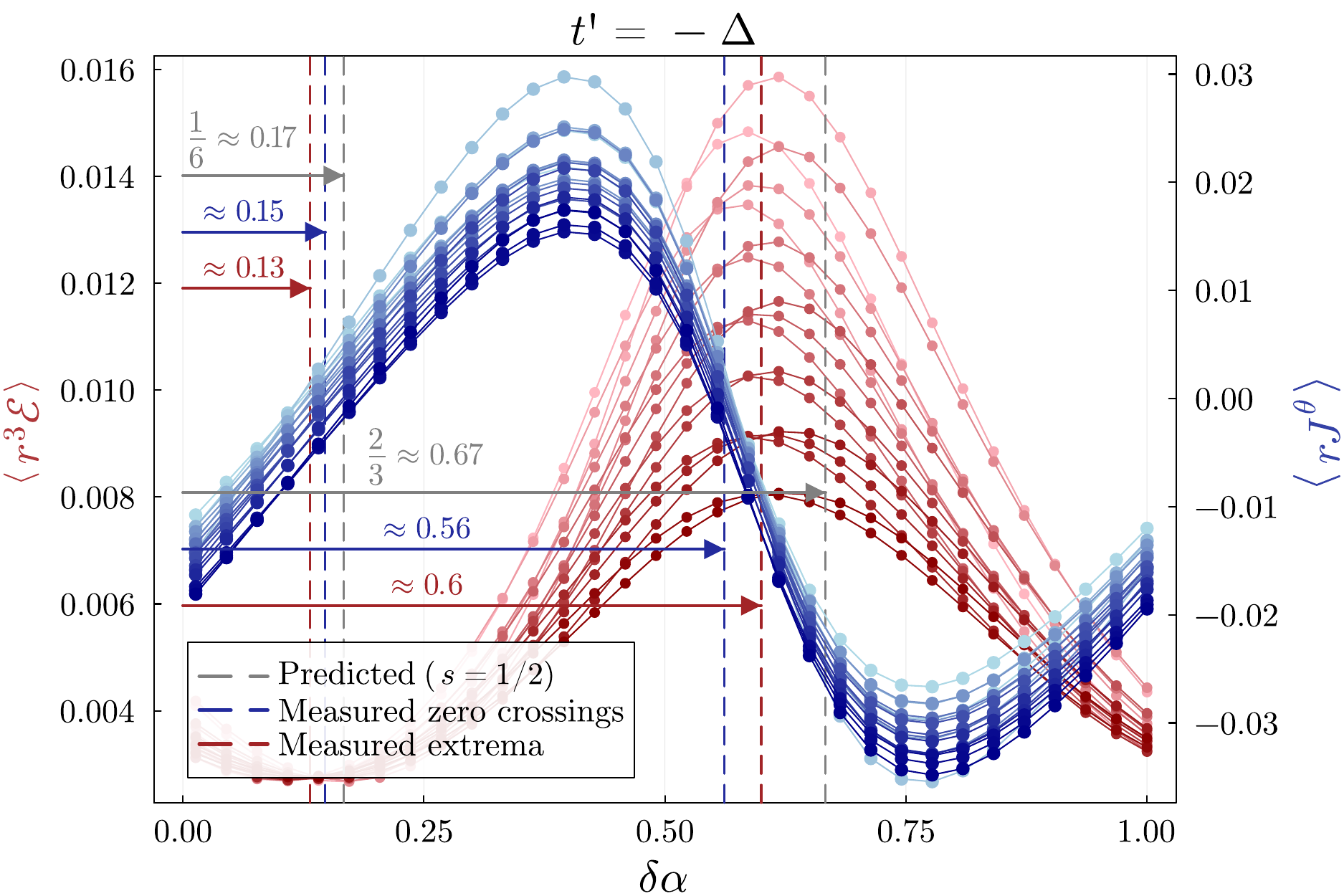}%
        \qquad\includegraphics[width=.38\linewidth]{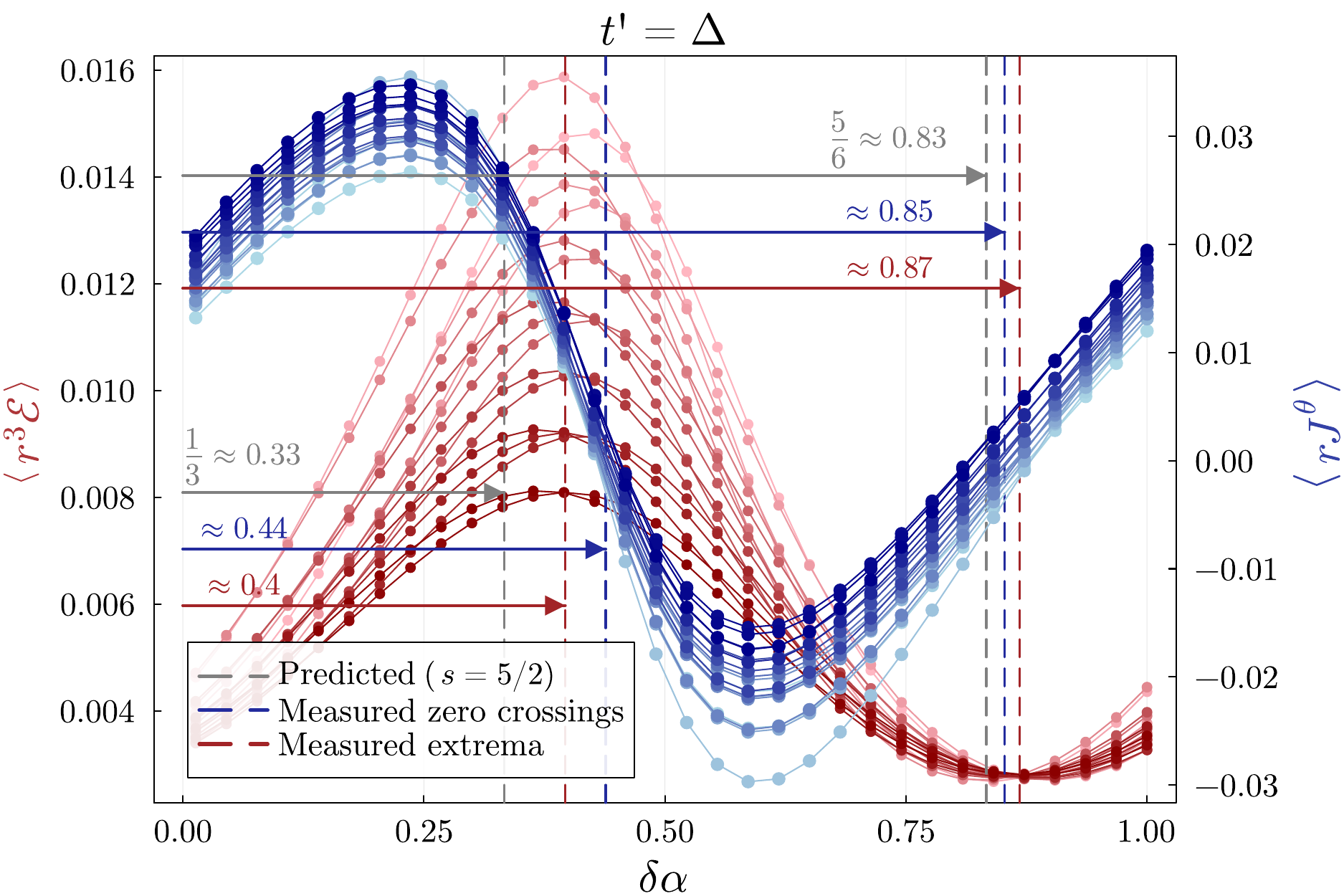}%
        }
        \caption{Haldane model with $\beta=1\pm 1/3$ disclinations. Here we plot the scaled current and energy density at the $t'=\pm\Delta$ critical points. Notice that the emanant flux is of opposite sign for the two different defects, which correspond to $\beta=1\pm1/3$. The shading of each curve in the plot indicates the distances from the defect, with lighter curves corresponding to lattice sites closer to the defect.}
        \label{fig:extra_wedge}
\end{figure*}

In this section, we study the Haldane model \cite{haldane1988}, the prototypical example of a Chern insulator, with a staggered on-site potential. Unlike the previous models we have discussed, this model has a threefold rotational symmetry. It is given by
\begin{equation}
\begin{gathered}
    H=\sum_{\langle \vi,\vj\rangle}c^\dagger_\vi c_\vj+\frac{it'}{4}\sum_{\llangle \vi,\vj\rrangle}\nu_{\vi\vj}c^\dagger_\vi c_\vj+\Delta\sum_{\vi}{\mu_\vi}c^\dagger_\vi c_\vi,
\end{gathered}
\end{equation}
where $\nu_{\vi\vj}$ assigns signs to the next-nearest neighbor hoppings such that they are always clockwise and $\mu_\vi\approx\pm1.3$ on the two sublattices of the honeycomb lattice. There are a pair of critical points at $t'=\pm \Delta$, each of which has a single Dirac cone. 

This model has a threefold rotation symmetry about the three inequivalent rotation centers labeled $A$, $B$, and $C$ in \cref{fig:honeycomb_clean_hams}. In each case, the accompanying gauge transformation is trivial, so the symmetries are ordinary rotations, not magnetic rotations. We also have a translation symmetry by the lattice vectors $\vec{v}_1=(1,0)$ and $\vec{v}_2=(\frac{1}{2},\frac{\sqrt{3}}{2})$, where the distance between neighboring sites on the same sublattice is $1$. Similarly, the accompanying gauge transformation is trivial, so this is an ordinary translation symmetry. 

Measuring the eigenvalues of the symmetry operators in the space of zero-energy single particle states, we find
\begin{equation}\label{eq:symm_eigs_honeycomb_staggered}
\begin{split}
    t'=-\Delta:&\quad s_A=\frac{1}{2},\text{ } s_B=\frac{5}{2},\text{ } s_C=\frac{3}{2},\text{ } \vec{k}_\star=\pi\left(\frac{2}{3},\frac{4
    }{3}\right),\\
    t'=\Delta:&\quad s_A=\frac{5}{2},\text{ }  s_B=\frac{1}{2},\text{ } s_C=\frac{3}{2},\text{ }\vec{k}_\star=\pi\left(\frac{4}{3},\frac{2
    }{3}\right).\\
\end{split}
\end{equation}
 The invariants obtained by instead performing partial rotations in the gapped phases and forming the effective Lagrangians are given in \cref{tab:honeycomb_staggered_invariants}. Notice once again that the invariants at each transition exactly match those obtained via rotation eigenvalues. We present results for the azimuthal current and local energy density in the presence of lattice defects in \cref{fig:extra_wedge,fig:staggered_honeycomb_currents}. 
 
 In particular, in \cref{fig:extra_wedge} we present results for the same rotation center, but with positive and negative disclination angles, i.e., one wedge removed or one wedge added. We find that the curves are shifted along the $\delta\alpha$ axis by equal and opposite values of emanant flux. We also see that the energy density is shifted above zero for all values of $\delta\alpha$ for $\beta=4/3$, which is consistent with the theoretical prediction. In the raw $\mathcal{E}$ data for the $\beta=4/3$ example, we found that there was no point of coalescence of the various curves, as there was in other models. This suggests that the energy density never crosses zero, and we define the zero point to be that which minimizes the variance in $\langle r^3\mathcal{E}\rangle$ across the various curves near their minima. In \cref{fig:staggered_honeycomb_currents}, we also see agreement with the predicted emanant flux due to either disclinations or dislocations.

\section{Discussion}

We have shown that on general grounds, the existence of a $C_M$ rotational symmetry where $(C_M)^M = +1$ requires that lattice disclinations induce an azimuthal current when the continuum theory contains Dirac fermions. This is a direct consequence of the emanant flux induced by the lattice disclination, which must be non-zero because the Dirac fermions carry spin-1/2. 
Similarly, lattice dislocations induce an emanant flux determined by the location of the Dirac cone in the Brillouin zone. 

In the case of a single valley (single Dirac cone), the azimuthal current is the total electrical current. In the case where there are multiple valleys (multiple Dirac cones), then each Dirac cone could see a different emanant flux, which leads to a different current for each valley. In particular, in time-reversal symmetric systems with multiple Dirac cones, such as in graphene, the total current vanishes at a disclination but there is a non-zero valley current. It would be interesting to develop experimental methods to detect this valley current in graphene, especially given that there have been experimental advances in manipulating valley degrees of freedom in two-dimensional materials  \cite{schaibley2016valleytronics, vitale2018valleytronics}. 

An interesting generalization of the discussion here is to a situation where the associated flux is that of a dynamical gauge field, in which case the magnetic flux could be partially or completely screened by the induced currents that we have found. We leave a detailed analysis of the case where the gauge field is dynamical to future work. Such an analysis would also be relevant for various Chern-Simons-Dirac quantum critical points enriched with crystalline symmetry, which describes quantum critical points of quantum Hall systems in the presence of a lattice \cite{chen1993, PhysRevB.89.235116,grover2013}.

While we have tested a number of theoretical predictions from the defect CFT in our numerics, we have not extracted the defect scaling exponents $\Delta_s^\pm$. This is particularly challenging as the irrelevant defect operators approach marginality when $\alpha \rightarrow 1/2$, which leads to significant finite-size corrections. It would be interesting to tackle such a difficulty with proper numerical methods and measure $\Delta_s^\pm$. The scaling dimensions provide an alternative way to physically measure the topological quantum numbers of the Dirac fermions without tuning an applied magnetic field. 

One remaining question is regarding the total charge in the vicinity of lattice defects at criticality. 
We found that the charge density vanishes away from the defect. This is a consequence of conformal invariance.  However, there could be localized charges. We leave this question for future work.

In \cref{sec:effective_lagr}, we assumed that rotations and translations commute and can therefore be simultaneously diagonalized. This is not the case when, for example, there are multiple Dirac cones which map into one another under rotations. We remarked above that this situation can largely be dealt with by choosing the basis which diagonalizes one or the other, depending on which invariant you wish to derive. However, the invariant $\vec{\mathscr{Q}}_o$ is defined via the cross term $\omega_o \wedge d\vec{R}$ and cannot be derived in this manner when rotations and translations don't commute. We expect that the general case can be considered by treating $\omega_o$ and $\vec{R}$ as non-abelian background gauge fields and leave this to future work. We remark that the situation where the magnetic flux per unit cell is non-zero modulo $2\pi$ and the two primitive translations do not commute is analogous and should be treated similarly.

We also point out that in the study of the crystalline topological phases, corners and boundaries exhibit phenomena closely related to the bulk disclinations and dislocations \cite{Benalcazar2019HOTI,zhang2024bdy}. It would be interesting to study the fate of the charges and currents at criticality due to corners and boundaries more systematically. 

Interesting open questions were also raised in our DCFT analysis. For monodromy defects in flat space, it seems a natural conjecture that $g\geq1 $ for the IR-stable fixed point, yet we were not able to prove it. On the other hand, we do not know if the $g$-function is an RG monotone in the presence of conical singularities, nor do we know if it is well-defined. It is tempting to formally compute the $g$-function from the partition function on hyperbolic space, and the analog to Eq.\eqref{eq_g function formula} reads
\begin{equation}
    \partial_\alpha \log g=\text{vol}(\mathbb{H}^2)\sum_{s\in \mathbb{Z}+\frac{1}{2}}\langle\bar{\tilde{\Psi}}_s\tilde{\Psi}_s\rangle=\frac{2\pi}{\beta}C_J(\alpha,\beta)~.
\end{equation}
Similarly, we can also obtain $\partial_\beta \log g$. We leave the investigation of these generalized $g$-functions to future works.

We conclude by pointing out that the general discussions of this paper can be applied to quantum critical points more broadly. For example, any CFT appearing as the low energy limit of a lattice model tuned to a critical point can be enriched by crystalline symmetry. The associated UV-IR homomorphism then determines how disclinations and dislocations induce monodromy and conical defects in the CFT. A particularly relevant example to study in the context of quantum magnetism would be the $O(N)$ fixed point, enriched by crystalline symmetries.  Magnetic transitions involving ferromagnets, anti-ferromagnets, or altermagnets \cite{vsmejkal2022emerging} could potentially furnish examples of distinct symmetry-enriched $O(N)$ fixed points. Finally, there are lattice models where the sites support full-fledged gauge theories, such as in~\cite{Razamat:2021jkx} (and see references therein). It would be interesting to study crystalline defects in that context.

\begin{acknowledgments}

We thank Yuxuan Zhang, Naren Manjunath, Gautam Nambiar, Xie Chen, Adar Sharon, Shu-Heng Shao, Jie Wang, Yifan Wang, and Yunqin Zheng for discussions. 
M.B. and C.F. are supported by NSF DMR-2345644. C.F. is also supported by the NSF-STAQ progam. Z.K. and S.Z. are supported in part by the Simons Foundation Grant 488657 (Simons Collaboration on the Non-Perturbative Bootstrap), the BSF Grant No. 2018204, and NSF Award Number 2310283.

\end{acknowledgments}
 
\appendix

\section{Further lattice model details}
\subsection{QWZ UV-IR homomorphism}\label{ap:QWZ_sym_eigs}
In this section, we explain in greater detail how to obtain the Hamiltonian in the vicinity of each Dirac point and use it to derive explicitly the UV-IR homomorphism $\rho$ as it acts on spatial rotations.
\subsubsection{\texorpdfstring{$m=-1$}{m=-1}}
When $m=-1$, there is a single Dirac point at $\vec{k}_\star:=(0,0)$. We expand around this critical point by defining $\vec{p}$ via $\vec{k}=\vec{k}_\star+\vec{p}$ and keeping only first-order terms in the components of $\vec{p}$.

The resultant Hamiltonian in the neighborhood of the Dirac point is given by
\begin{equation}
    h_{\vec{p}}=\left(\begin{array}{cc}
        0 & p_x-ip_y \\
        p_x+ip_y & 0
    \end{array}\right)=\vec{p}\cdot\vec{\sigma}.
\end{equation}

We therefore have a $2\times 2$ Dirac Hamiltonian:
\begin{equation}
    H=\sum_{\vec{p}}\Psi_{\vec{p}}^\dagger\left(\vec{p}\cdot\vec{\sigma}\right)\Psi_{\vec{p}},
\end{equation}
where $\Psi_{\vec{p}}=(c_{\vec{p},\uparrow}\text{ } c_{\vec{p},\downarrow})^T$ and $c_{\vec{p},\alpha}$ is shorthand for $c_{\vec{k}_\star+\vec{p},\alpha}$. 

Using the magnetic rotation operators defined in \cref{eq:QWZ_vert_sym,eq:QWZ_plaq_sym}, we can derive the transformation properties of the $\vec{k}$-space creation and annihilation operators. In particular, we find
\begin{equation}\label{eq:k_space_rot}
\begin{gathered}
    C_{4,B} c_{\vec{k},\uparrow}C_{4,B}^{-1}=ic_{R_{\pi/2}^B(\vec{k}),\uparrow},\\ C_{4,B} c_{\vec{k},\downarrow}C_{4,B}^{-1}=c_{R_{\pi/2}^B(\vec{k}),\downarrow},\\ C_{4,A} c_{\vec{k},\uparrow}C_{4,A}^{-1}=ie^{-ik_x}c_{R_{\pi/2}^B(\vec{k}),\uparrow},\\ C_{4,A} c_{\vec{k},\downarrow}C_{4,A}^{-1}=e^{-ik_x}c_{R_{\pi/2}^B(\vec{k}),\downarrow}.
\end{gathered}
\end{equation}
Notice that $R_{\pi/2}^B(\vec{k}_\star+\vec{p})=\vec{k}_\star+R_{\pi/2}^B(\vec{p})$ and $e^{-i k^x_\star}=1$, so $C_{4,A}$ and $C_{4,B}$ have the same action on the spinor at the Dirac point:
\begin{equation}
    \left(\begin{array}{c}
         c_{\vec{k}_\star,\uparrow} \\
         c_{\vec{k}_\star,\downarrow} \\
    \end{array}
    \right)\overset{C_{4_{A/B}}}{\longrightarrow}\left(\begin{array}{c}
         i c_{\vec{k}_\star,\uparrow} \\
         c_{\vec{k}_\star,\downarrow} \\
    \end{array}\right)=\left(\begin{array}{cc} 
     i & 0 \\
     0 & 1 \\
    \end{array}\right)\left(\begin{array}{c}
         c_{\vec{k}_\star,\uparrow} \\
         c_{\vec{k}_\star,\downarrow} \\
    \end{array}\right).
\end{equation}
In other words, we have
\begin{equation}
    \Psi_{\vec{p}}\overset{C_{4_{A/B}}}{\longrightarrow}\left(\begin{array}{cc} 
     i & 0 \\
     0 & 1 \\
    \end{array}\right)\Psi_{R^{A,B}_{\pi/2}(\vec{p})}= e^{i\frac{\pi}{4}}\Lambda(\pi/2)\Psi_{R^{A,B}_{\pi/2}(\vec{p})},
\end{equation}

where $\Lambda(\pi/2)=e^{i\frac{\pi}{4}\sigma_z}$ is the spinor transformation which accompanies a spatial $\pi/2$ transformation for a spin-1/2 particle. Matching to the form of \cref{eq:C_tranform_k_space}, we see that $s_v=s_p=\frac{1}{2}$. This matches the value of $s$ for the massless fermion at the transition which was determined in \cref{sec:QWZ_invariants} by measuring topological invariants and computing the effective Lagrangian.

\subsubsection{\texorpdfstring{$m=0$}{m=0}}
When $m=0$, there is a pair of Dirac points, one at $\vec{k}_\star^x:=(\pi,0)$ and one at $\vec{k}_\star^y:=(0,\pi)$. Once again, we let $\vec{k}=\vec{k}_\star^{x/y}+\vec{p}$ and expand to first-order in the components of $\vec{p}$. We find an effective Hamiltonian in the vicinity of each Dirac point: 
\begin{equation}
\begin{split}
    h_{\vec{p}}^x=\left(\begin{array}{cc}
        0 & -p_x-ip_y \\
        -p_x+ip_y & 0
    \end{array}\right),\\
    h_{\vec{p}}^y=\left(\begin{array}{cc}
        0 & p_x+ip_y \\
        p_x-ip_y & 0
    \end{array}\right).
\end{split}
\end{equation}
Rotating the basis of $h_{\vec{p}}^x$ by $-Y$ and $h_{\vec{p}}^y$ by $X$, we find
\begin{equation}
\begin{split}
    \tilde h_{\vec{p}}^x=\tilde h_{\vec{p}}^y=\left(\begin{array}{cc}
        0 & p_x-ip_y \\
        p_x+ip_y & 0
    \end{array}\right).
\end{split}
\end{equation}

We can combine the Hamiltonian in each valley into a single $4\times 4$ Hamiltonian:
\begin{equation}\label{eq:first_QWZ_Dirac}
    H=\sum_{\vec{p}}\Psi_{\vec{p}}^\dagger\left[(\vec{p}\cdot\vec{\sigma})\otimes I\right]\Psi_{\vec{p}},
\end{equation}
where $\Psi_{\vec{p}}=(ic_{\vec{p},\downarrow,x}\text{ } -ic_{\vec{p},\uparrow,x}\text{ }c_{\vec{p},\downarrow,y}\text{ } c_{\vec{p},\uparrow,y})^T$ and $c_{\vec{p},\alpha,x}$ is shorthand for $c_{\vec{k}_\star^x+\vec{p},\alpha}$ and similarly for $c_{\vec{p},\alpha,y}$. 

Although \cref{eq:first_QWZ_Dirac} is in the form of a Dirac Hamiltonian, we will see that it is still not in the basis which diagonalizes the spinor action of the microscopic $C_{4,B}$. Notice first that $R^B_{\pi/2}(\vec{k}_\star^x+\vec{p})=\vec{k}_\star^y+R^B_{\pi/2}(\vec{p})$ and $R^B_{\pi/2}(\vec{k}_\star^x+\vec{p})=\vec{k}_\star^y+R^B_{\pi/2}(\vec{p})$, so using \cref{eq:k_space_rot} we have
\begin{equation}
    \Psi_{\vec{p}}\overset{C_{4,B}}{\longrightarrow}\left(\begin{array}{cccc} 
    0 & 0 & i & 0 \\
    0 & 0 & 0 & 1 \\
    -i & 0 & 0 & 0 \\
    0 & -1 & 0 & 0 \\
    \end{array}\right)\Psi_{R^B_{\pi/2}(\vec{p})}.
\end{equation}

We can recover a diagonal spinor transformation by rotating the basis by
\begin{equation}
\frac{1}{\sqrt{2}}\left(
\begin{array}{cccc}
 -i & 0 & 1 & 0 \\
 0 & -i & 0 & 1 \\
 i & 0 & 1 & 0 \\
 0 & i & 0 & 1 \\
\end{array}
\right).
\end{equation}
In this new basis, we have
\begin{equation}
    H=\sum_{\vec{p}} \tilde\Psi_{\vec{p}}^\dagger\left[(\vec{p}\cdot\vec{\sigma})\otimes I\right] \tilde\Psi_{\vec{p}},
\end{equation}
where
\begin{equation}
    \tilde\Psi_{\vec{p}}=\frac{i}{\sqrt{2}}\left(\begin{array}{c}
         c_{\vec{p},\downarrow,x}+c_{\vec{p},\downarrow,y} \\
         -c_{\vec{p},\uparrow,x}+c_{\vec{p},\uparrow,y} \\
         -c_{\vec{p},\downarrow,x}+c_{\vec{p},\downarrow,y}  \\
         c_{\vec{p},\uparrow,x}+c_{\vec{p},\uparrow,y} \\
    \end{array}\right).
\end{equation}
Under the microscopic $C_{4,B}$, we now have
\begin{equation}
    \tilde\Psi_{\vec{p}}\overset{C_{4,B}}{\longrightarrow}\left(\begin{array}{cccc} 
    1 & 0 & 0 & 0 \\
    0 & -i & 0 & 0 \\
    0 & 0 & -1 & 0 \\
    0 & 0 & 0 & i \\
    \end{array}\right)\tilde\Psi_{R^B_{\pi/2}(\vec{p})}.
\end{equation}
We see that there is a different transformation matrix for each Dirac fermion:
\begin{equation}
\left(\begin{array}{cc}
 1 & 0 \\
 0 & -i \\
\end{array}\right)=e^{i\frac{7\pi}{4}}\Lambda(\pi/2),\quad \left(\begin{array}{cc}
 -1 & 0 \\
 0 & i \\
\end{array}\right)=e^{i\frac{3\pi}{4}}\Lambda(\pi/2).
\end{equation}
The corresponding invariants are therefore $s=\frac{7}{2}$ and $s=\frac{3}{2}$.

Let us now work out the homomorphism under the plaquette-centered rotation. Applying \cref{eq:k_space_rot}, we find
\begin{equation}
    \Psi_{\vec{p}=\vec{0}}\overset{C_{4,A}}{\longrightarrow}\left(\begin{array}{cccc} 
    0 & 0 & -i & 0 \\
    0 & 0 & 0 & -1 \\
    -i & 0 & 0 & 0 \\
    0 & -1 & 0 & 0 \\
    \end{array}\right)\Psi_{\vec{p}=\vec{0}}.
\end{equation}
Once again, we see that we are not in the basis which diagonalizes the spinor action of $C_{4,A}$. We can recover a diagonal spinor transformation by rotating the basis by
\begin{equation}
\frac{1}{\sqrt{2}}\left(
\begin{array}{cccc}
 1 & 0 & 1 & 0 \\
 0 & 1 & 0 & 1 \\
 -1 & 0 & 1 & 0 \\
 0 & -1 & 0 & 1 \\
\end{array}
\right).
\end{equation}
In this new basis, we have
\begin{equation}
    H=\sum_{\vec{p}} \tilde\Psi_{\vec{p}}^\dagger\left[(\vec{p}\cdot\vec{\sigma})\otimes I\right] \tilde\Psi_{\vec{p}},
\end{equation}
where
\begin{equation}
    \tilde\Psi_{\vec{p}}=\frac{1}{\sqrt{2}}\left(\begin{array}{c}
         ic_{\vec{p},\downarrow,x}+ c_{\vec{p},\downarrow,y} \\
         -ic_{\vec{p},\uparrow,x}+c_{\vec{p},\uparrow,y} \\
         -ic_{\vec{p},\downarrow,x}+c_{\vec{p},\downarrow,y}  \\
         ic_{\vec{p},\uparrow,x}+c_{\vec{p},\uparrow,y} \\
    \end{array}\right).
\end{equation}
Under the microscopic $C_{4,A}$, we now have
\begin{equation}
    \tilde\Psi_{\vec{p}=\vec{0}}\overset{C_{4,A}}{\longrightarrow}\left(\begin{array}{cccc} 
    -i & 0 & 0 & 0 \\
    0 & -1 & 0 & 0 \\
    0 & 0 & i & 0 \\
    0 & 0 & 0 & 1 \\
    \end{array}\right)\tilde\Psi_{\vec{p}=\vec{0}}.
\end{equation}
We see that there is a different transformation matrix for each Dirac fermion:
\begin{equation}
\left(\begin{array}{cc}
 -i & 0 \\
 0 & -1 \\
\end{array}\right)=e^{i\frac{5\pi}{4}}\Lambda(\pi/2),\quad \left(\begin{array}{cc}
 i & 0 \\
 0 & 1 \\
\end{array}\right)=e^{i\frac{\pi}{4}}\Lambda(\pi/2).
\end{equation}
The corresponding invariants are therefore $s=\frac{5}{2}$ and $s=\frac{1}{2}$.

\subsubsection{\texorpdfstring{$m=1$}{m=1}}
When $m=1$, there is a single Dirac point at $\vec{k}_\star:=(\pi,\pi)$. Expanding around this Dirac point in small momentum, we find
\begin{equation}
    h_{\vec{p}}=\left(\begin{array}{cc}
        0 & -p_x+ip_y \\
        -p_x-ip_y & 0
    \end{array}\right).
\end{equation}
Rotating the basis by $Z$, we have
\begin{equation}
    \tilde h_{\vec{p}}=\left(\begin{array}{cc}
        0 & p_x-ip_y \\
        p_x+ip_y & 0
    \end{array}\right)=\vec{p}\cdot\vec{\sigma}.
\end{equation}
We therefore have a $2\times 2$ Dirac Hamiltonian:
\begin{equation}
    H=\sum_{\vec{p}}(c_{\vec{p},\uparrow}^\dagger\text{ } -c_{\vec{p},\downarrow}^\dagger)\left(\vec{p}\cdot\vec{\sigma}\right)\left(\begin{array}{c}
         c_{\vec{p},\uparrow} \\
         -c_{\vec{p},\downarrow} \\
    \end{array}\right),
\end{equation}
where $c_{\vec{p},\alpha}$ is shorthand for $c_{\vec{k}_\star+\vec{p},\alpha}$. 

In this case, $R_{\pi/2}^B(\vec{k}_\star+\vec{p})=\vec{k}_\star+R_{\pi/2}^B(\vec{p})$ and $e^{-i k^x_\star}=-1$, so using \cref{eq:k_space_rot} we have:
\begin{equation}
\begin{gathered}
    \Psi_{\vec{p}}
    \overset{C_{4,B}}{\longrightarrow}\left(\begin{array}{cc} 
     i & 0 \\
     0 & 1 \\
    \end{array}\right)\Psi_{R^B_{\pi/2}(\vec{p})},\\
    \Psi_{\vec{p}}\overset{C_{4,A}}{\longrightarrow}\left(\begin{array}{cc} 
     -i & 0 \\
     0 & -1 \\
    \end{array}\right)\Psi_{R^B_{\pi/2}(\vec{p})}.
\end{gathered}
\end{equation}
The transformation matrices are therefore given by:
\begin{equation}
\left(\begin{array}{cc}
 i & 0 \\
 0 & 1 \\
\end{array}\right)=e^{i\frac{\pi}{4}}\Lambda(\pi/2),\quad \left(\begin{array}{cc}
 -i & 0 \\
 0 & -1 \\
\end{array}\right)=e^{i\frac{5\pi}{4}}\Lambda(\pi/2).
\end{equation}
The corresponding invariants are therefore $s_v=\frac{1}{2}$ and $s_p=\frac{5}{2}$.

\subsection{Topological invariants from partial rotations}\label{ap:partial_rotations}
In this section, we review the method of partial rotations introduced in \cite{zhang2023complete,manjunath2024Characterization}. Suppose we have a tight binding model with an $M$-fold magnetic rotation symmetry about the high-symmetry point $o$, which is realized by $C_{M,o}$. We can define an operator
\begin{equation}
    C_{M_o}^-:=e^{-i\frac{\pi}{M_o}\hat N} C_{M,o},
\end{equation}
where $N$ is the fermion number operator $\hat N=\sum_{\vi}c_\vi^\dagger c_\vi$. Because $C_{M_o}$ is a symmetry, $C_{M_o}^-$ will also be a symmetry. However, it satisfies $(C_{M_o}^-)^M=(-1)^F$, in contrast to $(C_{M_o})^M=1$, where $(-1)^F$ is fermion parity. We may then define two families of operators
\begin{equation}
    C_{M,o,\chi}:=e^{i\chi \frac{2\pi}{M}\hat N}C_{M,o},\quad C_{M,o,\chi}^-:=e^{i\chi \frac{2\pi}{M}\hat N}C_{M,o}^-,
\end{equation}
where $\chi=0,\ldots,M-1$. These will also be symmetries. 

Next, we fix a rotation-invariant region of lattice sites $D$, for a lattice on either open or periodic boundary conditions. In order to extract topological data, we measure expectation values in the many-body ground state $\ket{\Psi}$ of rotation operators restricted to the region $D$, which are called partial rotations. Letting $|_D$ denote the restriction of an operator to sites in $D$, we find
\begin{equation}
   \bra{\Psi}C_{M,o,\chi}^{(-)}|_D\ket{\Psi}=e^{\frac{2\pi i}{M}l_{D,o,\chi}^{(-)}-\gamma_{D,o,\chi}^{(-)}},
\end{equation}
where $\gamma_{D,o,\chi}^{(-)}\propto \partial D$ sets the real-valued amplitude and $l_{D,o,\chi}^{(-)}$ encodes the topological data. In particular, $l_{D,o,\chi}^{(-)}$ satisfies
\begin{equation}
    l_{D,o,\chi}^{(-)}= \frac{C \chi^2}{2} + \mathscr{S}_o^{(-)}\chi + K_o^{(-)}\text{ mod }M.
\end{equation}
For a given set of $l_{D,o,\chi}^{(-)}$, we find $C$, $\mathscr{S}_o^{(-)}$, and $K_o^{(-)}$ by fitting to the measured values of $l_{D,o,\chi}^{(-)}$, keeping in mind the quantization of each invariant. Once $C$, $\mathscr{S}_o^{(-)}$, and $K_o^{(-)}$ are obtained, we define the \textit{real-space invariants} 
\begin{equation}
\begin{aligned}
    \Theta_o:=&\begin{cases}
        K_o\text{ mod }\frac{M}{2},\text{ }M\text{ even}\\
        K_o\text{ mod }M,\text{ }M\text{ odd}
    \end{cases},\\
    \Theta_o^{-}:=&\begin{cases}
        K_o^{-}\text{ mod }M,\text{ }M\text{ even}\\
        K_o^{-}\text{ mod }\frac{M}{2},\text{ }M_o\text{ odd}
    \end{cases}.
\end{aligned}
\end{equation}
We can finally calculate $\ell_o$ via
\begin{equation}
    \ell_o=\begin{cases}
        \frac{11}{12}C+2\Theta_o\text{ mod }3,\text{ } M=3\\
        \frac{5}{4}C+2\Theta_o\text{ mod }4,\text{ } M=4\\
    \end{cases}.
\end{equation}
Following this procedure, we find the values of $C$, $\mathscr{S}_o$, and $\ell_o$ reported in the main text. 

\onecolumngrid

\section{Free boson model}
\label{appendix_free boson model}

As a supplement to the Dirac fermion model \eqref{eq_fermion action} in the main text, we discuss the conical monodromy defect in a $(2+1)d$ free boson model. The scalar field $\Phi_\text{b}\in \mathbb{C}$ is subject to a $U(1)$ 0-form symmetry, and we couple it to $A_\mu=\delta_{\mu \theta}\alpha\in [0,1)$ that implements the monodromy. We consider the bulk action in the conical space \eqref{eq_cone metric} 
\begin{equation}
\label{eq_appendix_free massless boson action}
    \begin{aligned}
S_\text{boson}=&\int_{\text{Cone}} d^3 X|\nabla_\mu\Phi_\text{b}|^2~,
    \end{aligned}
\end{equation}
where $d^3X$ is the covariant integral measure. Let $\sigma\in \mathbb{S}^1$ be the phonon dual to the connection $A_\mu$, we find the local gauge-invariant field operator $\Phi=e^{i\sigma}\Phi_{\text{b}}$. An angular mode expansion to $\Phi$ reads
\begin{equation}
\label{eq_appendix_boson mode expansion}
    \begin{aligned}
\Phi(t,r,\theta)=&\frac{1}{\sqrt{2\pi r}}\sum_{n\in\mathbb{Z}}e^{i(n+\alpha)\theta}\Tilde{\Phi}_n(t,r)~. 
    \end{aligned}
\end{equation}
We now apply the same trick we used in the main text \eqref{eq_fermion map to AdS} that maps the cone \eqref{eq_cone metric} to $\text{AdS}_2\times \mathbb{S}^1_{\beta}$. The action of angular Kaluza-Klein modes on $\text{AdS}_2$ is such that
\begin{equation}
\label{eq_appendix_boson map to AdS}
S_\text{boson}=\beta\sum_{n\in \mathbb{Z}}\int_{\text{AdS}_2}d^2 \Tilde{X}\Tilde{\Phi}^*_n\left[-\tilde{\Box}^2-\frac{1}{4}+\frac{(n+\alpha)^2}{\beta^2}\right]\Tilde{\Phi}_n~.
\end{equation}

DCFT data regarding \eqref{eq_appendix_free massless boson action} can be obtained by studying the asymptotic scaling of $\text{AdS}_2$ propagators \eqref{eq_appendix_ads boson Green function} and \eqref{eq_appendix_ads boson singular solution}. We note the expansion
\begin{equation}
\label{eq_appendix_boson DCFT data}
\tilde{\Phi}_n=r^{\Delta_n^\pm}\phi_n(t)+O\left(r^{\Delta_n^\pm+1}\right)\text{, where } \Delta_n^\pm=\frac{1}{2}\pm\frac{|n+\alpha|}{\beta}>0.
\end{equation}
\eqref{eq_appendix_boson DCFT data} is also known as the bulk-to-defect OPE of the standard ($+$) and alternative ($-$) quantization. The defect primary $\phi_n$ is of the $SL(2,\mathbb{R})$ conformal dimension $\Delta_n^\pm$ and $SO(2)$ transversal rotation spin $n+\alpha$. Remarkably, the defect bilinear deformation $\int dt |\phi_0|^2$ when $\alpha=0$ is marginal at the tree level. It is similar to the fermion sliding operator $\int d t \bar{\psi}_{-\frac{1}{2}}\psi_{-\frac{1}{2}}$ at $\alpha=\frac{1}{2}$, except that the fermion bilinear deformation is exactly marginal while $\int dt |\phi_0|^2$ is marginally irrelevant. Such a fact can be checked either from a conformal perturbation calculation or directly from the $\log$-divergence in the singular solution \eqref{eq_appendix_ads boson singular solution}.

Now let us consider the bulk one-point functions of the $U(1)$ current $J_\mu=i(\Phi\partial_\mu \Phi^*-\Phi^*\partial_\mu \Phi)$ and stress energy tensor $T_{\mu \nu}=\partial_\mu \Phi^*\partial_\nu \Phi$. $\langle J_\mu \rangle$ and $\langle T_{\mu \nu } \rangle$ of the free massless boson are respectively of the tensor structure \eqref{eq_def azimuthal current} and \eqref{eq_def energy density}, which is fixed by the residual conformal symmetry. We find 
\begin{equation}
\label{eq_appendix_boson current density expansion}
    \begin{aligned}
C_J=&2\sum_{n \in \mathbb{Z}}(n+\alpha)\langle |\tilde{\Phi}_n|^2 \rangle~,\\
C_T=&\sum_{n\in \mathbb{Z}}\left(\frac{n+\alpha}{\beta}\right)^2\langle |\tilde{\Phi}_n|^2 \rangle~.
    \end{aligned}
\end{equation}
More specifically, \eqref{eq_appendix_boson current density expansion} in the standard quantization admits an integral representation similar to \eqref{eq_conical CJ integral} and \eqref{eq_conical CT integral}. For the azimuthal current density, we find that 
\begin{equation}
\label{eq_boson current integral}
    \begin{aligned}
C_{J}=\frac{1}{\pi \beta}\left[\int_{0}^1\frac{(1-\alpha)\left(u^{\frac{\alpha}{\beta}}-u^{-\frac{\alpha}{\beta}}\right)+\alpha\left(u^{\frac{\alpha-1}{\beta}}-u^{\frac{1-\alpha}{\beta}}\right)}{\left(u^{\frac{1}{2\beta}}-u^{-\frac{1}{2\beta}}\right)^2(1-u)^{v-1}}u^{\frac{v-3}{2}}du\right]_{v\to 2}~.
    \end{aligned}
\end{equation}
On the other hand, the energy density is such that
\begin{equation}
\mathtoolsset{multlined-width=0.93\displaywidth}
\begin{multlined}
C_{T}=\frac{1}{2\pi \beta^3}\left\{\int_{0}^1\left[(2\alpha^2-2\alpha-1)\left(u^{\frac{\alpha}{\beta}-\frac{1}{2\beta}}+u^{-\frac{\alpha}{\beta}+\frac{1}{2\beta}}\right)-(1-\alpha)^2\left(u^{\frac{\alpha}{\beta}+\frac{1}{2\beta}}+u^{-\frac{\alpha}{\beta}-\frac{1}{2\beta}}\right)\right.\right.\\
\left.\left.-\alpha^2\left(u^{\frac{\alpha}{\beta}-\frac{3}{2\beta}}+u^{-\frac{\alpha}{\beta}+\frac{3}{2\beta}}\right)\right]\frac{u^{\frac{v-3}{2}}du}{\left(u^{\frac{1}{2\beta}}-u^{-\frac{1}{2\beta}}\right)^3(1-u)^{v-1}}\right\}_{v\to 2}~.
\end{multlined}
\end{equation}
In addition to the azimuthal current and energy density, the boson field develops a non-zero particle density profile
\begin{equation}
\langle |\Phi|^2\rangle=\frac{1}{2\pi r}\sum_{n\in \mathbb{Z}}\langle |\tilde{\Phi}_n|^2\rangle \equiv \frac{C_{|\Phi|^2}}{2\pi r}~, \text{where }C_{|\Phi|^2}=-\frac{1}{2\pi \beta}\left(\int_0^1\frac{u^{-\frac{\alpha }{ \beta}+\frac{1}{2 \beta}}+u^{\frac{\alpha }{ \beta}-\frac{1}{2 \beta}}}{u^{\frac{1}{2\beta}
   }-u^{-\frac{1}{2\beta} }}\frac{u^{\frac{v-3}{2}}du}{(1-u)^{v-1}}\right)_{v\to 2}~.
\end{equation}
At alternative fixed points, we also find a relation similar to \eqref{eq_alternaive fixed point current and energy}, such that
\begin{equation}
    \begin{aligned}
    {C}'_{|\Phi|^2}=&{C}_{|\Phi|^2}+\frac{1}{2\beta}\sum_{{n}'}\tan (\pi \frac{|{n}'+\alpha|}{\beta})~,\\
{C}'_J=&C_J+\sum_{{n}'}\frac{{n}'+\alpha}{\beta}\tan(\pi\frac{|{n}'+\alpha|}{\beta})~,\\
{C}'_T=&C_T+\frac{1}{2\beta}\sum_{{n}'}\frac{({n}'+\alpha)^2}{\beta^2}\tan(\pi\frac{|{n}'+\alpha|}{\beta})~.
    \end{aligned}
\end{equation}

\begin{figure}[thb]
\centering
\includegraphics[width=.23\textwidth ]{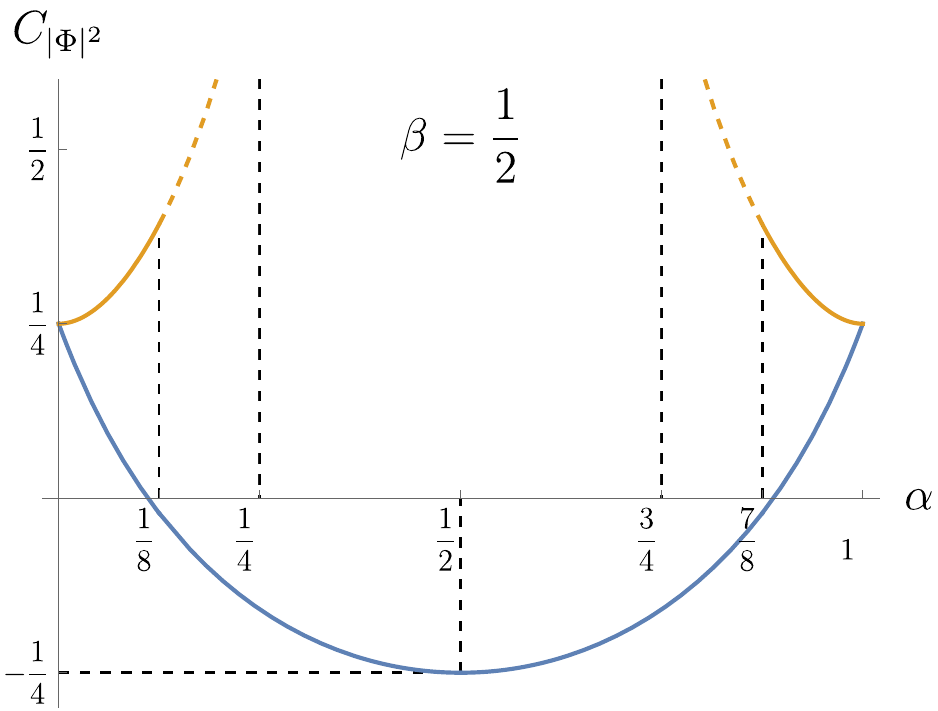}
\hspace{.1 \textwidth }
\includegraphics[width=.23\textwidth ]{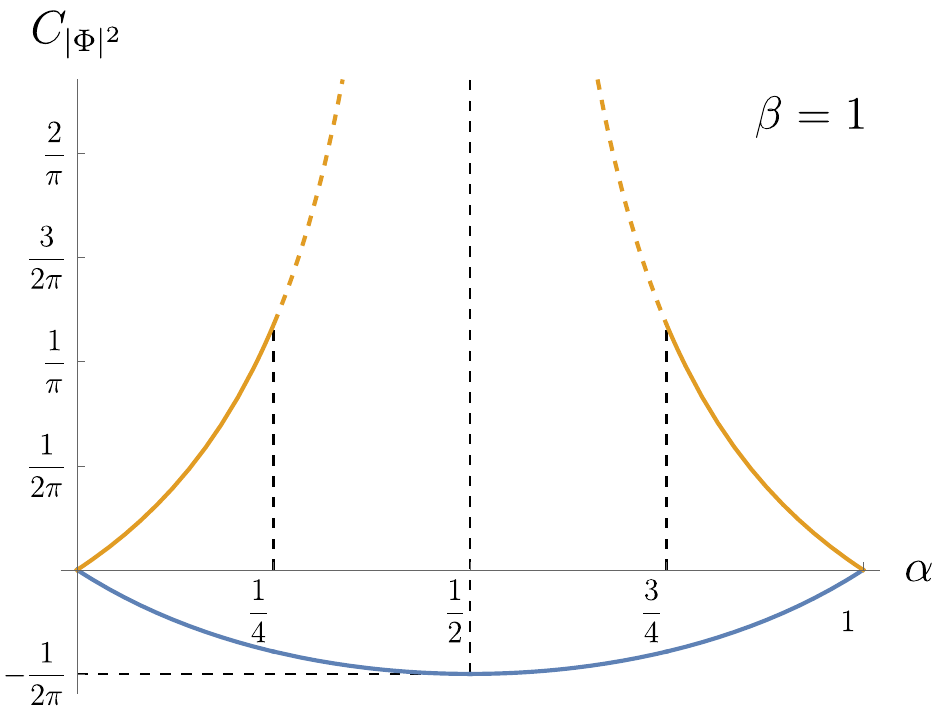}
\hspace{.1 \textwidth }
\includegraphics[width=.23\textwidth ]{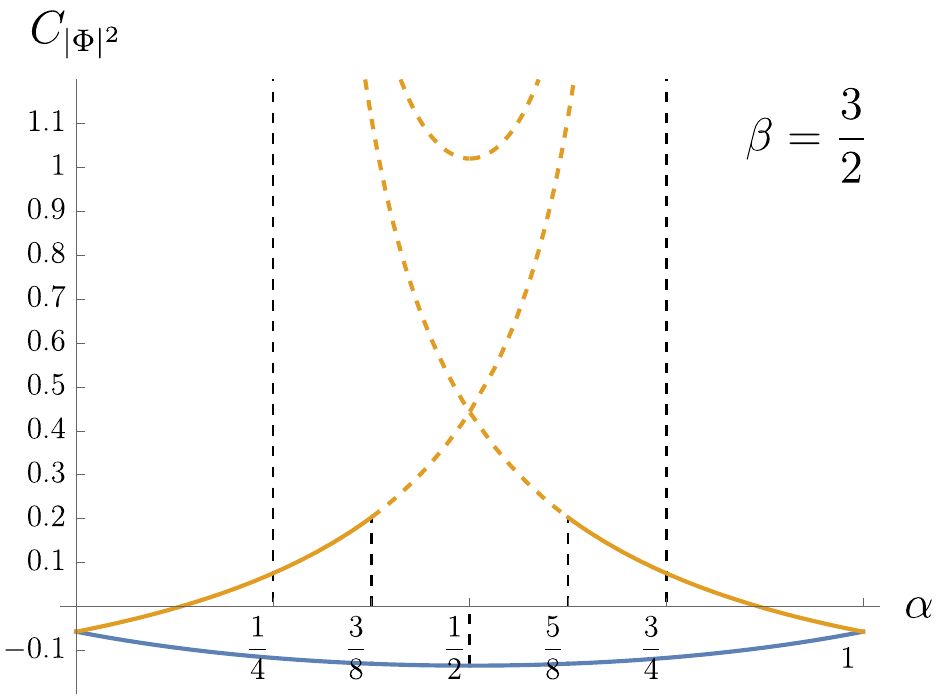}
\includegraphics[width=.23\textwidth ]{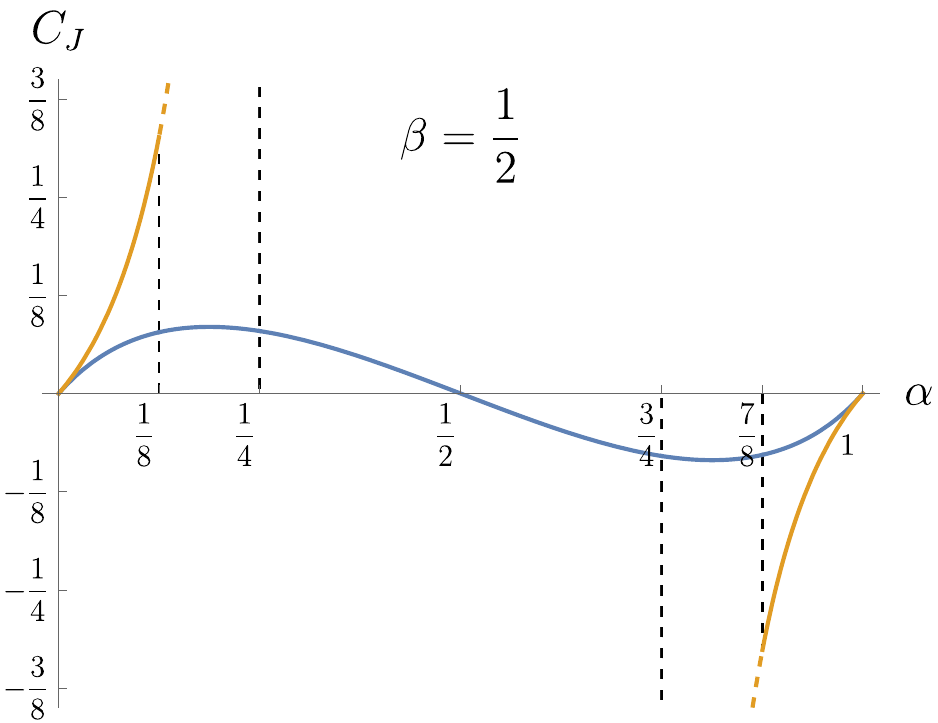}
\hspace{.1 \textwidth }
\includegraphics[width=.23\textwidth ]{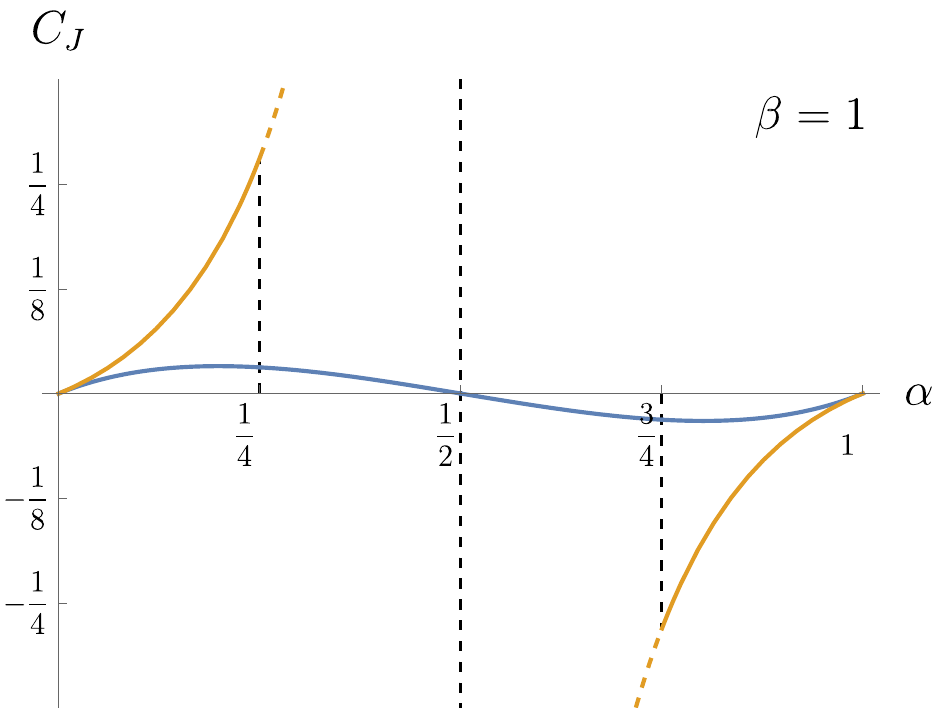}
\hspace{.1 \textwidth }
\includegraphics[width=.23\textwidth ]{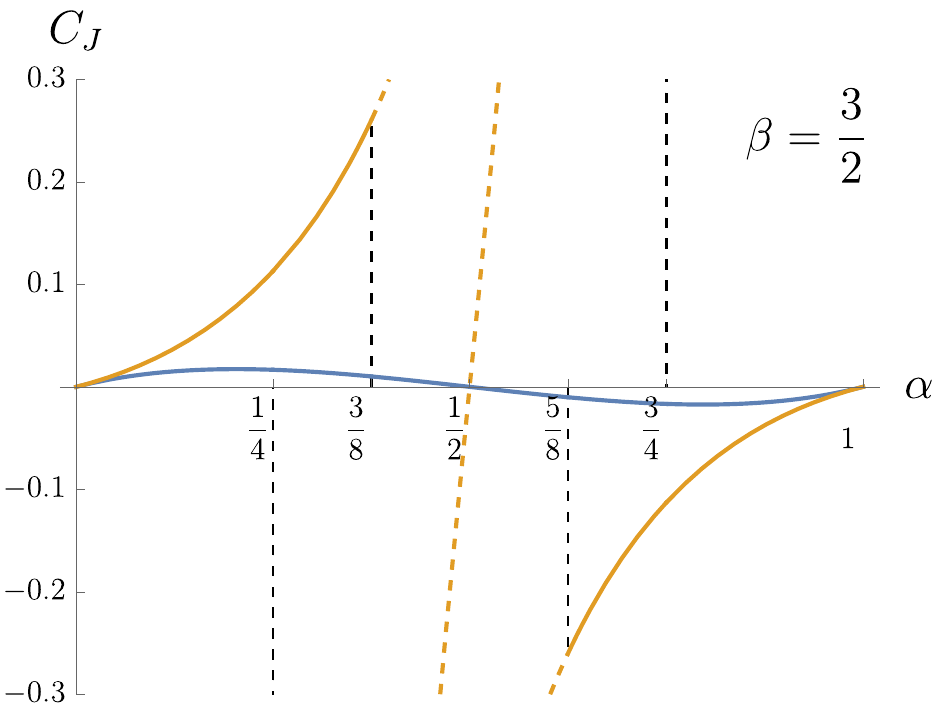}
\includegraphics[width=.23\textwidth ]{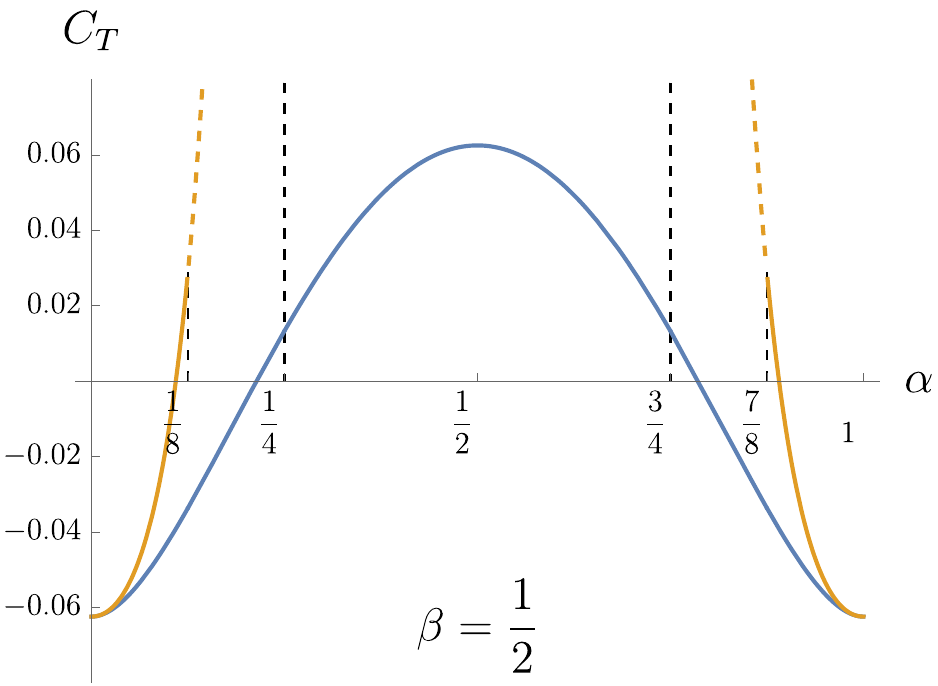}
\hspace{.1 \textwidth }
\includegraphics[width=.23\textwidth ]{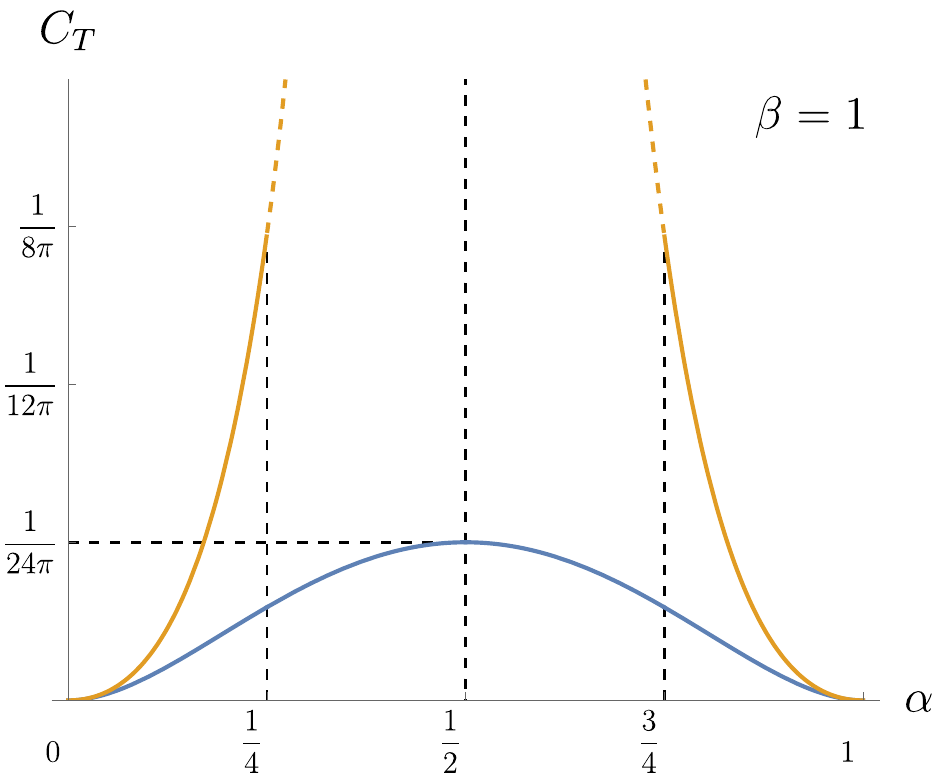}
\hspace{.1 \textwidth }
\includegraphics[width=.23\textwidth ]{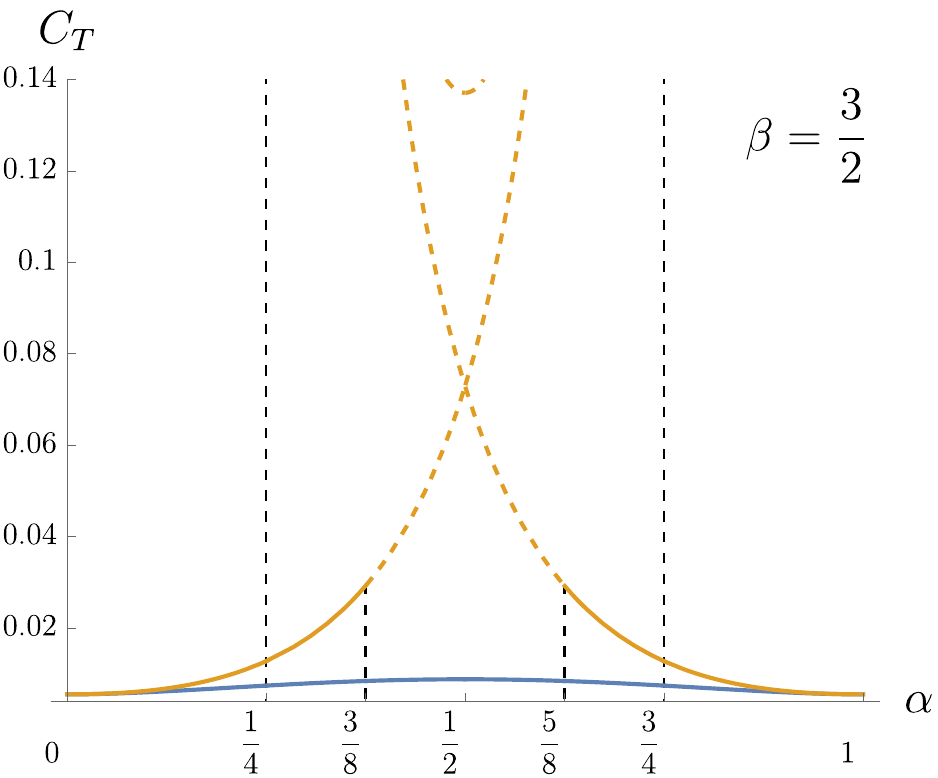}
  \caption{\label{pic_appendix_boson gallery} Particle density $C_{|\Phi|^2}$, azimuthal current $C_J$, and conformal weight $C_T$ of the free massless boson defect. From left to right: defect with a conical deficit ($\beta=\frac{1}{2}$), defect in flat space ($\beta=1$), and defect with a conical excess ($\beta=\frac{3}{2}$).}
\end{figure}

When $\beta=1$ and there is no conical singularity, we find for $\alpha\in [0,1)$ that 
\begin{equation}
\label{appendix_eq_boson one point functions}
    \begin{aligned}
    {C}_{|\Phi|^2}=&-\frac{1}{4} (1-2 \alpha ) \tan (\pi  \alpha )~,\\
C_J=&\frac{1}{8} (1-2 \alpha )^2 \tan (\pi  \alpha )~,\\
C_T=&\frac{1}{12} (1-\alpha ) \alpha  (1-2 \alpha) \tan (\pi
    \alpha )~,
    \end{aligned}
\end{equation}
in agreement with \cite{Bianchi:2021snj,Giombi:2021uae}. A gallery of $C_{|\Phi|^2}$, $C_J$, and $C_T$  can be found in figure \ref{pic_appendix_boson gallery}. We remark on two distinctions between the free boson and the free fermion defect: firstly, free boson observables $C_J$ and $C_T$ as a function of $\alpha$ is differentiable at $\alpha=\frac{1}{2}$. There exists no defect conformal manifold at $\pi$-flux, and the current vanishes $C_J(\alpha=\frac{1}{2})=0$ at the standard fixed point. Secondly, the bulk energy density of the free boson at $\pi$-flux is lower than that at the $0$-flux. We find the energy density difference between time-reversal invariant DCFTs where $C_J=0$, such that
\begin{equation}
\label{eq_appendix_boson energy difference}
C_{T}(\alpha=\frac{1}{2})-C_{T}(\alpha=0)=-\frac{1}{2\pi \beta^2}\int_0^1\frac{u^{\frac{1}{2\beta}  }-u^{-\frac{1}{2\beta}
   }}{\left(u^{\frac{1}{2\beta}  }+u^{-\frac{1}{2\beta}
   }\right)^3}\frac{du}{1-u^2}>0
\end{equation}
which is similar to \eqref{eq_fermion energy difference}.

\section{AdS propagators}

\label{appendix_AdS Propagators}

In this appendix, we discuss the analytical properties of free field propagators in $\text{AdS}_d$. Let the coordinate system be $x=(r,\vec{x})$, where $r$ is the distance to the $\text{AdS}_d$ boundary and $\vec{x}$ are transversal directions. The metric reads
\begin{equation}
    \begin{aligned}
(ds)^2=\frac{(dr)^2+(d\vec{x})^2}{r^2}~, 
    \end{aligned}
\end{equation}
and the connected isometry group is $SO(d,1)$. We note that the geodesic distance between two points $x_{1,2}\in \text{AdS}_d$ is fixed by the isometry to depend solely on the variable 
\begin{equation}
    \begin{aligned}
\xi \equiv \frac{2r_1r_2}{r_1^2+r_2^2+(\vec{x}_{12})^2}
    \end{aligned}
\end{equation}
with $\vec{x}_{12}\equiv\vec{x}_{1}-\vec{x}_{2}$, which we will use intensively in the following.

Following \cite{burges1986supersymmetry,d2004supersymmetric,Giombi:2021uae}, let us first discuss the free complex boson $\phi$ of the action
\begin{equation}
\label{eq_AdS boson Action}
    \begin{aligned}
\Tilde{S}_\text{boson}=\int_{\text{AdS}_{d}}d^d \Tilde{X}\phi^*\left[-\tilde{\Box}-\frac{(d-1)^2}{4}+m^2\right]\phi~.
    \end{aligned}
\end{equation}
The Green's function $\tilde{G}_\phi(x_1,x_2)\equiv \langle\phi^*(x_1)\phi(x_2)\rangle$ is subject to the 2nd order differential equation
\begin{equation}
\label{eq_appendix_ads boson differential equation}
    \begin{aligned}
\left[-\tilde{\Box}_1-\frac{(d-1)^2}{4}+m^2\right]\tilde{G}_\phi(x_1,x_2)=\delta_{\text{AdS}_d}(x_1-x_2)~, 
    \end{aligned}
\end{equation}
which generally admits two linearly independent solutions. Noticing that the isometry fixes $\tilde{G}_\phi(x_1,x_2)=\tilde{G}_\phi(\xi)$, it is obtained that when $m\neq 0$: 

\begin{equation}
\label{eq_appendix_ads boson Green function}
\tilde{G}_\phi(\xi)=\frac{ \Gamma (\Delta )(\xi/2)^\Delta}{2\pi ^{\frac{d-1}{2}}\Gamma \left(\Delta +\frac{3-d}{2}\right)}\, _2F_1\left(\frac{\Delta }{2},\frac{\Delta+1}{2};\Delta +\frac{3-d}{2};\xi ^2\right)~,
\end{equation}
where $\Delta=\frac{d-1}{2}\pm |m|$ marks the 
asymptotic scaling as one of the correlation points approaches the $\text{AdS}_d$ boundary at $r=0$:
\begin{equation}
\label{eq_appendix_ads boson boundary scaling}
    \begin{aligned}
\tilde{G}_\phi(\xi \ll 1)=\frac{ \Gamma (\Delta )\left(\xi/2\right)^\Delta}{2 \pi ^{\frac{d-1}{2}}\Gamma \left(\Delta +\frac{3-d}{2}\right)}+O(\xi^{\Delta+2})~.
    \end{aligned}
\end{equation}
When $m=0$, one solution follows \eqref{eq_appendix_ads boson Green function} and \eqref{eq_appendix_ads boson boundary scaling} with $\Delta=\frac{d-1}{2}$, while the other one reads
\begin{equation}
\label{eq_appendix_ads boson singular solution}
\tilde{G}^{\text{log}}_\phi(\xi)=\frac{\xi ^{\frac{d-1}{2}} \left(1-\xi ^2\right)^{1-\frac{d}{2}}}{4\pi ^{\frac{d-2}{2}}  \sin \left(\frac{\pi  d}{2}\right)\Gamma \left(2-\frac{d}{2}\right)} \, _2F_1\left(\frac{3-d}{4},\frac{5-d}{4};2-\frac{d}{2};1-\xi ^2\right)~.
\end{equation}

We remark that the solution \eqref{eq_appendix_ads boson singular solution} yields logarithmic divergence when one point is close to the $\text{AdS}_d$ boundary:
\begin{equation}
\label{eq_appendix_ads boson zero mode log}
    \begin{aligned}
\tilde{G}^{\text{log}}_\phi(\xi\ll 1)=\frac{(\xi/2\pi)^{\frac{d-1}{2}}\log (\xi )}{(d-1) \sin \left(\frac{\pi  d}{2}\right) \Gamma
   \left(\frac{1-d}{2}\right)}+O\left((\log \xi)^0,\xi^{\frac{d+1}{2}}\right)~.
    \end{aligned}
\end{equation}
In the main text, as we calculate the one-point functions we will also need the propagator \eqref{eq_appendix_ads boson Green function} at the coincident point:
\begin{equation}
\label{eq_appendix_boson coincident point function}
\tilde{G}_\phi(\xi=1)=\frac{ \Gamma \left(1-\frac{d}{2}\right) \Gamma (\Delta )}{(4\pi)^{d/2}\Gamma (\Delta +2-d)}=\frac{1}{2 \pi^{\frac{d-1}{2}}\Gamma \left(\frac{3-d}{2}\right)}\int_{0}^{1}\frac{u^{\Delta-1}}{(1-u)^{d-1}}du~,
\end{equation}
which follows Gauss's summation theorem. Particularly, the coincident point is singular when $\Delta$ approaches zero:
\begin{equation}
\label{eq_appendix_alternative divergence}
    \begin{aligned}
\tilde{G}_\phi(\xi=1,|\Delta|\ll 1)=\frac{1}{2 \pi^{\frac{d-1}{2}} \Gamma \left(\frac{3-d}{2}\right)\Delta}+O\left(\Delta^0\right)~.
    \end{aligned}
\end{equation}

Now let us follow \cite{muck2000spinor,giombi2022fermions,Giombi:2021uae,giombi2023line} and move on to discuss the free Dirac fermion:
\begin{equation}
    \begin{aligned}
\Tilde{S}_\text{Dirac}=\int_{\text{AdS}_{d}}d^d \Tilde{X}\Bar{\psi}\left(\Tilde{\gamma}\cdot\Tilde{\nabla}-m\right)\psi~.
    \end{aligned}
\end{equation}
The spinor Green's function $\langle \Bar{\psi}(x_1) \psi(x_2)\rangle=\tilde{G}_\psi(x_1,x_2)$, similar to the boson case, is subject to the Dirac equation
\begin{equation}
\label{eq_appendix ads fermion Dirac equation}
    \begin{aligned}
(\Tilde{\gamma}\cdot\Tilde{\nabla}_{1}-m)\tilde{G}_\psi(x_1,x_2)=\delta_{\text{AdS}_d}(x_1-x_2)~.
    \end{aligned}   
\end{equation}
Explicitly, the Dirac operator on $\text{AdS}_d$ reads
\begin{equation}
    \begin{aligned}
\tilde{\gamma}\cdot \tilde{\nabla}=r \tilde{\gamma}^\mu \partial_\mu-\frac{d-1}{2}\tilde{\gamma}_r
    \end{aligned}
\end{equation}
where $\tilde{\gamma}_\mu$ are the flat-space Clifford matrices. \eqref{eq_appendix ads fermion Dirac equation} is a first-order differential equation of two variables $\psi=(\psi_{\text{u}},\psi_{\text{d}})$, which also admits two solutions distinguished by their asymptotic scaling dimension $\Delta=\frac{d-1}{2}\pm |m|$. When $m\neq 0$, it is found that
\begin{equation}
\label{eq_appendix ads fermion complete green function}
\mathtoolsset{multlined-width=0.93\displaywidth}
\begin{multlined}
\tilde{G}_\psi(x_1,x_2)=\frac{\Gamma(\frac{d}{2}\pm|m|)}{2\pi^{\frac{d-1}{2}}\Gamma(\frac{1}{2}\pm |m|)}\left(\frac{\xi}{2-2\xi}\right)^{\pm |m|}\times\\
\left[\frac{\vec{\tilde{\gamma}}\cdot\vec{x}_{12}+\tilde{\gamma}_r(r_1-r_2)}{\sqrt{r_1r_2}}\left(\frac{\xi }{2-2\xi }\right)^{\frac{d}{2}} \, _2F_1\left(\pm |m|,\frac{d}{2}\pm |m|;1\pm 2|m|;\frac{2 \xi }{\xi -1}\right)\right.\\
\hfill\left. \pm \text{sgn}(m)\frac{\tilde{\gamma}_r\vec{\tilde{\gamma}}\cdot\vec{x}_{12}-(r_1+r_2)}{\sqrt{r_1r_2}}\left(\frac{\xi }{2+2\xi }\right)^{\frac{d}{2}} \, _2F_1\left(\pm|m|,\frac{2-d}{2}\pm |m|;1\pm 2|m|;\frac{2 \xi }{\xi -1}\right) \right]~. 
\end{multlined}
\end{equation}
Let one of the correlation points $x_1$ approach the $\text{AdS}_d$ boundary, we obtain an asymptotic expansion to \eqref{eq_appendix ads fermion complete green function} as
\begin{equation}
\label{eq_appendix_ads fermion green function asy 1}
\tilde{G}_\psi(r_1\ll r_2, |\vec{x_{12}}|)=(1\pm \tilde{\gamma}_r  \text{sgn}(m)) \frac{(\vec{\tilde{\gamma}}\cdot\vec{x}_{12}-\tilde{\gamma}_r r_2) \Gamma \left(\frac{d}{2}\pm |m|\right)}{2\pi^{\frac{d-1}{2} }\Gamma \left(\frac{1}{2}\pm |m|\right)}\frac{ 
   \left(r_1 r_2\right)^{\frac{d-1}{2}\pm |m|}}{\left[(\vec{x}_{12})^2+r_2^2\right]^{\frac{d}{2}\pm |m|}}+O\left(r_1^{\frac{d+1}{2}\pm |m|}\right)~.
\end{equation}

On the other hand, the two solutions when $m=0$ reads
\begin{equation}
\label{eq_appendix_ads fermion deg solutions}
\tilde{G}_\psi(x_1,x_2)=\frac{\Gamma \left(\frac{d}{2}\right) }{2 \pi ^{\frac{d}{2}} } \left[\frac{\vec{\tilde{\gamma}}\cdot\vec{x}_{12}+\tilde{\gamma}_r(r_1-r_2)}{\sqrt{r_1r_2}}\left(\frac{\xi }{2-2\xi }\right)^{\frac{d}{2}}\pm \frac{\tilde{\gamma}_r\vec{\tilde{\gamma}}\cdot\vec{x}_{12}-(r_1+r_2)}{\sqrt{r_1r_2}}\left(\frac{\xi }{2+2\xi }\right)^{\frac{d}{2}}\right]~.
\end{equation}
One important distinction between the bosonic case \eqref{eq_appendix_ads boson singular solution} and \eqref{eq_appendix_ads boson zero mode log} is that such two fermionic massless solutions are degenerate in their asymptotic scaling close to the boundary:
\begin{equation}
\label{eq_appendix_ads fermion green function asy 2}
    \begin{aligned}
\tilde{G}_\psi(r_1\ll r_2, |\vec{x_{12}}|)=(1\pm \tilde{\gamma}_r) (\vec{\tilde{\gamma}}\cdot\vec{x}_{12}-\tilde{\gamma}_r r_2)\frac{\Gamma \left(\frac{d}{2}\right)}{2\pi^{\frac{d}{2} }}\frac{ 
   \left(r_1 r_2\right)^{\frac{d-1}{2}}}{\left[(\vec{x}_{12})^2+r_2^2\right]^{\frac{d}{2}}}+O\left(r_1^{\frac{d+1}{2}}\right)~.
    \end{aligned}
\end{equation}

Finally, we note that coincident point functions
\begin{equation}
\label{eq_appendix_ads fermion coincident point function}
    \begin{aligned}
\tilde{G}_\psi(x_1=x_2)&=\mp\frac{\text{sgn}(m) \Gamma \left(1-\frac{d}{2}\right) \Gamma \left(\frac{d}{2}\pm |m|\right)}{(4\pi)^{\frac{d}{2}}\Gamma
   \left(\frac{2-d}{2}\pm |m|\right)}=\mp \frac{\text{sgn}(m)\Gamma\left(1-\frac{d}{2}\right)}{(4\pi)^{\frac{d}{2}}\Gamma\left(1-d\right)}\int_{0}^1\frac{u^{\pm |m|+\frac{d}{2}-1}}{{(1-u)^d}}du~.
    \end{aligned}
\end{equation}
For the $m=0$ case, one can generally choose a linear combination of the solutions in \eqref{eq_appendix_ads fermion deg solutions}, such that $-\Gamma(\frac{d}{2})/(4\pi)^{\frac{d}{2}}\leq \tilde{G}_\psi(x_1=x_2)\leq \Gamma(\frac{d}{2})/(4\pi)^{\frac{d}{2}}$. 

\section{Further numerical details}
In this appendix, we present additional technical details about how the numerical results presented throughout the paper were obtained. We also present numerical results for several models introduced in the main text.

\subsection{Defect Hamiltonians}
In this section, we present explicit representative Hamiltonians for each defect studied in this work. While the Hamiltonians we show here include a relatively small number of lattice sites for ease of visualization, the actual Hamiltonians used in numerics are qualitatively the same but on larger lattices.

\begin{figure*}[h!]
\includegraphics[width=.95\linewidth]{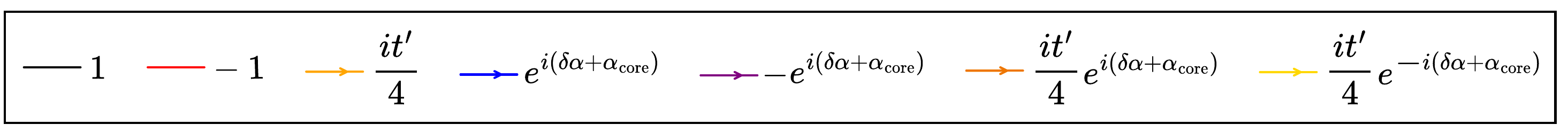}%
\caption{Legend defining the hoppings used in the following figures. $\alpha_\text{core}$ denotes a potentially nonzero flux at the defect core which depends on the definition of the symmetry operator used to create the defect, which varies from model to model.}
\end{figure*}

\begin{figure*}[h!]
        \hfill
        \subfloat[$A$-centered disclination, $\beta=3/4$]{%
            \includegraphics[width=.27\linewidth]{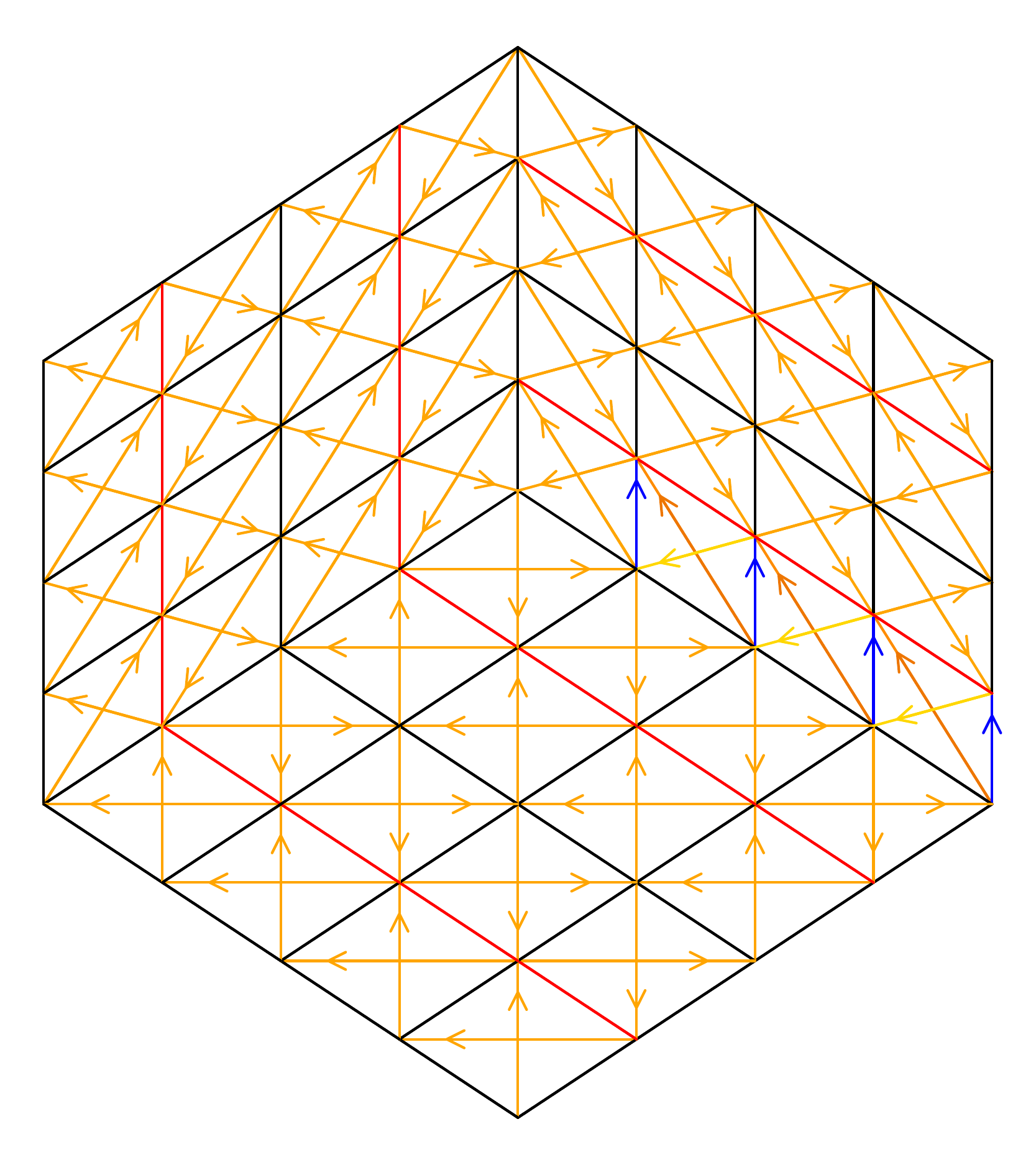}%
        }
        \hfill
        \subfloat[$B$-centered disclination, $\beta=3/4$]{%
        \includegraphics[width=.27\linewidth]{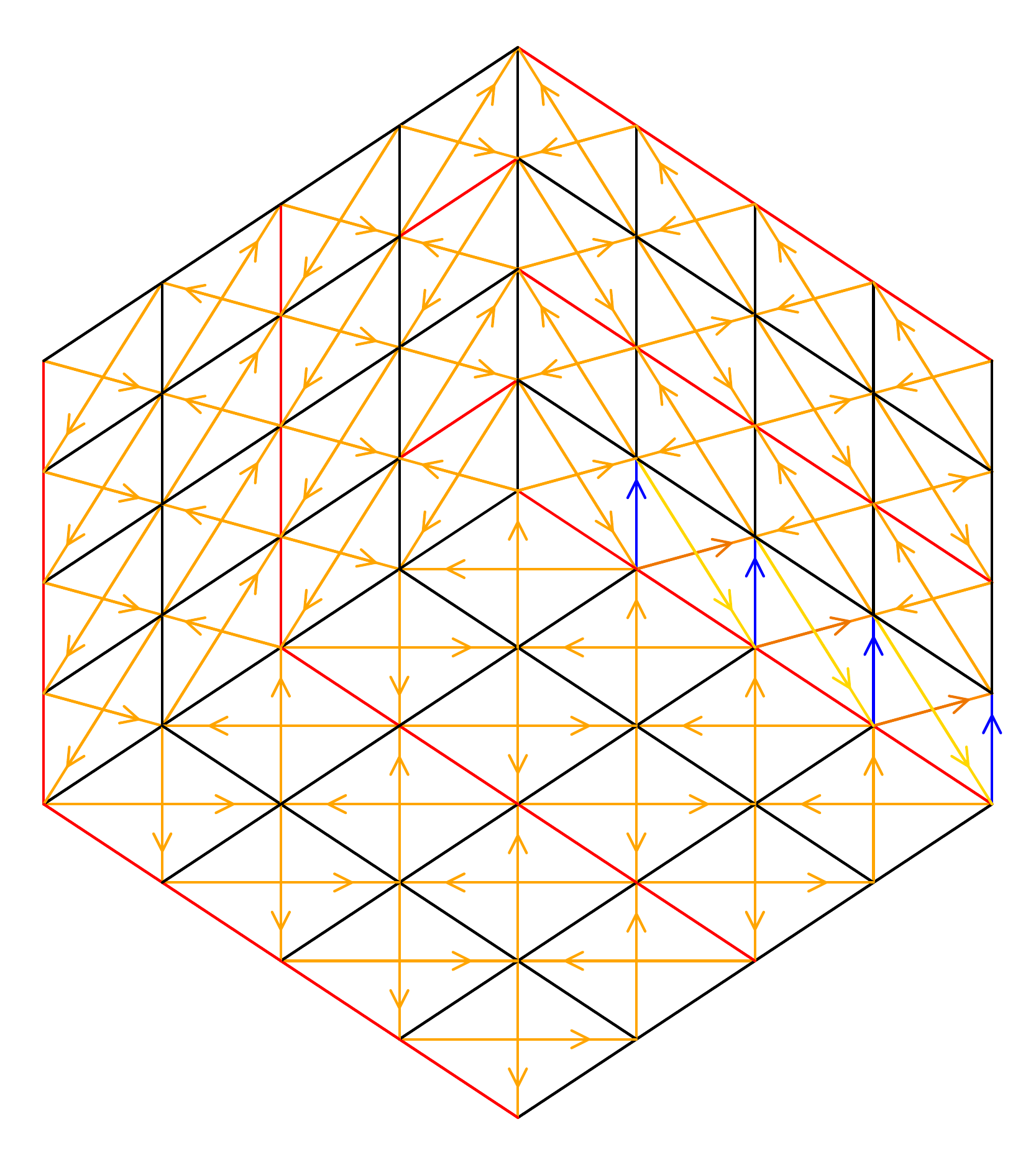}%
        }
        \hfill
\label{}
\caption{No on-site potential $\pi$-flux model}
\end{figure*}

\begin{figure*}[h!]
        \hfill
        \subfloat[$A$-centered disclination, $\beta=3/4$]{%
            \includegraphics[width=.27\linewidth]{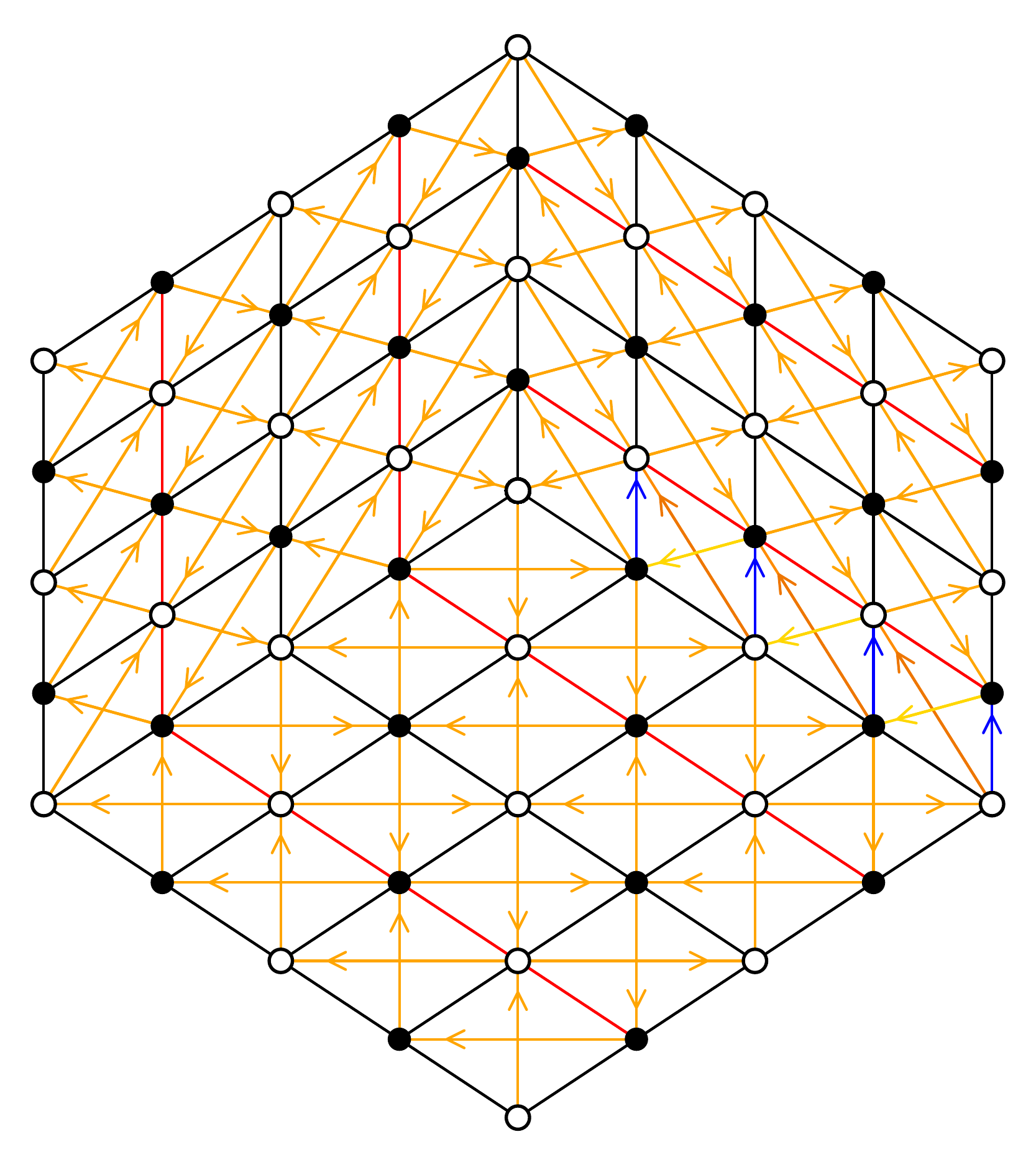}%
        }
        \hfill
        \subfloat[$B$-centered disclination, $\beta=3/4$]{%
        \includegraphics[width=.27\linewidth]{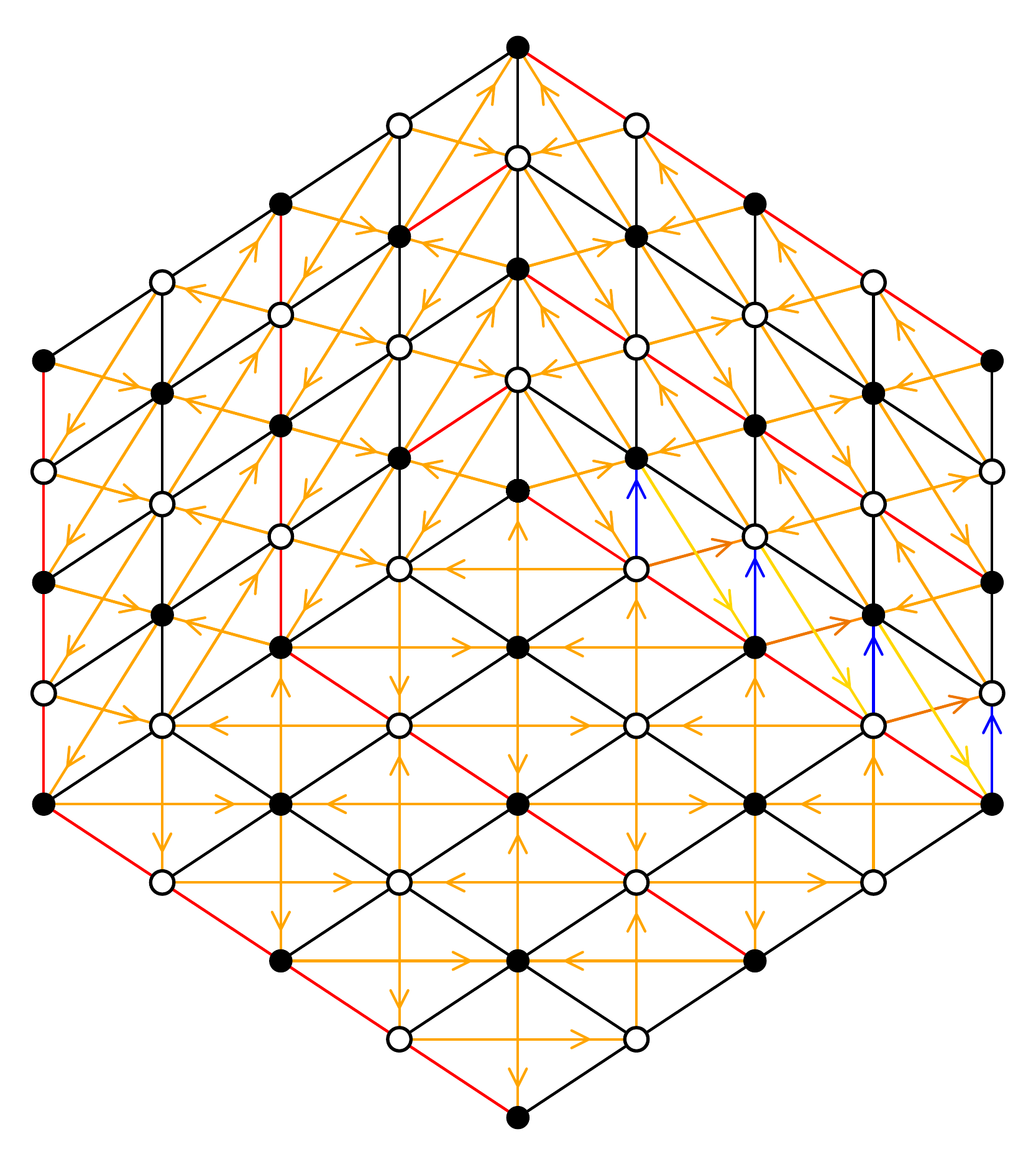}%
        }
        \hfill
        \subfloat[$\vec{b}=(1,1)$ dislocation]{%
        \includegraphics[width=.21\linewidth]{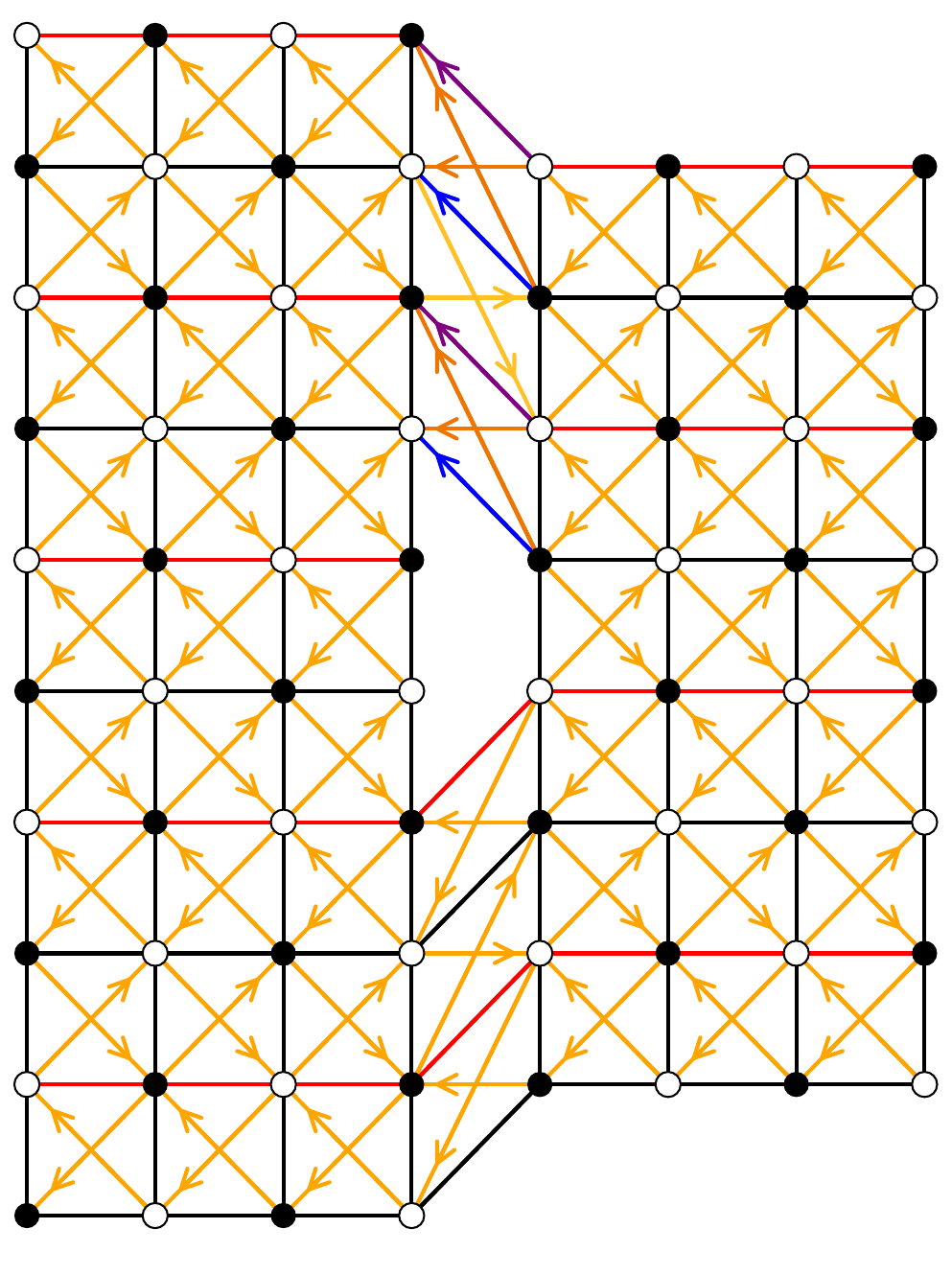}%
        }
        \hfill
\label{}
\caption{One-site staggered potential $\pi$-flux model}
\end{figure*}

\begin{figure*}[h!]
        \subfloat[$A$-centered disclination, $\beta=3/4$]{%
            \includegraphics[width=.27\linewidth]{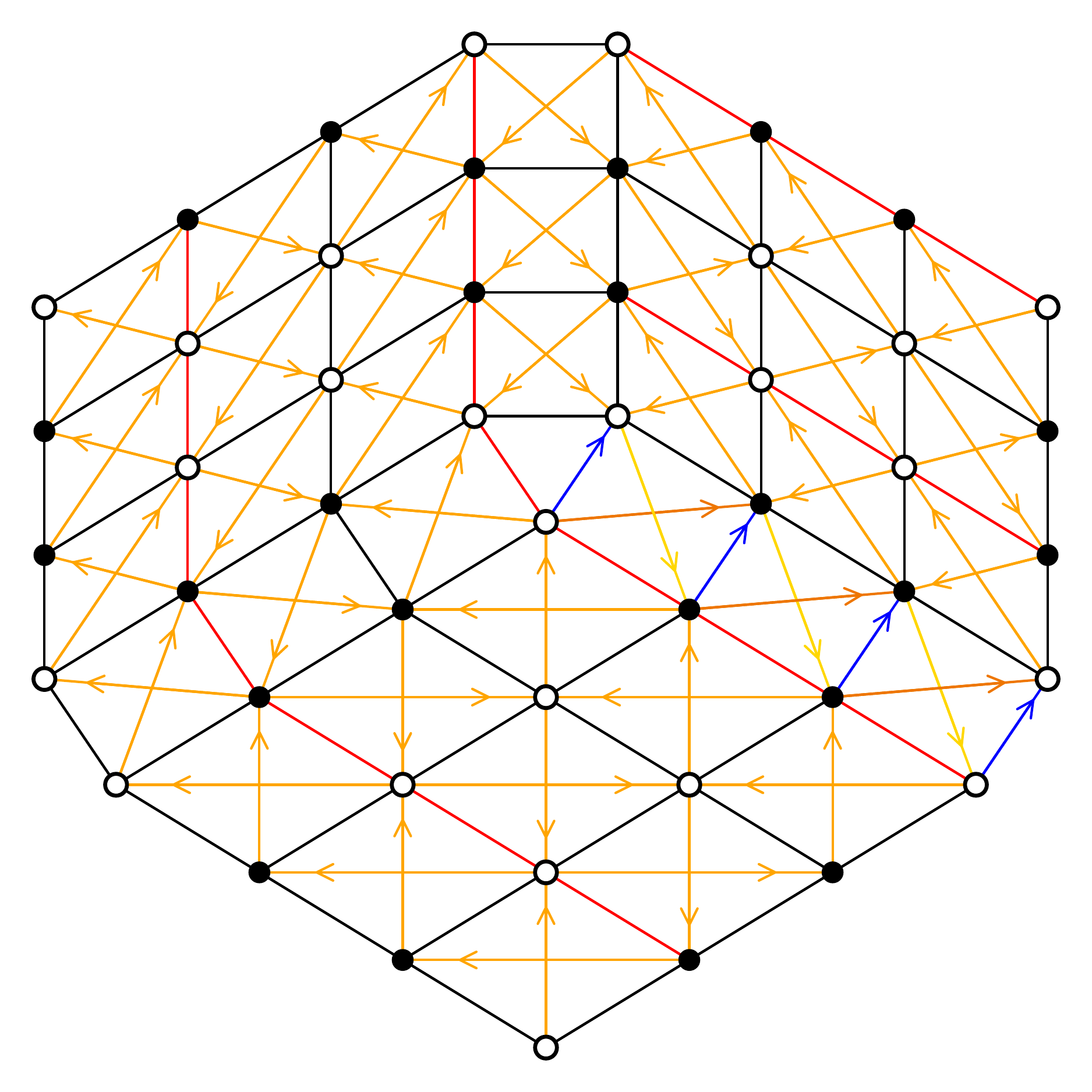}%
        }
        \qquad\qquad\qquad
        \subfloat[$B$-centered disclination, $\beta=3/4$]{%
        \includegraphics[width=.27\linewidth]{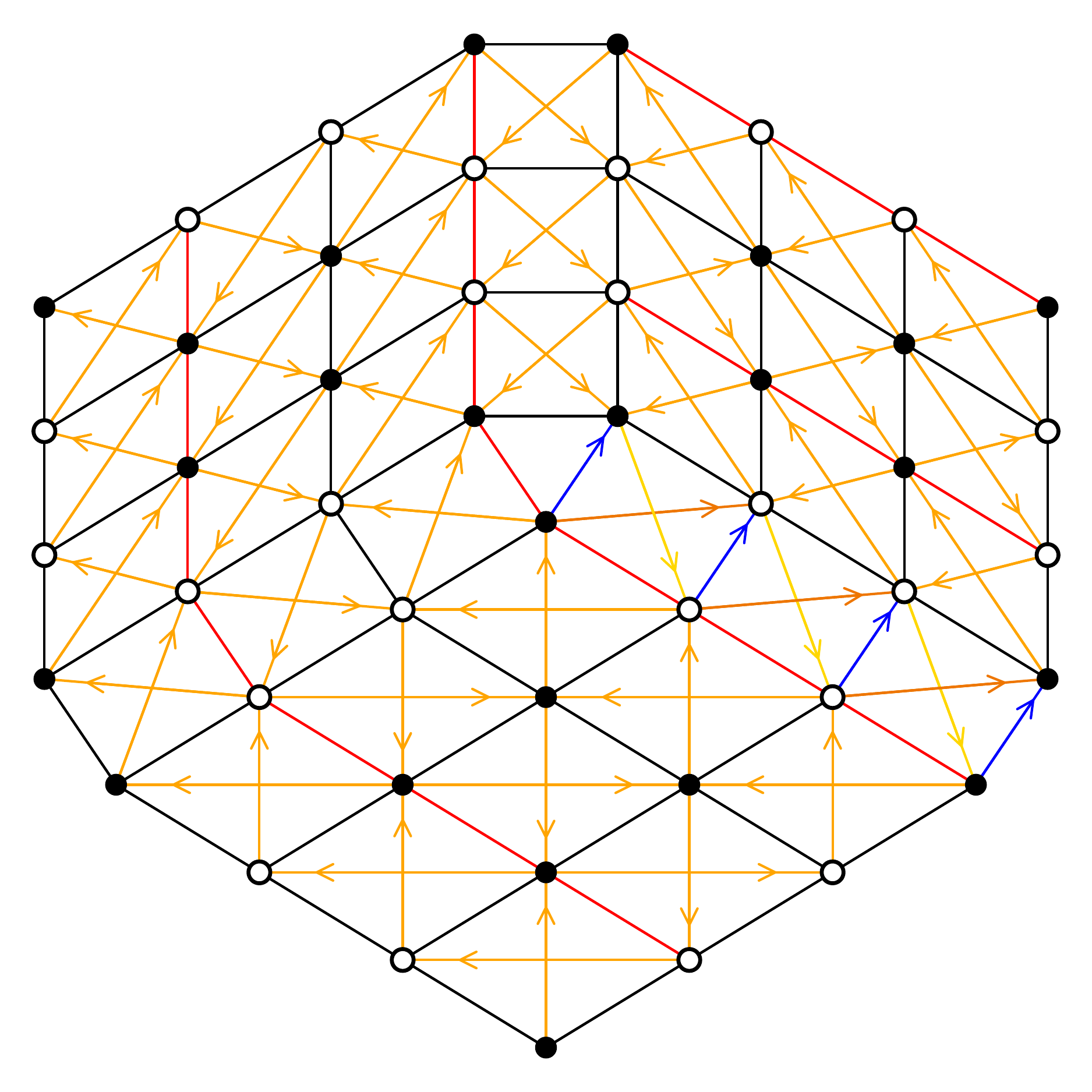}%
        }
\label{}
\caption{Two-site staggered potential $\pi$-flux model }
\end{figure*}

\begin{figure*}[h!]
        \hfill
        \subfloat[$A$-centered disclination, $\beta=2/3$]{%
            \includegraphics[width=.3\linewidth]{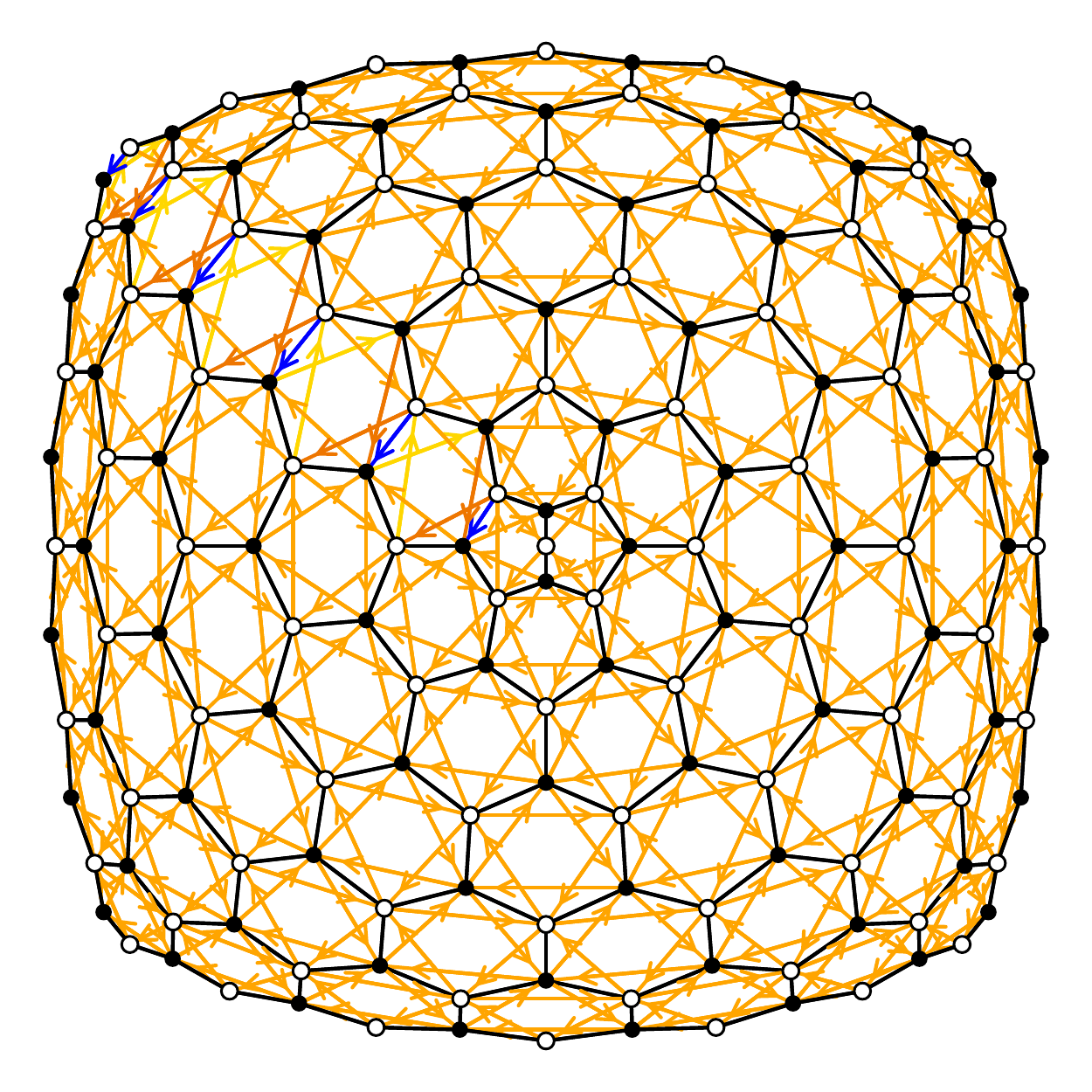}%
        }
        \hfill
        \subfloat[$B$-centered disclination, $\beta=2/3$]{%
        \includegraphics[width=.3\linewidth]{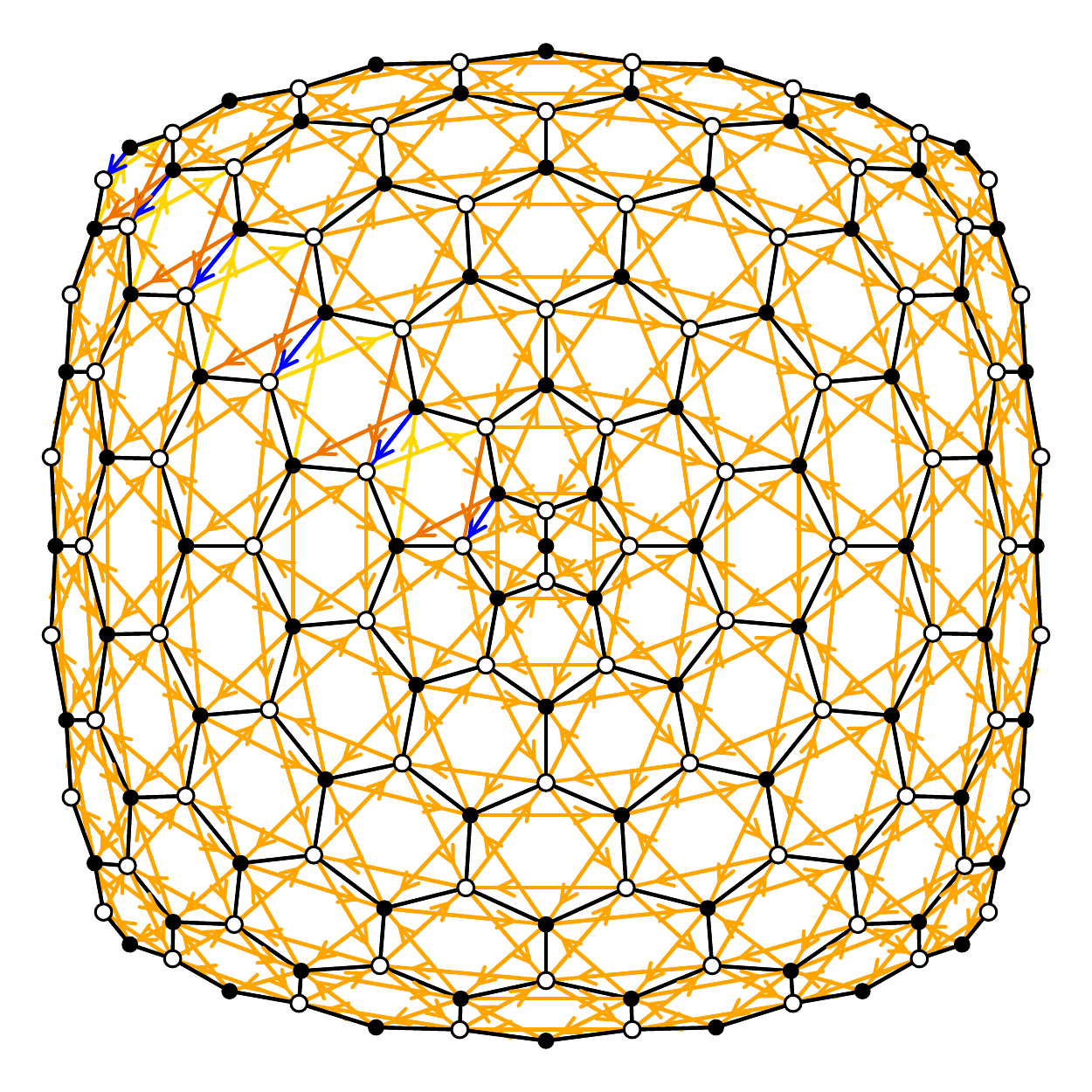}%
        }
        \hfill
        \subfloat[$C$-centered disclination, $\beta=2/3$]{%
        \includegraphics[width=.3\linewidth]{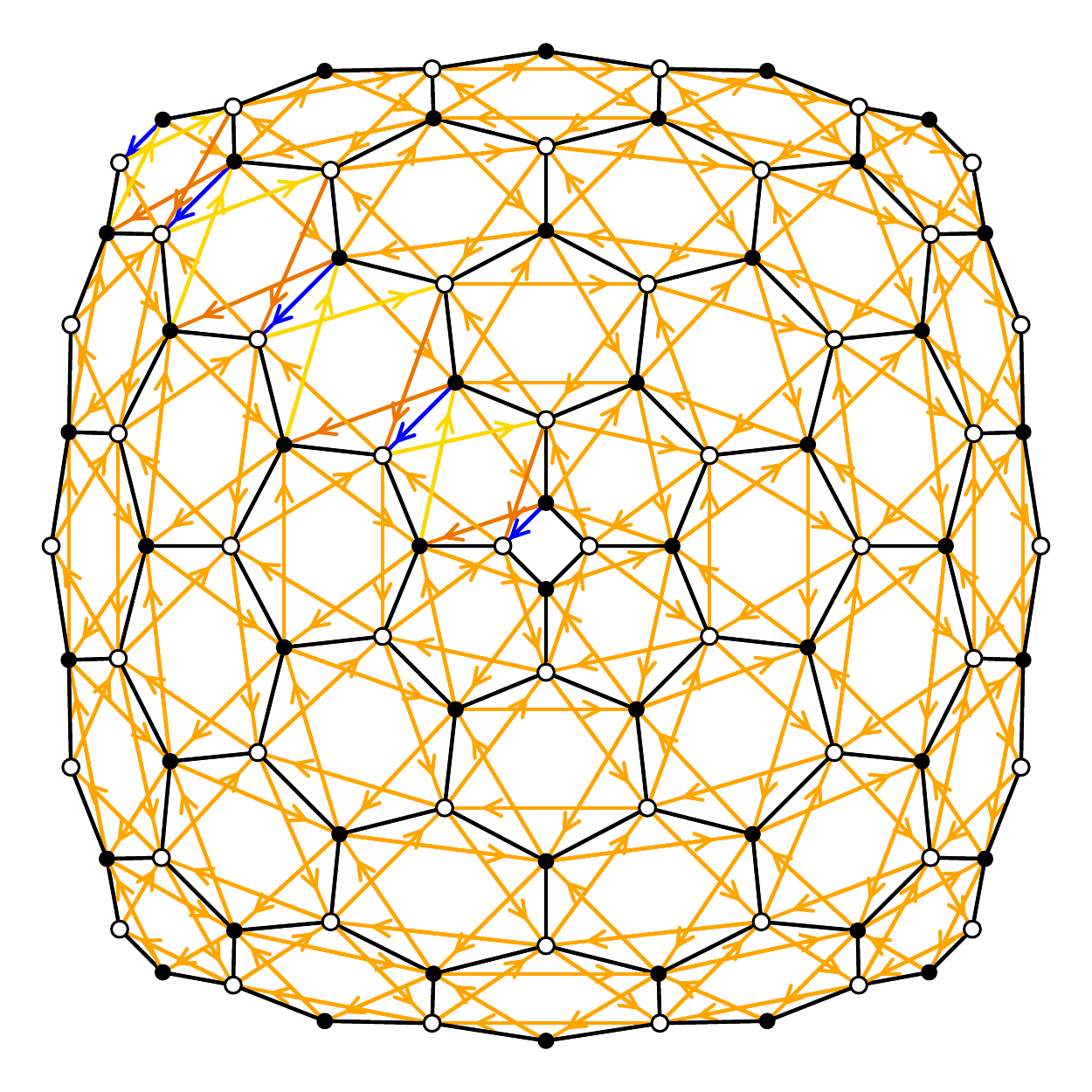}%
        }

        \subfloat[$A$-centered disclination, $\beta=4/3$]{%
            \includegraphics[width=.3\linewidth]{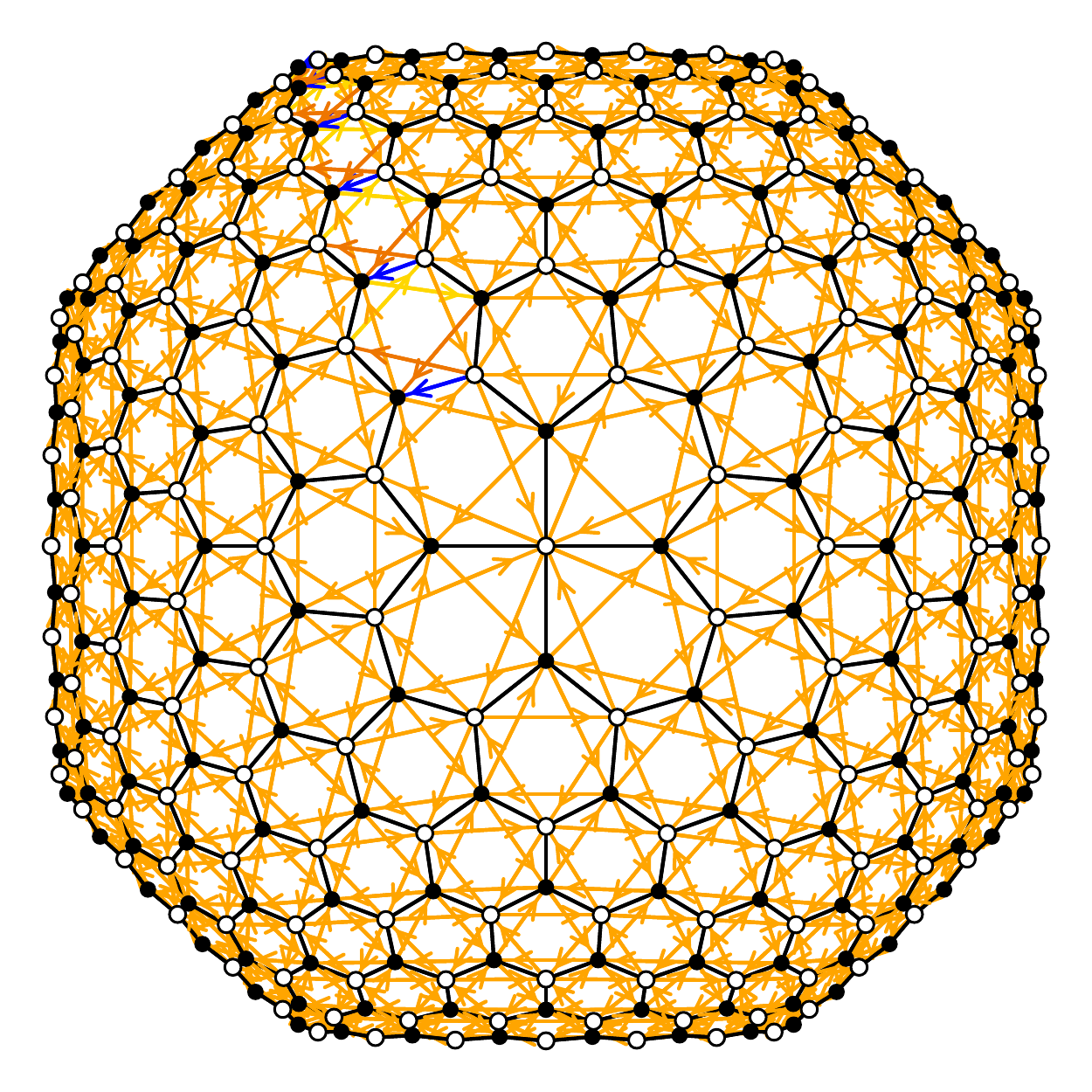}%
        }
        \qquad
        \subfloat[$\vec{b}=(-1,0)$ dislocation]{%
        \includegraphics[width=.25\linewidth]{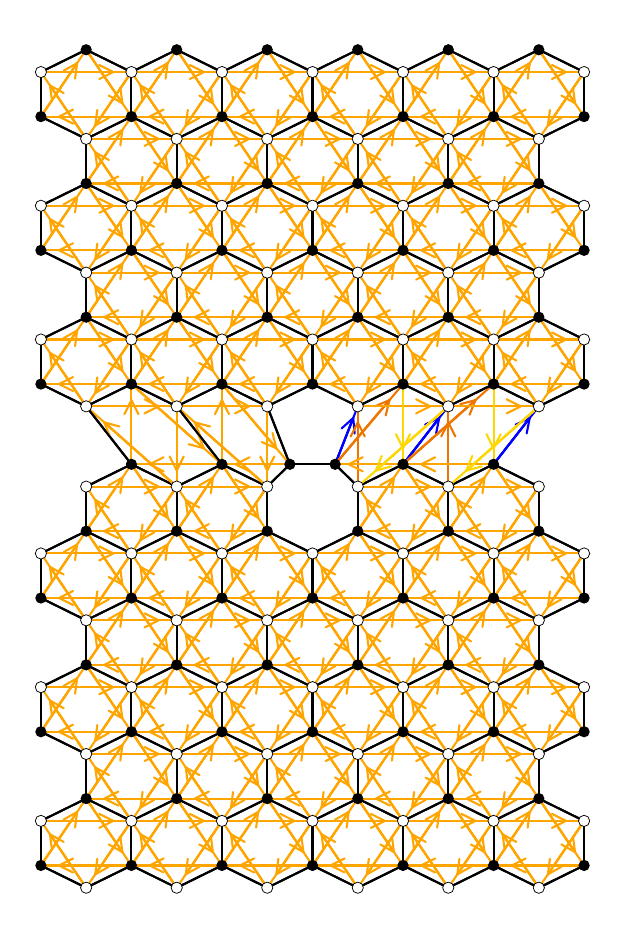}%
        }
\label{}
\caption{Haldane model}
\end{figure*}

\clearpage

\subsection{Additional numerical results}
\begin{figure}[h!]
        \subfloat[Lattice with $N=5419$ sites and $A$-centered $\beta=3/4$ disclination]{%
            \includegraphics[width=.45\linewidth]{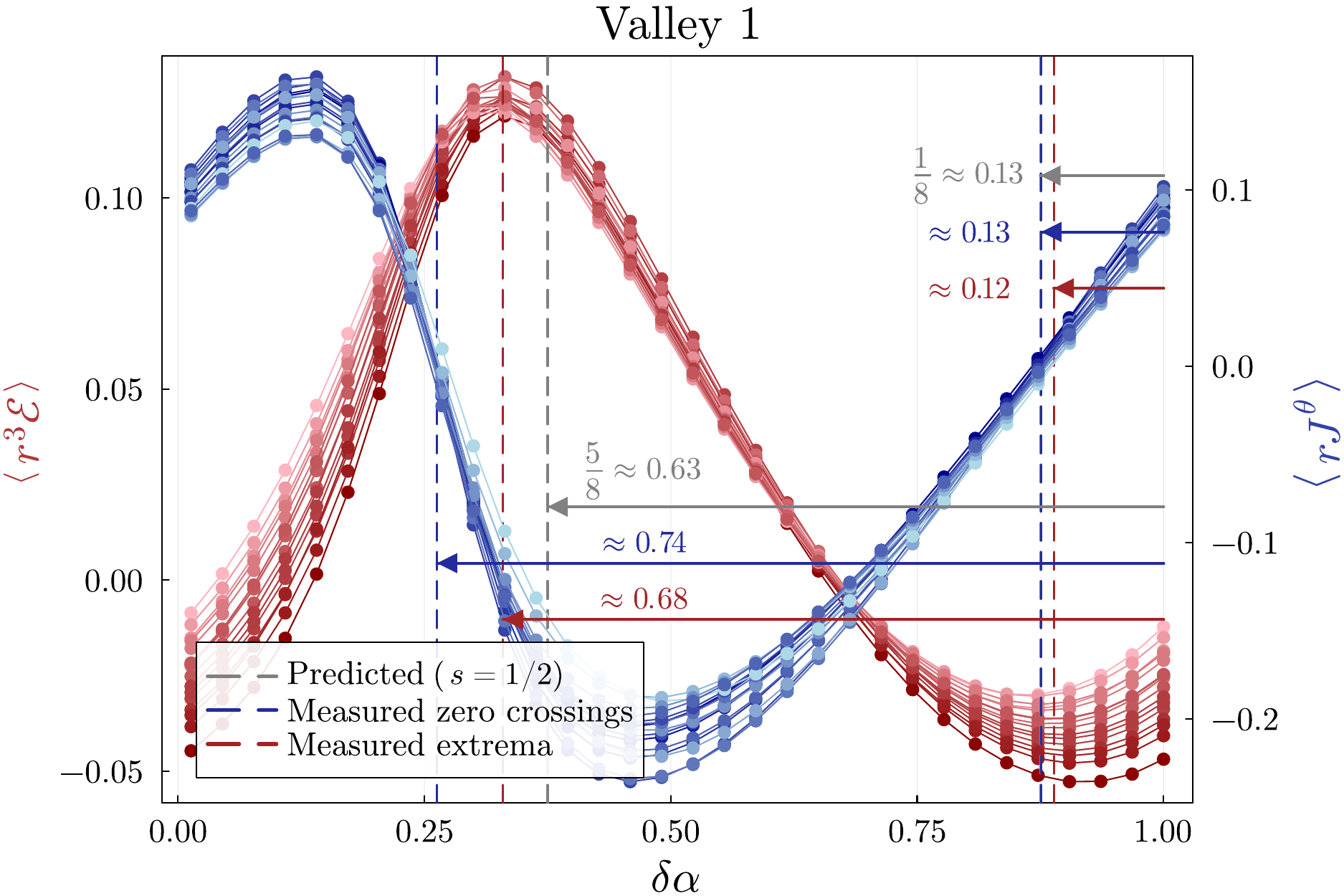}%
        \qquad\includegraphics[width=.45\linewidth]{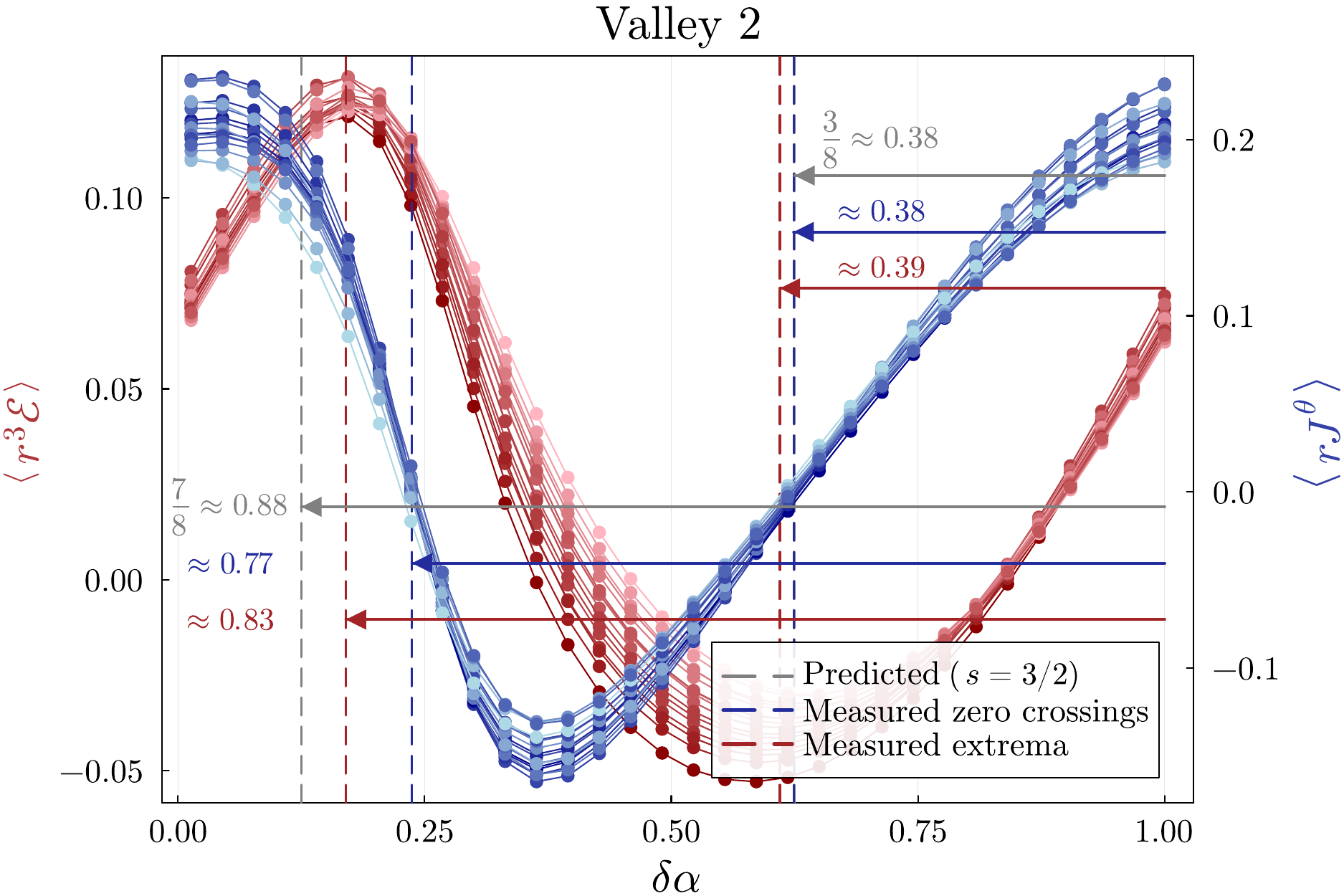}%
        }\\
        
        \subfloat[Lattice with $N=5419$ sites and $B$-centered $\beta=3/4$ disclination]{%
        
        \includegraphics[width=.45\linewidth]{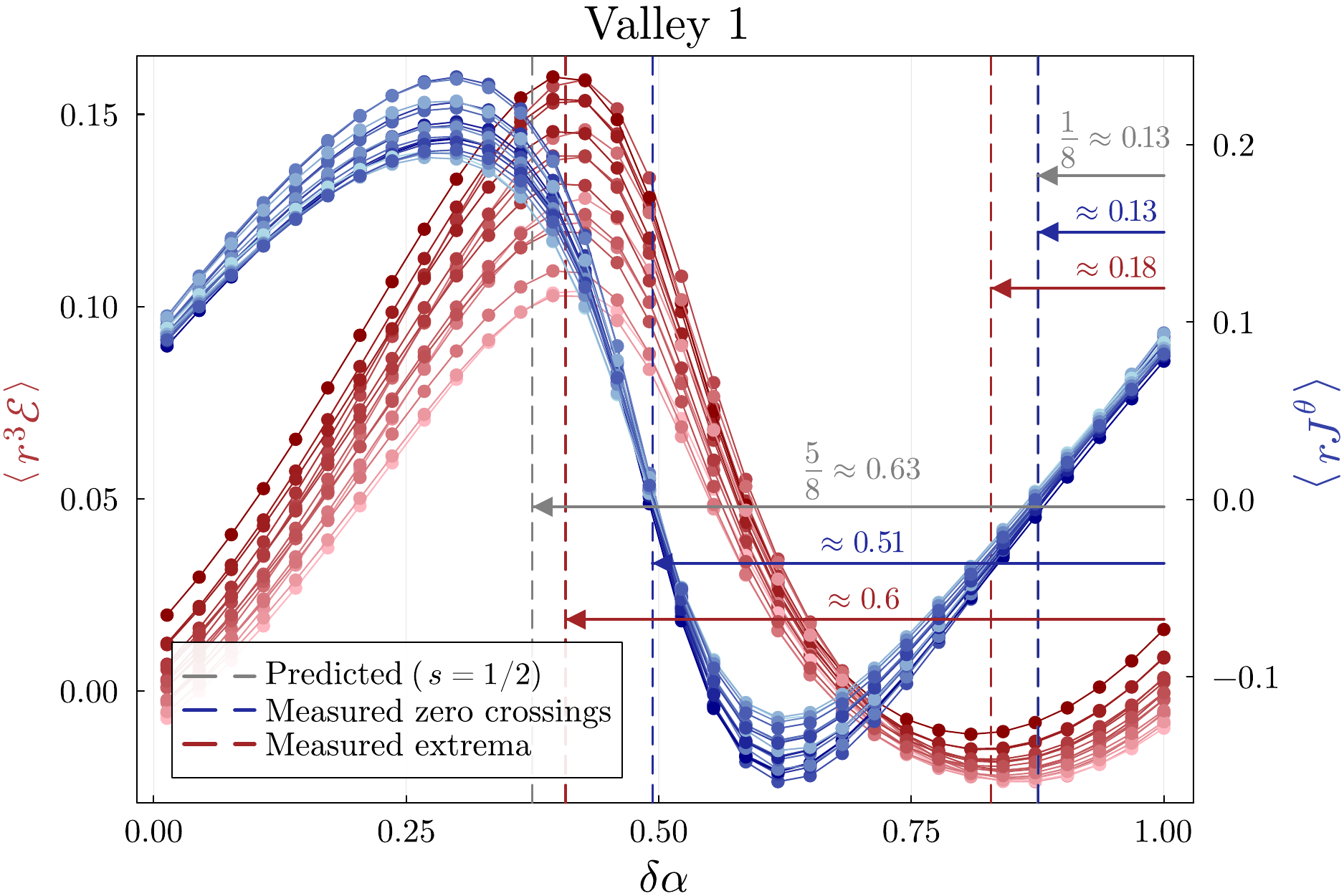}%
            \qquad\includegraphics[width=.45\linewidth]{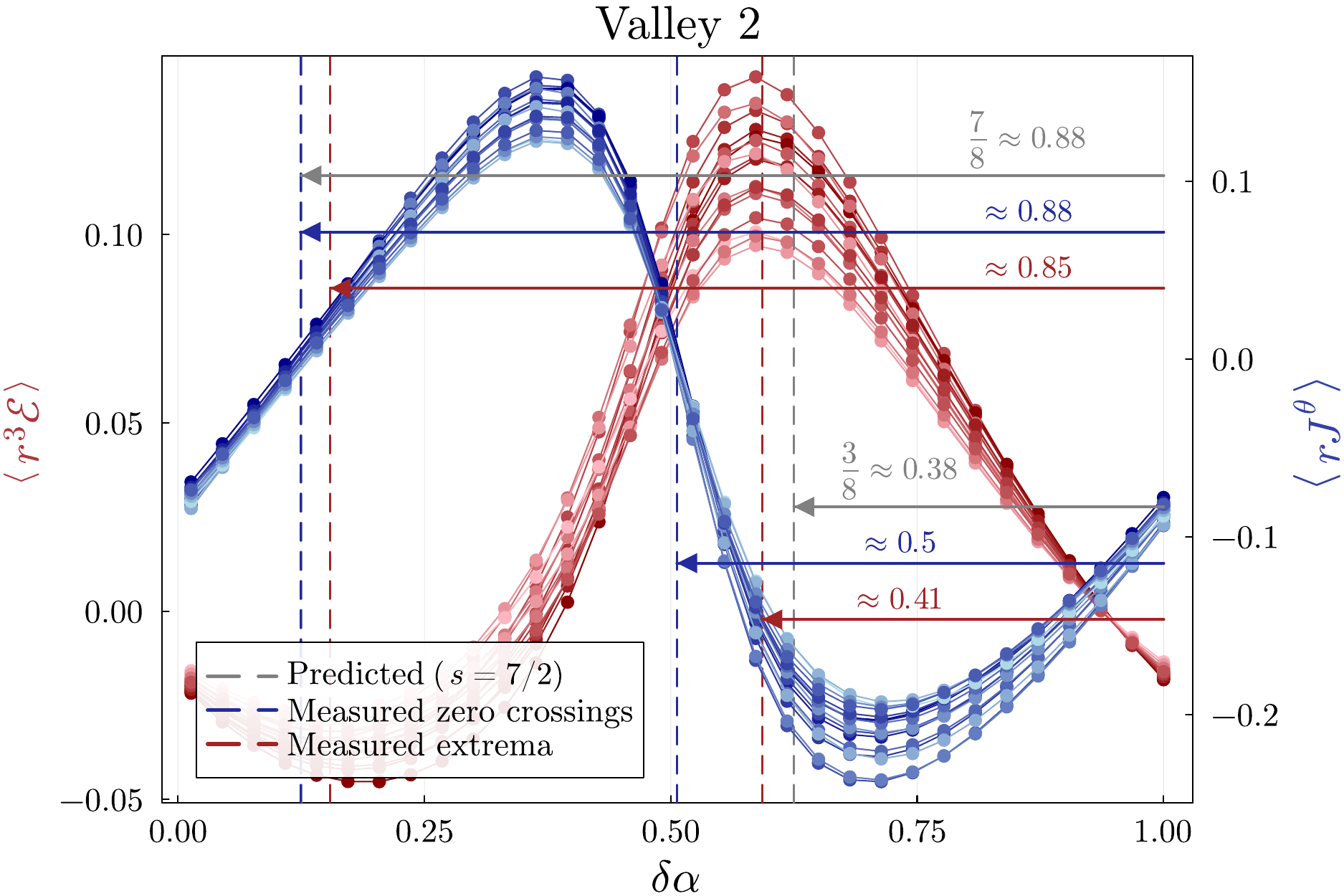}%
        }
        \caption{No on-site potential ${\pi}$-flux model with crystalline defects}
        \label{fig:valley_results}
    \end{figure}

\begin{figure}[h!]
        \subfloat[Lattice with $N=5419$ sites and $A$-centered $\beta=3/4$ disclination]{%
            \includegraphics[width=.45\linewidth]{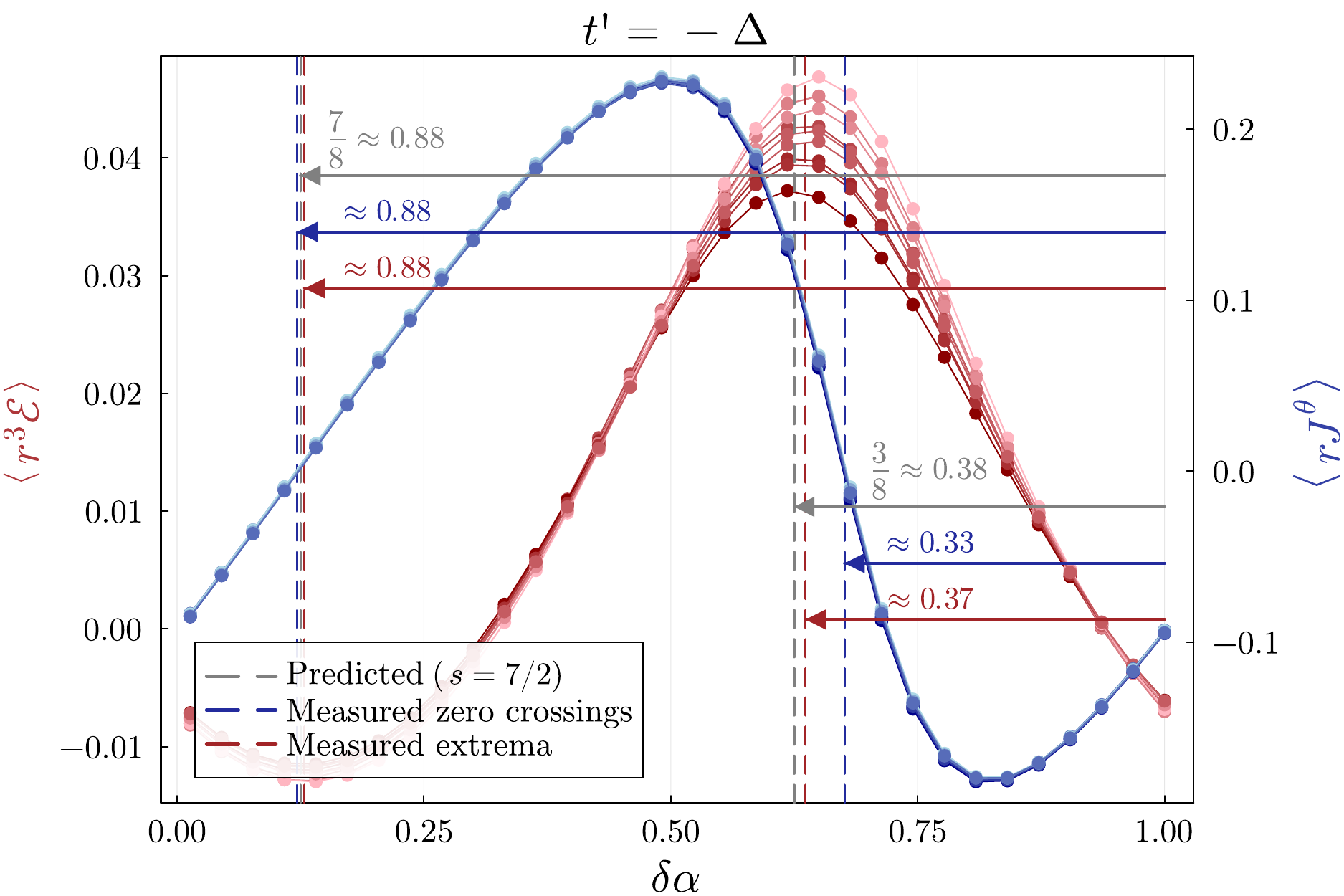}%
        \qquad\includegraphics[width=.45\linewidth]{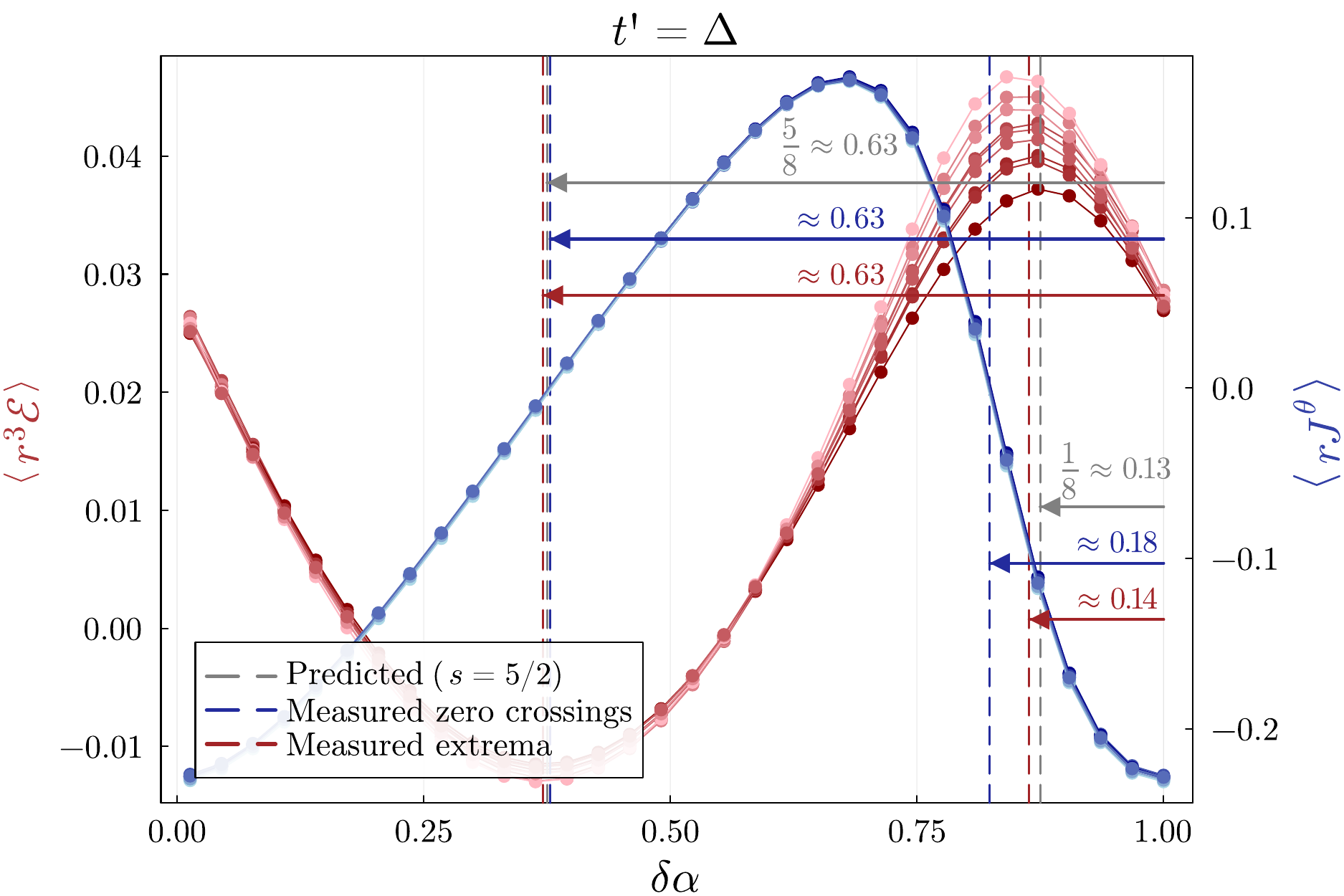}%
        }\\
        
        \subfloat[Lattice with $N=4921$ sites and $B$-centered $\beta=3/4$ disclination]{%
        
        \includegraphics[width=.45\linewidth]{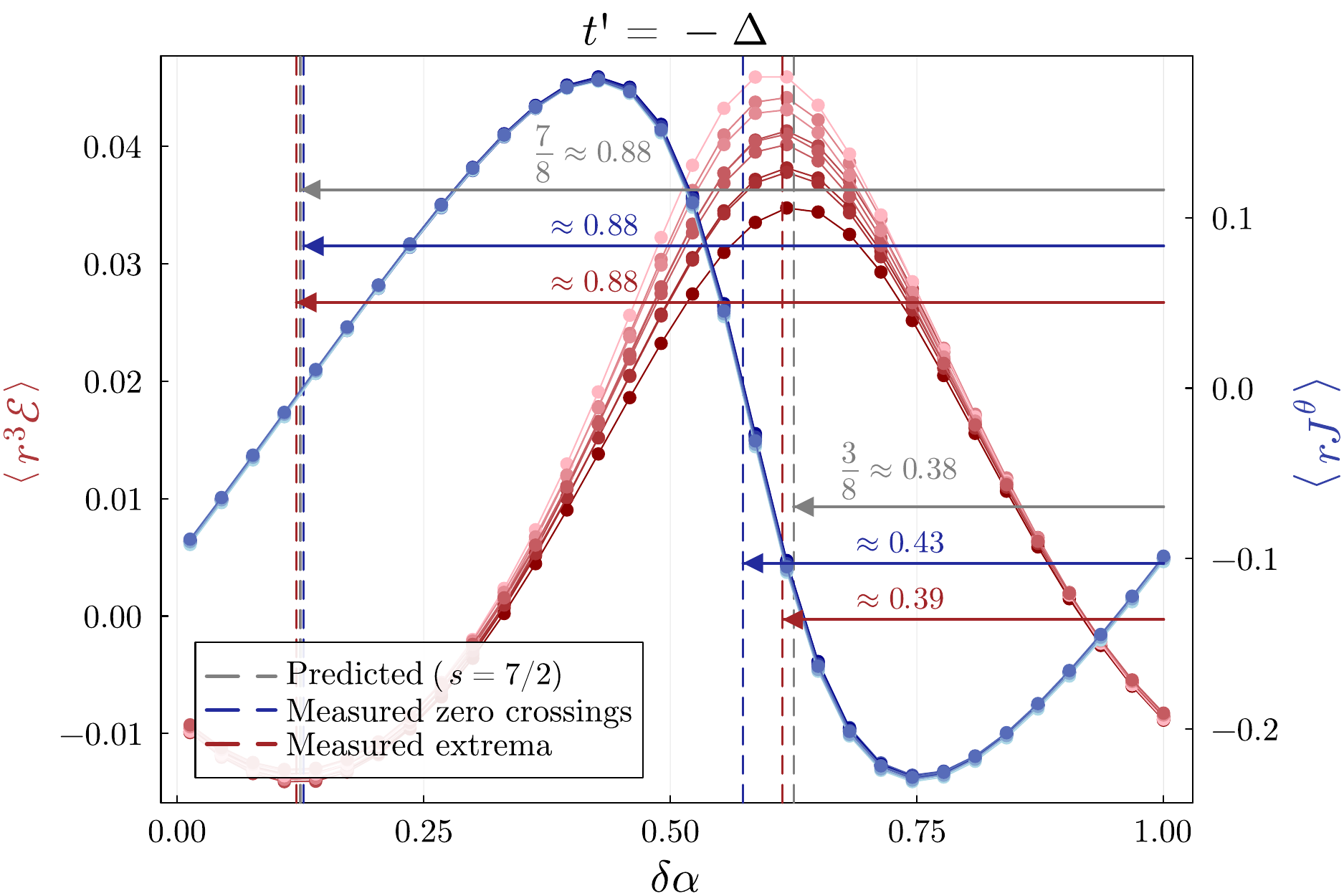}%
            \qquad\includegraphics[width=.45\linewidth]{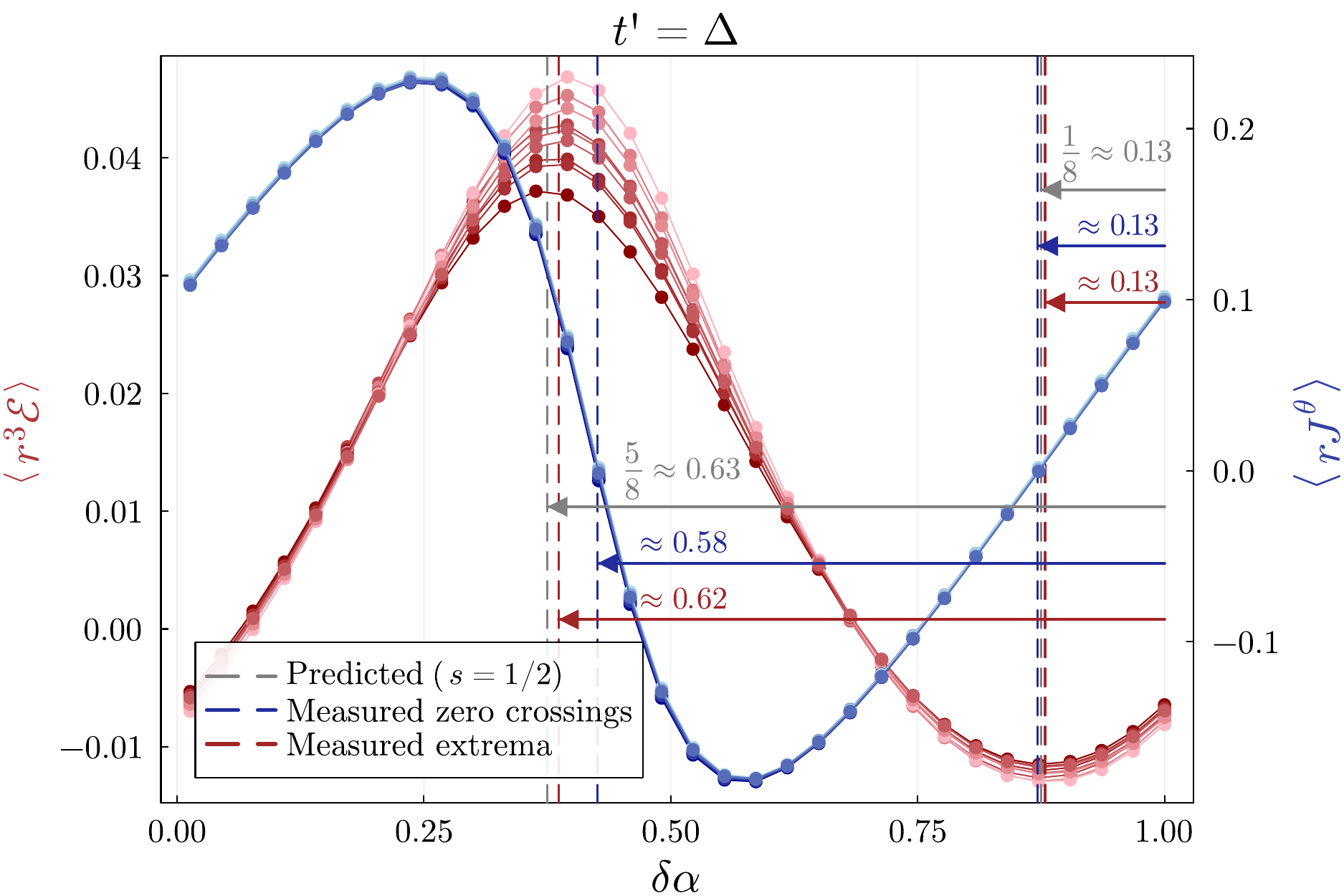}%
        }\\
        \subfloat[Lattice with $N=6440$ sites and $\vec{b}=(1,1)$ dislocation]{%
        
        \includegraphics[width=.45\linewidth]{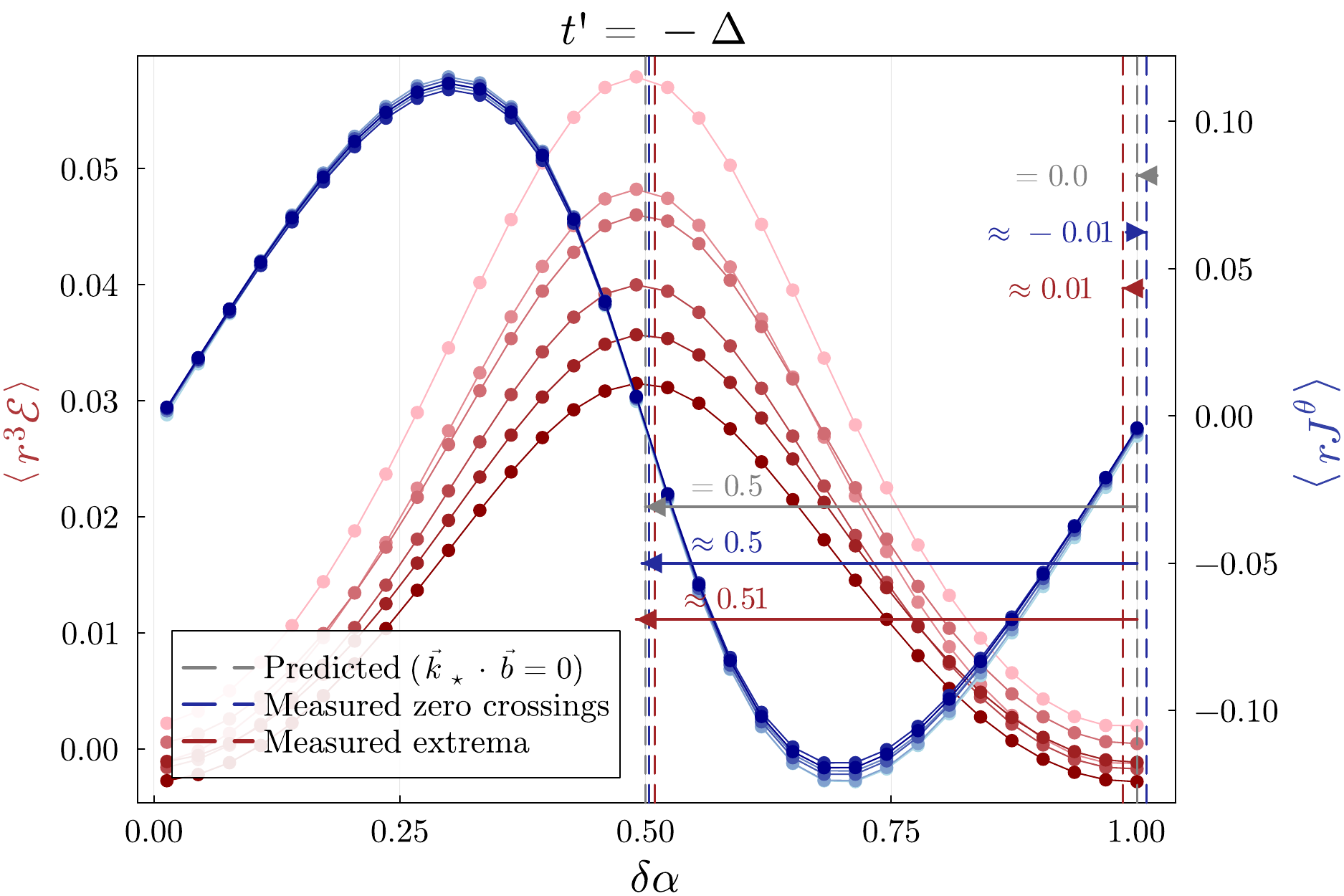}%
            \qquad\includegraphics[width=.45\linewidth]{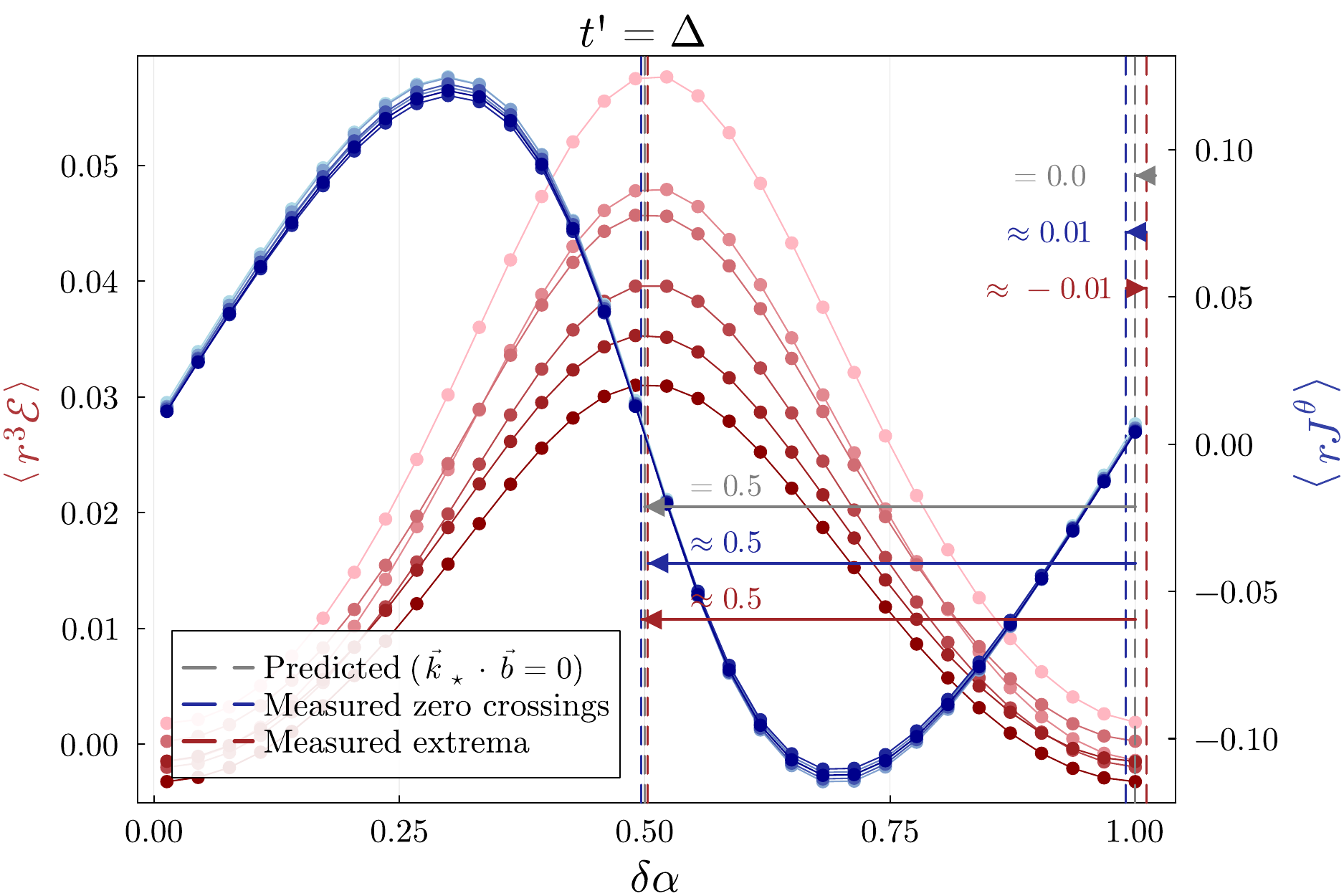}%
        }
        \caption{One-site staggered potential ${\pi}$-flux model with crystalline defects}
        \label{fig:pi_flux_4_currents}
    \end{figure}
    
    \begin{figure*}[h]
        \subfloat[$A$-centered disclination lattice with $N=4801$ sites]{%
            \includegraphics[width=.45\linewidth]{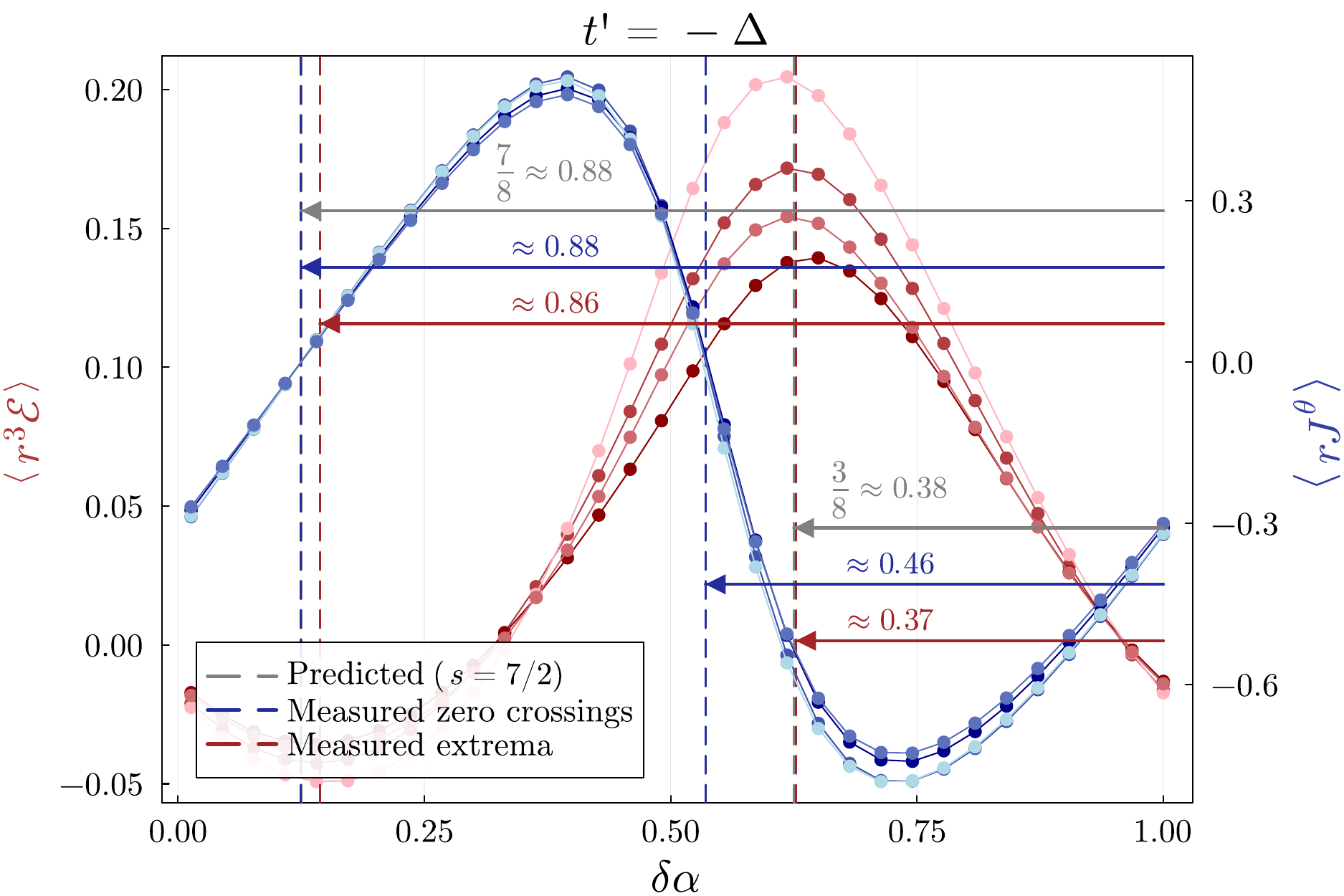}%
        \qquad\includegraphics[width=.45\linewidth]{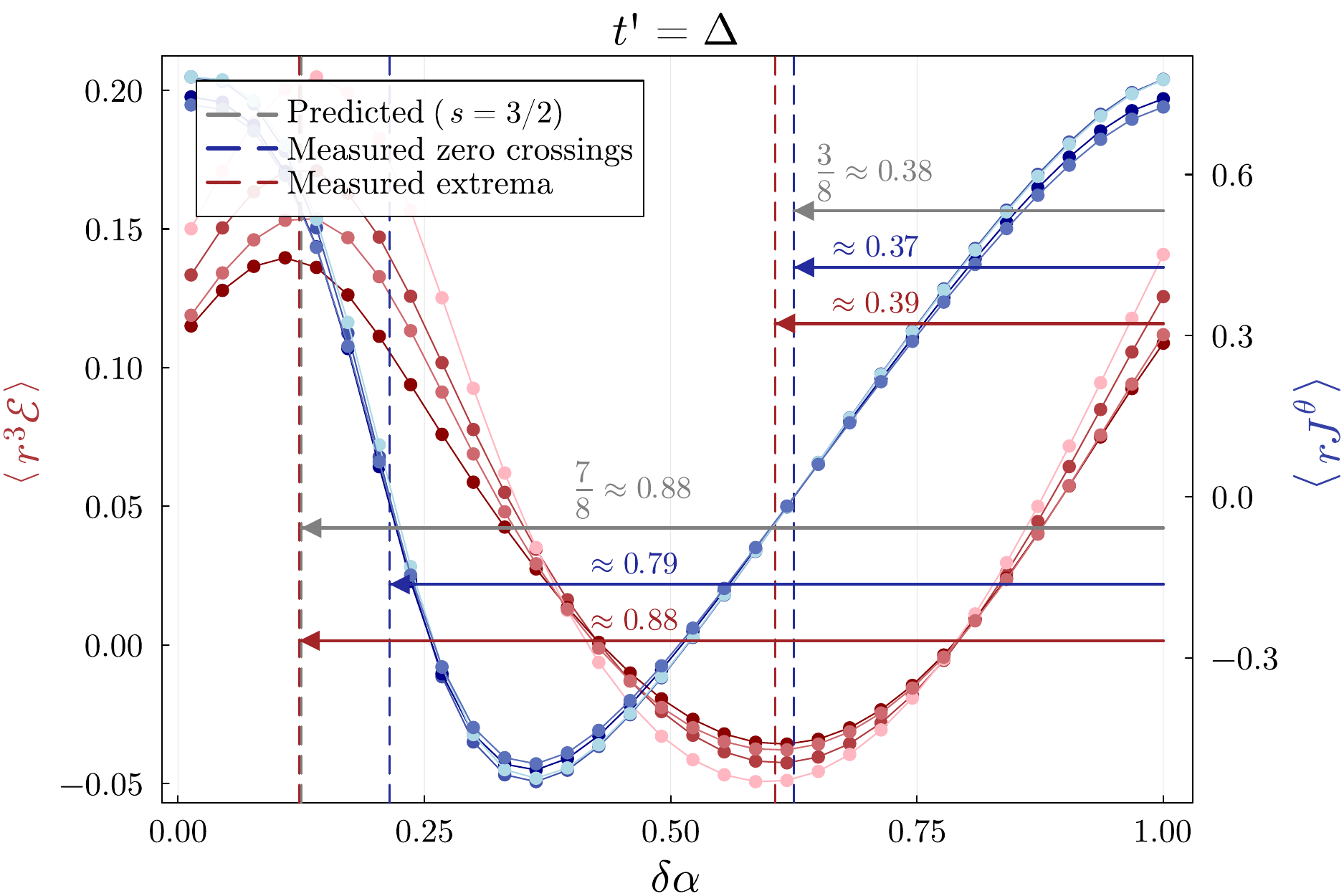}%
        }\\
        \subfloat[$B$-centered disclination lattice with $N=4801$ sites]{%
        
        \includegraphics[width=.45\linewidth]{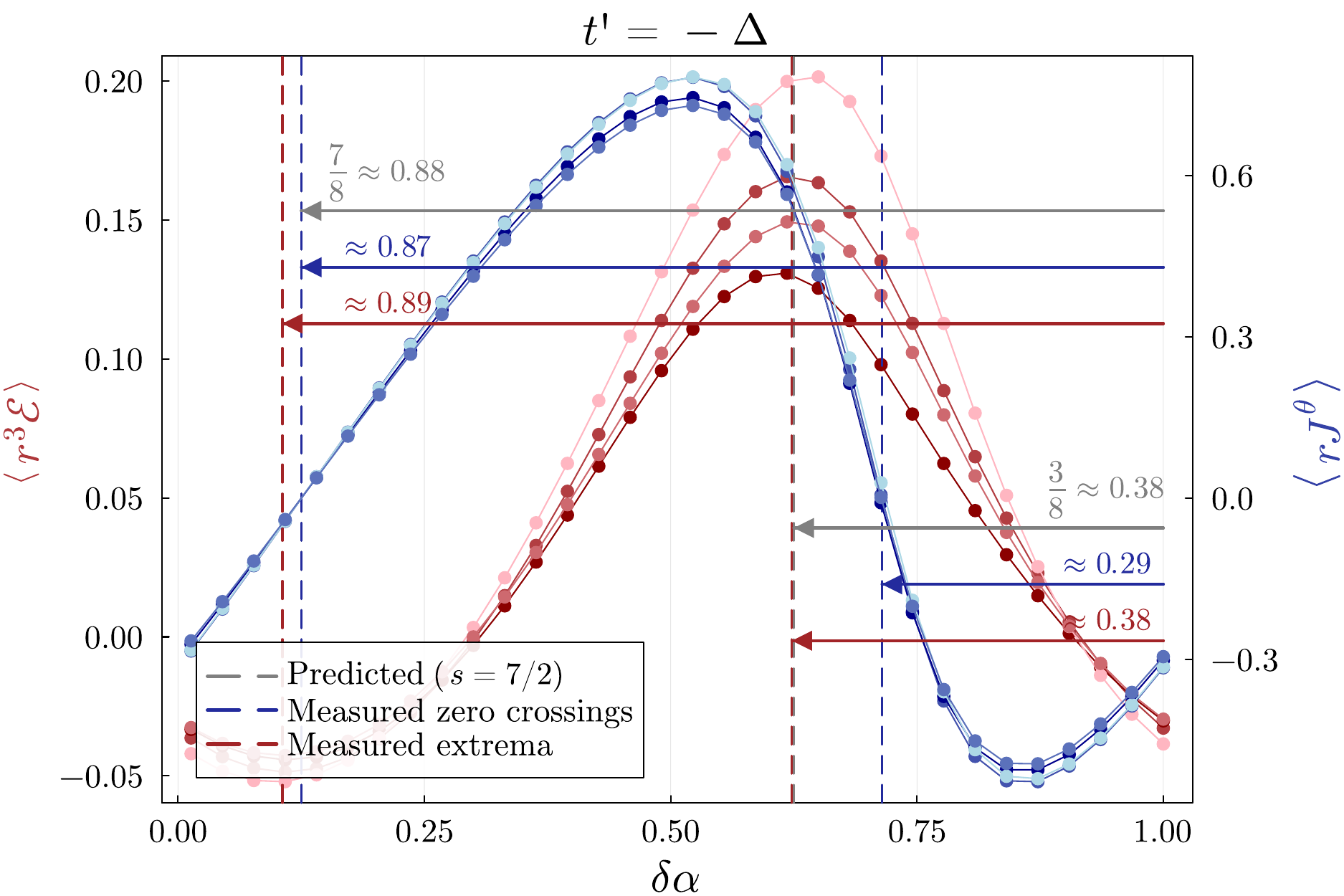}%
            \qquad\includegraphics[width=.45\linewidth]{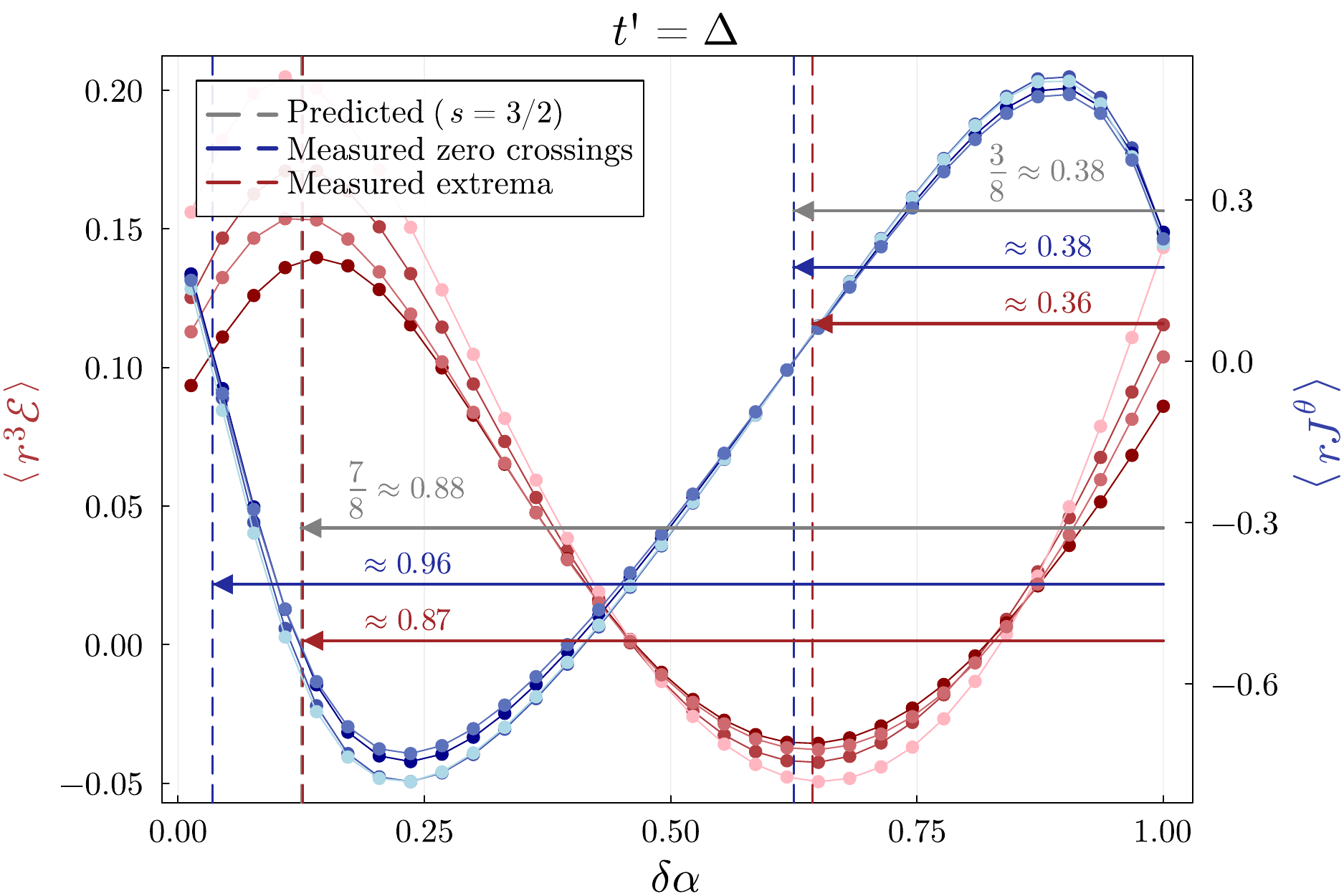}%
        }\\
        \caption{Two-site staggered potential $\pi$-flux model with crystalline defects}
        \label{fig:pi_flux_8_currents}
    \end{figure*}

\begin{figure*}[h!]
        \subfloat[Lattice with $N=4761$ sites and a $B$-centered $\beta=2/3$ disclination]{%
        
        \includegraphics[width=.45\linewidth]{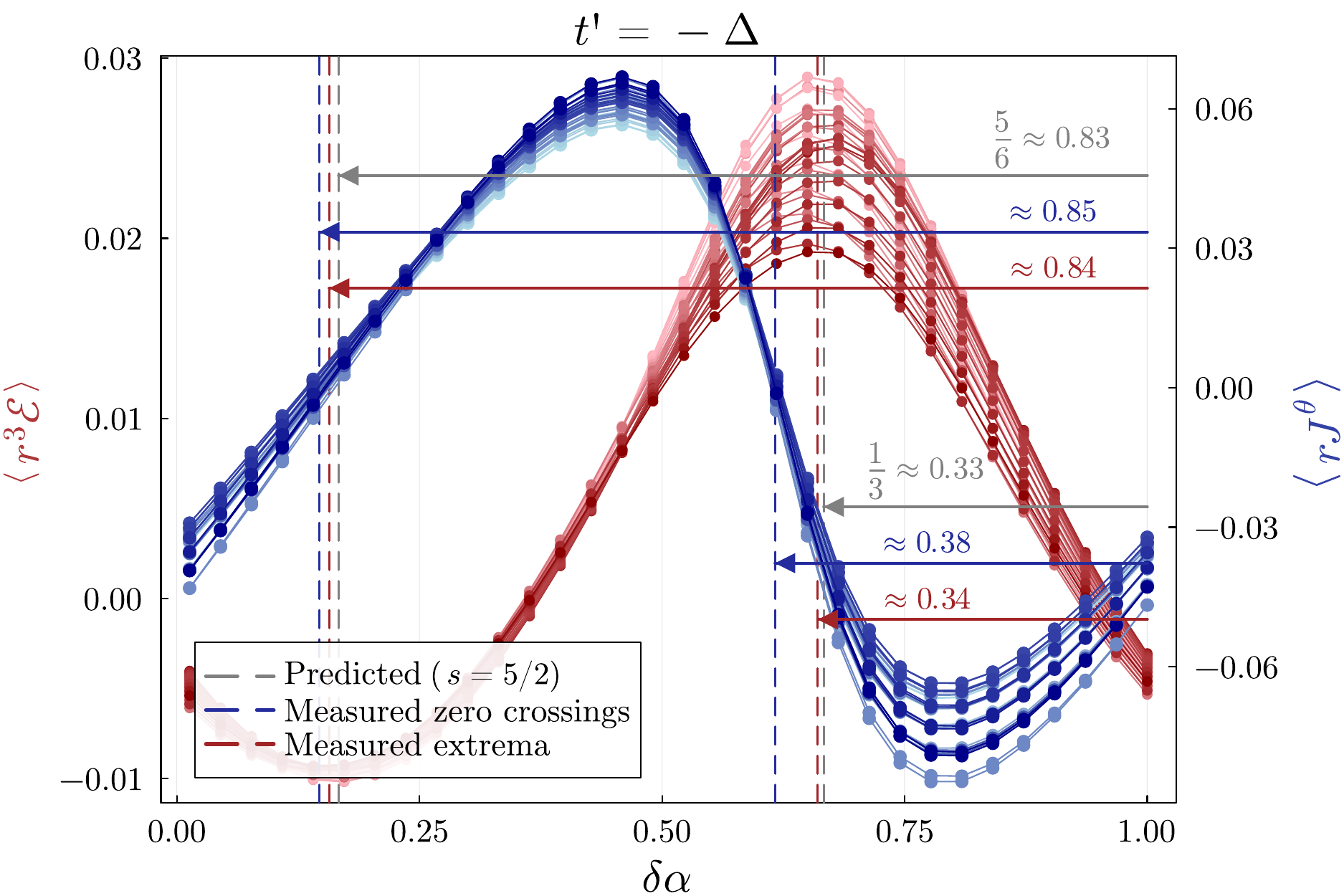}%
            \qquad\includegraphics[width=.45\linewidth]{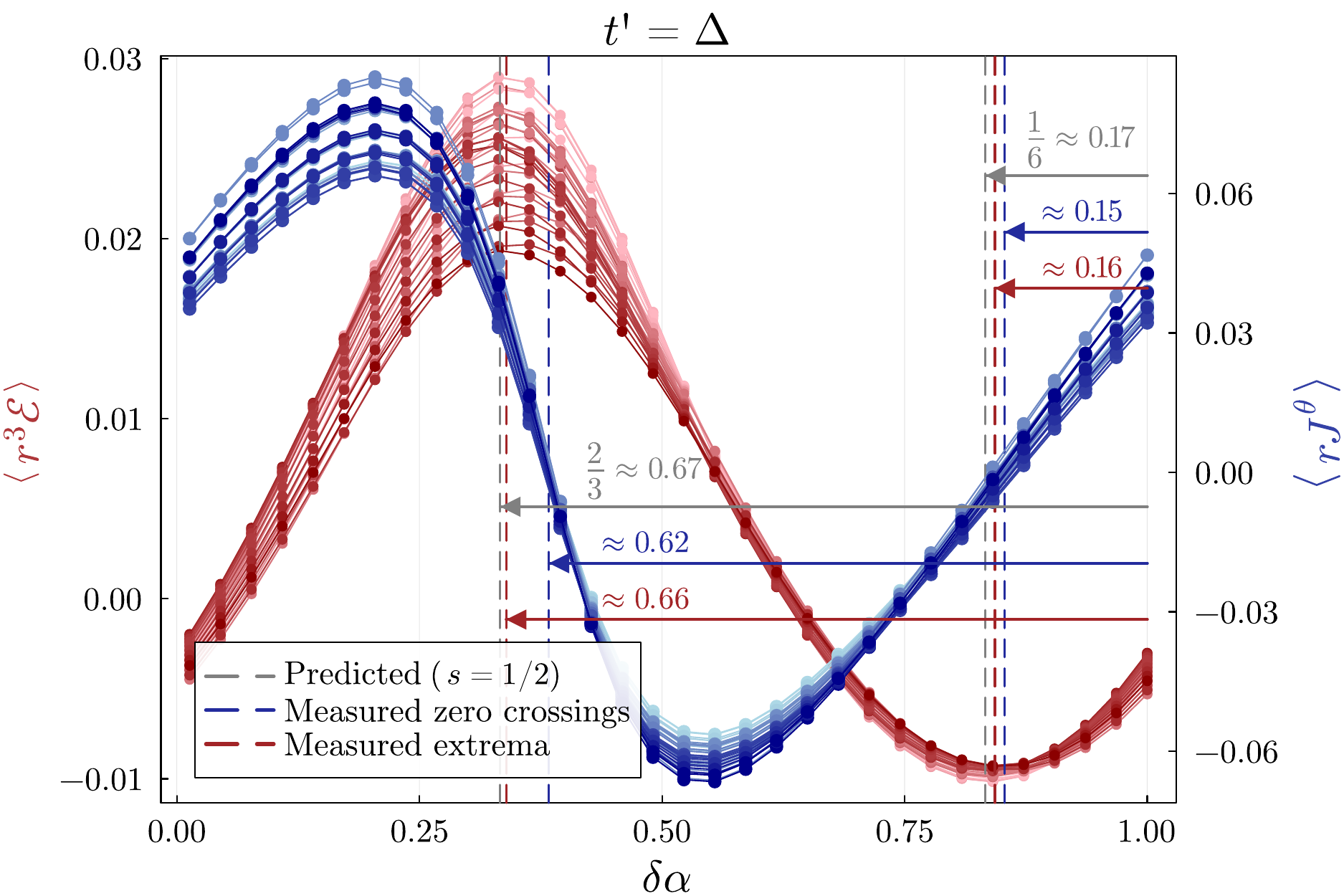}%
        }\\
        \subfloat[Lattice with $N=4356$ sites and a $C$-centered $\beta=2/3$ disclination]{%
        
        \includegraphics[width=.45\linewidth]{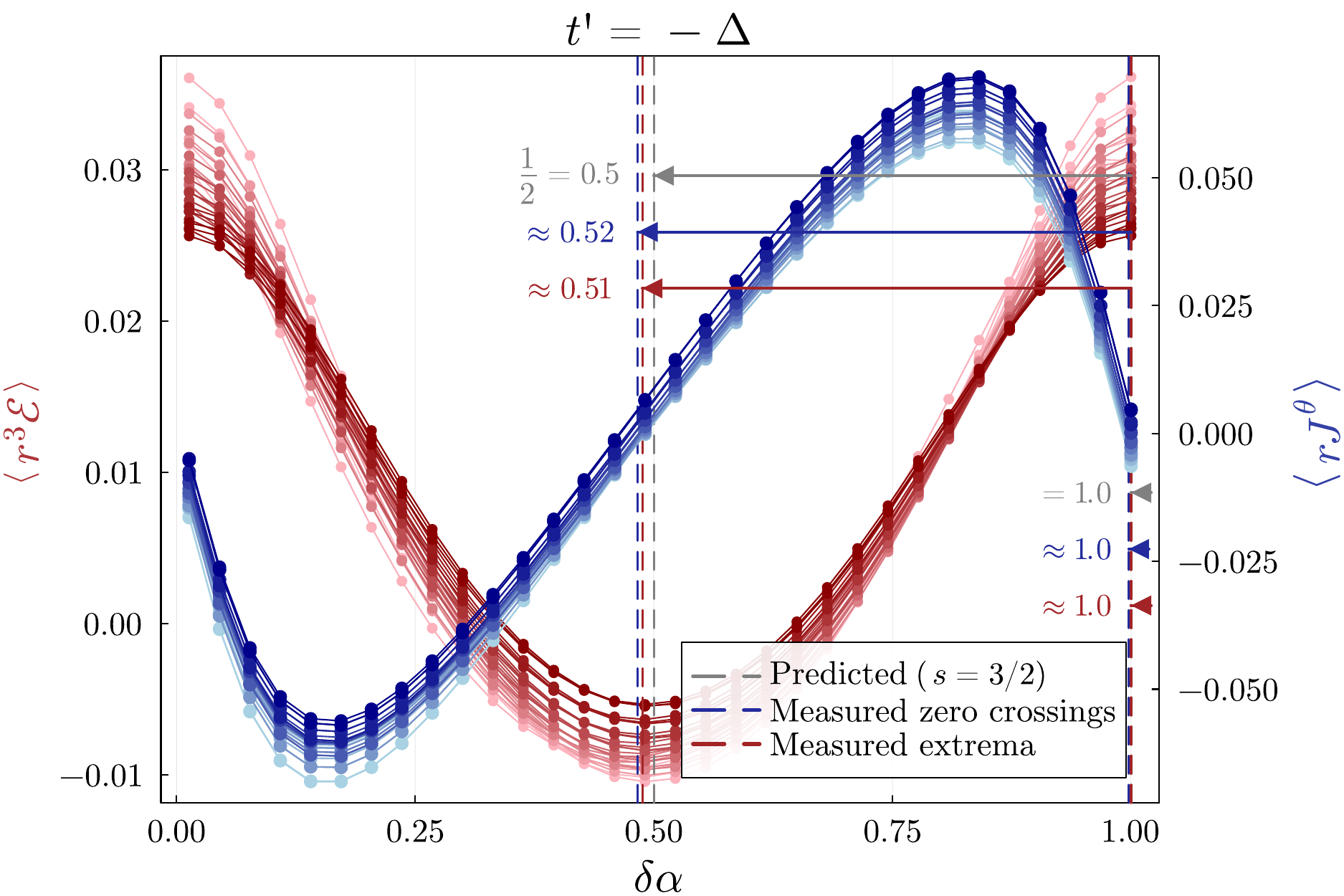}%
            \qquad\includegraphics[width=.45\linewidth]{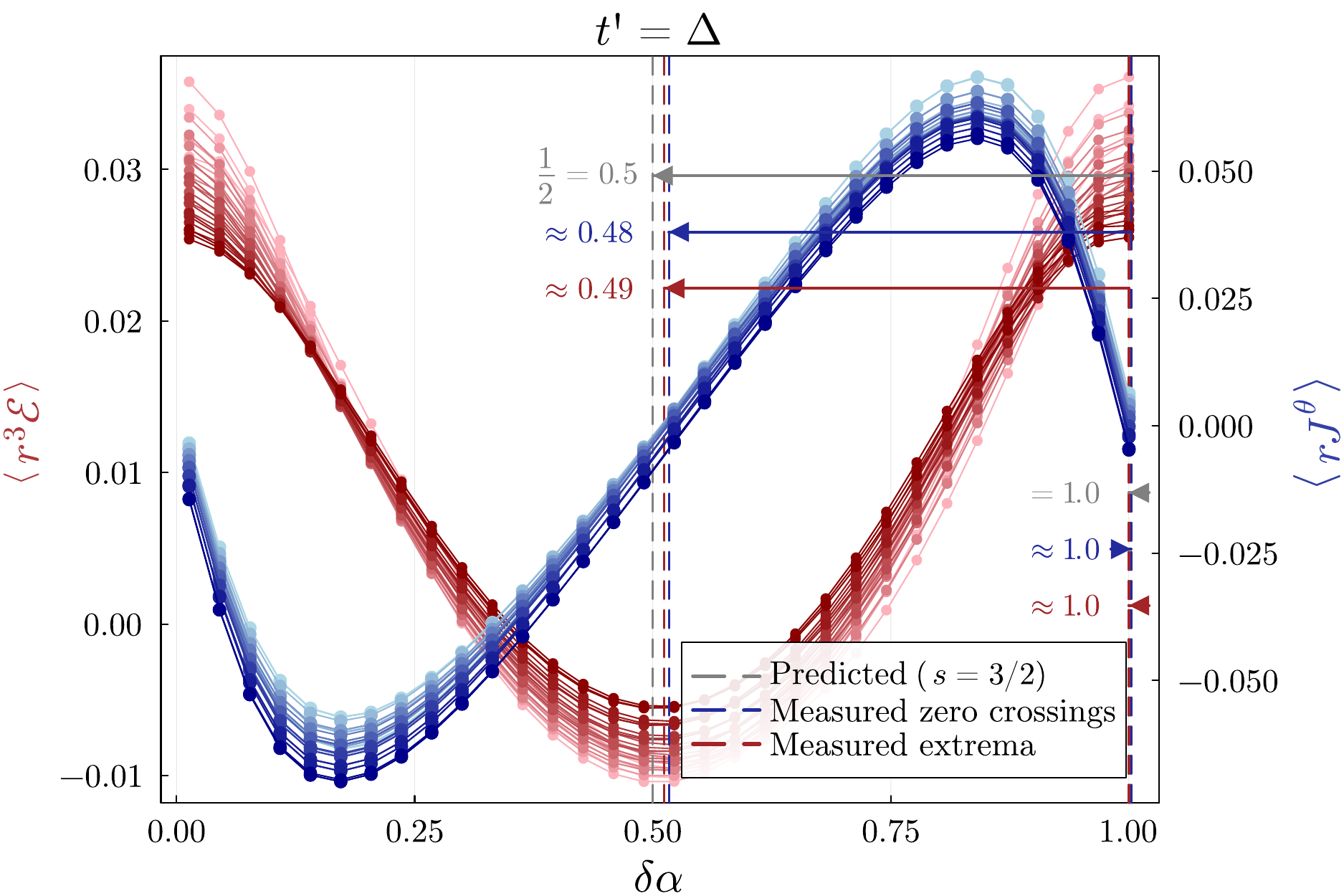}%
        }\\
        \subfloat[Lattice with $N=9408$ sites and a $\vec{b}=(-1,0)$ dislocation]{%
        \includegraphics[width=.45\linewidth]{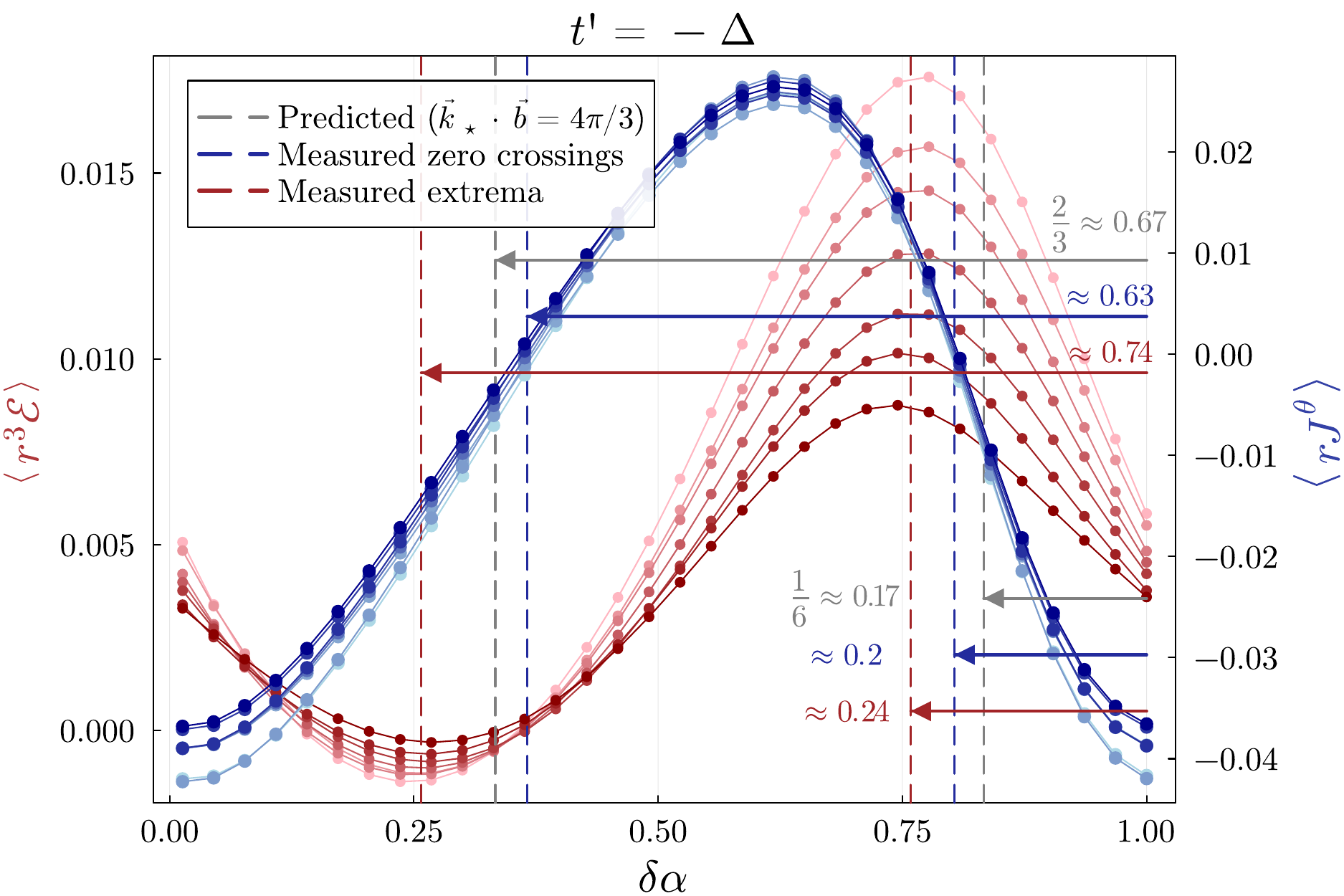}%
            \qquad\includegraphics[width=.45\linewidth]{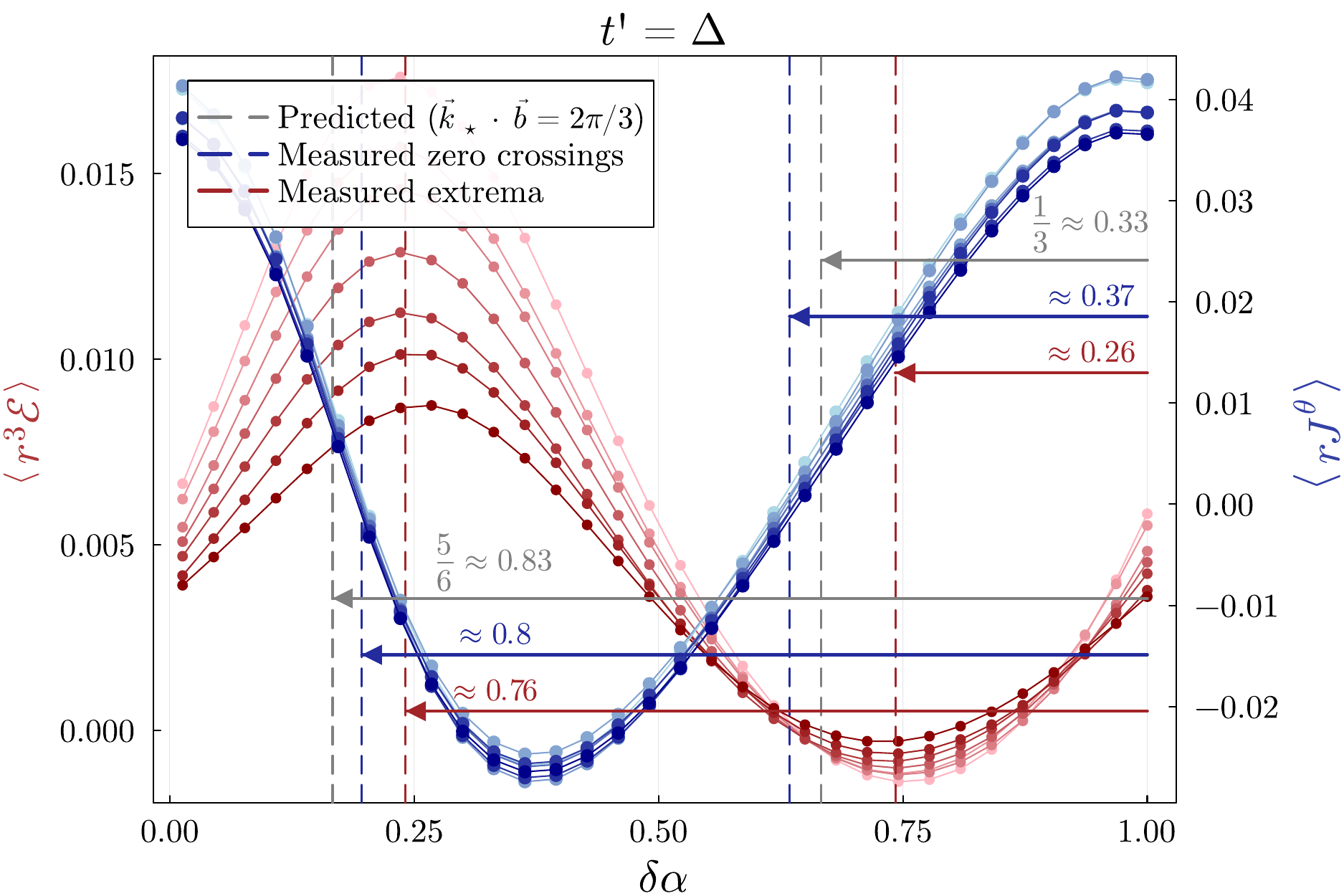}%
        }
        \caption{Haldane model with crystalline defects}
        \label{fig:staggered_honeycomb_currents}
    \end{figure*}

\clearpage

\subsection{Topological Invariants}\label{sec:abs_invariants}
In this section, we present the absolute values of the topological invariants measured through partial rotations.
\begin{table}[h!]
\def\arraystretch{1.3}
    {\begin{tabular}{cccccccccc}
         \hline\hline & $\text{ }$ & $C$ & $\mathscr{S}_A$ & $\mathscr{S}_B$ & $\ell_A$ & $\ell_B$   \\\hline
         $m<-1$ && $0$ & $0$ & $1$ & $0$ & $0$\\
         $-1<m<0$ && $-1$ & $\frac{7}{2}$ & $\frac{1}{2}$ & $\frac{15}{4}$ & $\frac{15}{4}$ \\
         $0<m<1$ && $1$ & $\frac{5}{2}$ & $\frac{3}{2}$ & $\frac{9}{4}$ & $\frac{9}{4}$ \\
         $m>1$ && $0$ & $0$ & $1$ & $0$ & $2$ \\ \hline\hline 
    \end{tabular}}
    \label{}
    \caption{Topological invariants in each phase of the QWZ model discussed in \cref{sec:QWZ_rot_eigs,sec:QWZ_invariants,sec:QWZ_observables}.}
\end{table}

\begin{table}[h!]
\def\arraystretch{1.3}
    \subfloat[No on-site potential]{
    \begin{tabular}{ccccccccccc}
         \hline\hline & $\text{ }$ & $C$ & $\mathscr{S}_A$ & $\mathscr{S}_B$ & $\ell_A$ & $\ell_B$  \\\hline
          $t'<0$ && $1$ & $\frac{3}{2}$ & $\frac{1}{2}$ & $\frac{9}{4}$ & $\frac{1}{4}$  \\
         $t'>0$ && $-1$ & $\frac{7}{2}$ & $\frac{1}{2}$ & $\frac{15}{4}$ & $\frac{15}{4}$  \\ \hline\hline
    \end{tabular}}\hfill
    \subfloat[One-site staggered potential]{
    \begin{tabular}{ccccccccccccccccccccccccccccc}
         \hline\hline & $\text{ }$ & $C$ & $\mathscr{S}_A$ & $\mathscr{S}_B$ & $\ell_A$ & $\ell_B$  \\\hline
          $t'<-\Delta$ && $1$ & $\frac{7}{2}$ & $\frac{1}{2}$ & $\frac{1}{4}$ & $\frac{1}{4}$ \\
          $-\Delta<t'<\Delta$ && $0$ & $0$ & $1$ & $0$ & $0$ \\
         $t'>\Delta$ && $-1$ & $\frac{3}{2}$ & $\frac{1}{2}$ & $\frac{7}{4}$ & $\frac{15}{4}$ \\ \hline\hline
    \end{tabular}}
    \hfill
    \subfloat[Two-site staggered potential]{\begin{tabular}{ccccccccccccccccccccc}
         \hline\hline & $\text{ }$ & $C$ & $\mathscr{S}_A$ & $\mathscr{S}_B$ & $\ell_A$ & $\ell_B$ \\\hline
          $t'<-\Delta$ && $1$ & $\frac{7}{2}$ & $\frac{7}{2}$ & $\frac{1}{4}$ & $\frac{1}{4}$ \\
          $-\Delta<t'<\Delta$ && $0$ & $0$ & $0$ & $0$ & $0$\\
         $t'>\Delta$ && $-1$ & $\frac{5}{2}$ & $\frac{5}{2}$ & $\frac{7}{4}$ & $\frac{7}{4}$\\ \hline\hline 
    \end{tabular}}
    \caption{Topological invariants in each phase of the three $\pi$-flux models discussed in \cref{sec:pi_flux_models}.}
\end{table}

\begin{table}
\def\arraystretch{1.3}
{\begin{tabular}{cccccccccccccccccccccccc}
         \hline\hline & $\text{ }$ & $C$ & $\mathscr{S}_A$ & $\mathscr{S}_B$ & $\mathscr{S}_C$ & $\ell_A$ & $\ell_B$ & $\ell_C$ \\\hline
          $t'<-\Delta$ && $1$ & $\frac{1}{2}$ & $\frac{1}{2}$ & $\frac{3}{2}$ & $\frac{1}{4}$ & $\frac{1}{4}$ & $\frac{9}{4}$ \\
          $-\Delta<t'<\Delta$ && $0$ & $0$ & $1$ & $0$ & $0$ & $0$ & $0$ \\
         $t'>\Delta$ && $-1$ & $\frac{1}{2}$ & $\frac{1}{2}$ & $\frac{3}{2}$ & $\frac{11}{4}$ & $\frac{11}{4}$ & $\frac{3}{4}$ \\ \hline\hline
    \end{tabular}}
    \caption{Topological invariants in each phase of the Haldane model discussed in \cref{sec:honeycomb_models}.}
\end{table}
\raggedbottom

\twocolumngrid

\bibliography{bibliography,chris_references,mb_bib}

\end{document}